\documentclass[sigconf]{acmart}

\usepackage{amsmath,amsfonts}
\usepackage[utf8]{inputenc}
\usepackage[T1]{fontenc}
\usepackage{textcomp}
\usepackage[font=small,skip=0pt]{caption}
\usepackage{threeparttable}
\usepackage{mdframed}
\usepackage{multirow}
\usepackage{todonotes}
\usepackage{enumitem}
\usepackage{ocr}
\usepackage{soul}
\usepackage{xcolor}
\AtBeginDocument{%
  \providecommand\BibTeX{{%
    \normalfont B\kern-0.5em{\scshape i\kern-0.25em b}\kern-0.8em\TeX}}}

\acmConference[ICSE 2022]{The 44th International Conference on Software Engineering}{May 21–29, 2022}{Pittsburgh, PA, USA}
\copyrightyear{2022}
\acmYear{2022}
\setcopyright{rightsretained}
\acmConference[ICSE '22]{44th International Conference on Software Engineering}{May 21--29, 2022}{Pittsburgh, PA, USA}
\acmBooktitle{44th International Conference on Software Engineering (ICSE '22), May 21--29, 2022, Pittsburgh, PA, USA}
\acmDOI{10.1145/3510003.3510043}
\acmISBN{978-1-4503-9221-1/22/05}

\begin{document}

\newcommand\TOOL{{\textbf{GLIMPS}}}
\newcommand\GT{{\textbf{G}}}
\newcommand\LT{{\textbf{L}}}
\newcommand\IT{{\textbf{I}}}
\newcommand\MT{{\textbf{M}}}
\newcommand\PT{{\textbf{P}}}
\newcommand\ST{{\textbf{S}}}

\newcommand\GL{\GT lobal}
\newcommand\LL{\LT ocal}
\newcommand\PIM{performance-\IT nfluence \MT odels}
\newcommand\CP{CPU \PT rofiling}
\newcommand\PS{program \ST licing}

\definecolor{pastelorange}{rgb}{1.0, 0.7, 0.28}
\definecolor{paleplum}{rgb}{0.8, 0.6, 0.8}
\definecolor{teagreen}{rgb}{0.82, 0.94, 0.75}

\colorlet{myyellow}{yellow!20}
\colorlet{myblue}{blue!20}
\colorlet{myred}{red!20}
\colorlet{mygreen}{green!20}

\DeclareRobustCommand{\hlyellow}[1]{{\sethlcolor{myyellow}\hl{#1}}}
\DeclareRobustCommand{\hlblue}[1]{{\sethlcolor{myblue}\hl{#1}}}
\DeclareRobustCommand{\hlred}[1]{{\sethlcolor{myred}\hl{#1}}}
\DeclareRobustCommand{\hlgreen}[1]{{\sethlcolor{mygreen}\hl{#1}}}

\newcommand\IO{{influencing options}}
\newcommand\OH{{option hotspots}}
\newcommand\CC{{cause-effect chain}}
\newcommand\UH{{user hotspots}}

%\newcommand\GC{{\hlblue{\GT lobal performance-influence models}}}
%\newcommand\LC{{\hlblue{\LT ocal performance-influence models}}}
%\newcommand\PC{{\hlblue{CPU \PT rofiling}}}
%\newcommand\SC{{\hlblue{program \ST licing}}}

% \title{On Debugging the Performance of Configurable Systems: User Studies and Tool Support}
%\title{Towards Helping Developers Debug the Performance of Configurable Systems}
%\title{Debugging the Performance of Configurable Systems: Information Needs and Support}
%\title{On Debugging the Performance of Configurable Systems: \\ Which, Where, and How do Options Influence Performance?}
%\title{On Debugging the Performance of Configurable Software Systems: \\ Which, Where, and How do Options Influence Performance?}
%\title{Helping to Answer ``Which, Where, and How Options \\ Influence Performance in Configurable Software Systems?''}
%\title{Satisfying Developers' Needs when Debugging the Performance of Configurable Software System}
%\title{On Performance Debugging of Configurable Software Systems: \\ Developer Needs and Tailored Tool Support}
\title{On Debugging the Performance of Configurable Software Systems: Developer Needs and Tailored Tool Support}

\author{Miguel Velez}
%\email{webmaster@marysville-ohio.com}
\affiliation{
  \institution{Carnegie Mellon University}
%  \streetaddress{P.O. Box 1212}
  \country{}
}

\author{Pooyan Jamshidi}
\affiliation{
  \institution{University of South Carolina}
  \country{}
}

\author{Norbert Siegmund}
\affiliation{
  \institution{Leipzig University}
  \country{}
}

\author{Sven Apel}
\affiliation{
  \institution{Saarland Informatics Campus - Saarland University}
  \country{}
}

\author{Christian Kästner}
\affiliation{
  \institution{Carnegie Mellon University}
  \country{}
}
  
%\author{Lars Th{\o}rv{\"a}ld}
%\affiliation{%
%  \institution{The Th{\o}rv{\"a}ld Group}
%  \streetaddress{1 Th{\o}rv{\"a}ld Circle}
%\email{larst@affiliation.org}

%\affiliation{%
%  \institution{Inria Paris-Rocquencourt}


\begin{abstract}
 Determining whether a configurable software system has a performance bug or
 %the system 
 it was misconfigured is often challenging.
 While there are numerous debugging techniques that can support developers in this task, there is limited empirical evidence of how useful the techniques are to address the actual needs that developers have when debugging the performance of configurable software systems; most techniques are often evaluated in terms of technical accuracy instead of their usability.
 In this paper, we take a human-centered approach to identify, design, implement, and evaluate a solution to support developers in the process of debugging the performance of configurable software systems.
 We first conduct an exploratory study with $19$ developers to identify the information needs that developers have during this process.
 Subsequently, we design and implement a tailored tool, adapting techniques from prior work,
 %building on relevant information provided by \GL\ and \LL\ \PIM, \CP, and \PS, 
 to support those needs.
 Two user studies, with a total of $20$ developers, validate and confirm that the information that we provide helps developers debug the performance of configurable software systems.
 \looseness=-1
\end{abstract}

%\begin{CCSXML}
%<ccs2012>
%  <concept_id>10010520.10010553.10010562</concept_id>
%  <concept_desc>Computer systems organization~Embedded systems</concept_desc>
%  <concept_significance>500</concept_significance>
%  <concept_id>10010520.10010575.10010755</concept_id>
%  <concept_desc>Computer systems organization~Redundancy</concept_desc>
%  <concept_significance>300</concept_significance>
%  <concept_id>10010520.10010553.10010554</concept_id>
%  <concept_desc>Computer systems organization~Robotics</concept_desc>
%  <concept_significance>100</concept_significance>
%  <concept_id>10003033.10003083.10003095</concept_id>
%  <concept_desc>Networks~Network reliability</concept_desc>
%  <concept_significance>100</concept_significance>
%</ccs2012>
%\end{CCSXML}

%causal chainsdesc[500]{Computer systems organization~Embedded systems}
%causal chainsdesc[300]{Computer systems organization~Redundancy}
%causal chainsdesc{Computer systems organization~Robotics}
%causal chainsdesc[100]{Networks~Network reliability}

%%
%% Keywords. The author\ST  should pick words that accurately describe
%\keywords{datasets, neural networks, gaze detection, text tagging}

\maketitle

\section{Introduction}

\begin{figure}[t]
\begin{center}
    \includegraphics[width=\columnwidth]{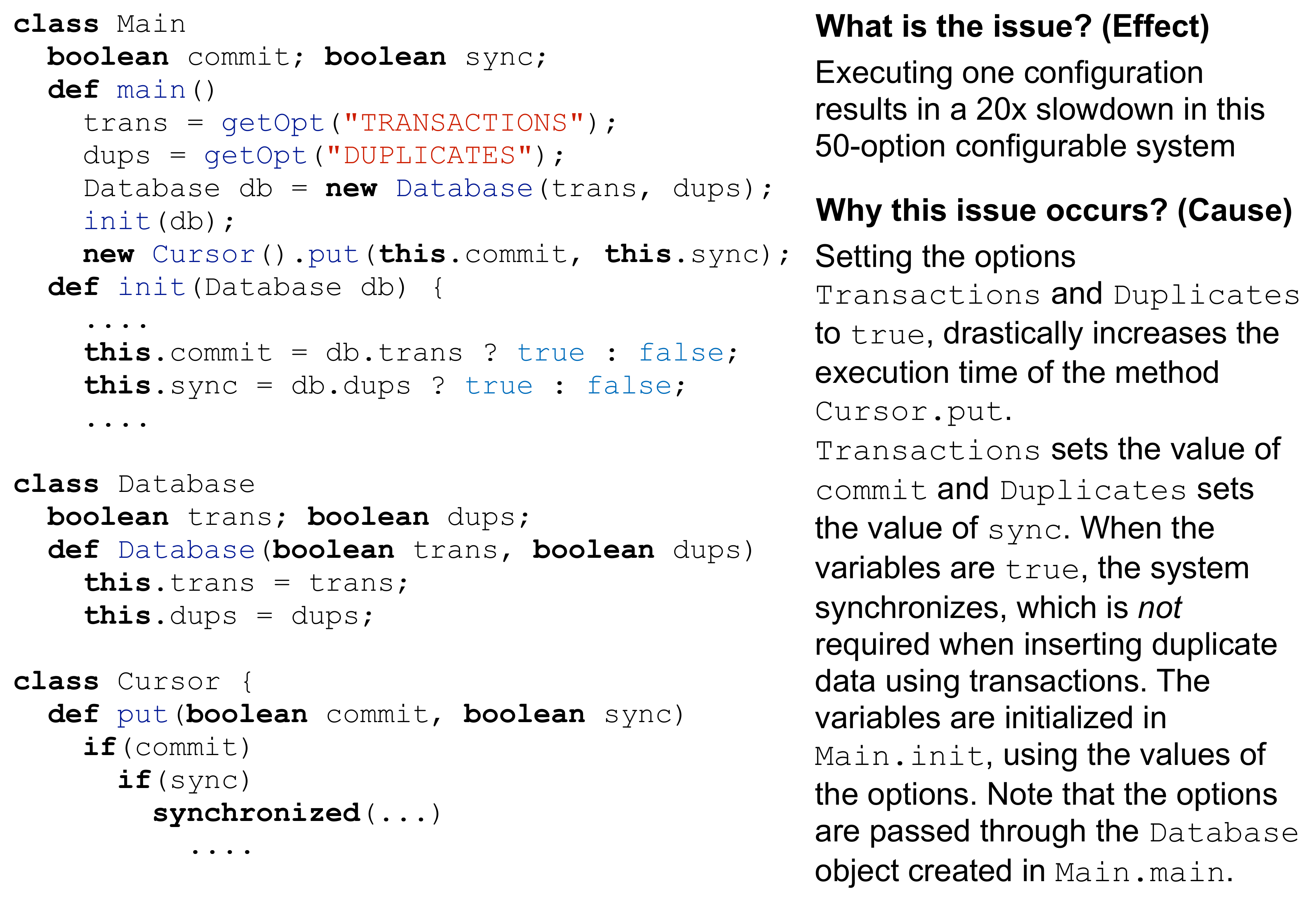}
\end{center}
\caption{
Artificial example
%of a system 
contrasting effects and causes when debugging the performance of configurable software systems.
\looseness=-1
}
\label{cause-effect-fig}
\end{figure}

Developers often spend a substantial amount of time diagnosing a configurable software system to localize and fix a performance bug, or to determine that the system was misconfigured~\cite{Z:WPF09, HY:ESEM16, PKRB:MSR12, PO:ISSTA11, JAH:OOPSLA11, CLZMDMBN:ESECFSE17, BPSZ:CSCW10, JSSSL:PLDI12, NJT:MSR13, IKJRK:EUROSYS22}.
This struggle is quite common when maintaining configurable software systems.
Some empirical studies find that $59$~percent of performance issues are related to configuration errors, $88$~percent of these issues require fixing the code~\cite{HY:ESEM16, HYL:ASE18}, of which $61$~percent take an average of $5$ weeks to fix~\cite{KIJRJ:CADET20}, and that $50$~percent of patches in open-source cloud systems and $30$~percent of questions in forums are related to configurations~\cite{WLHLSK:ASPLOS18}.  
Regardless of how developers find the root cause of the issue or misconfiguration, performance issues impair user experience, which often result in long execution times or increased energy consumption~\cite{HY:ESEM16, HJLXYYWL:ASE20, JSSSL:PLDI12, SL:ICSE17, LLGH:ICSE16, WRGPA:GCC13, IKJRK:EUROSYS22}.
%One of the primary challenges of this task is identifying and tracking how options and interactions are directly and indirectly used in the implementation~\cite{ABKS:FOSPL13, XZHZSYZP:SOSP13,XJFZPT:ESECFSE15, LKB:TSE18, MWKTS:ASE16, WMLK:OOPSLA18} and impact performance~\cite{VJSAK:ICSE21, VJSSAK:ASEJ20, WAS:ICSE21, HYP:ICPE21}.
\looseness=-1

When performance issues occur, developers typically use profilers to identify the locations of performance bottlenecks~\cite{CAPPJ:TACO15, YP:EMSE18, CLBKMG:ICSECP18, G:CAMC16, CB:ATC16}.
Unfortunately, locations where a system spends the most time executing are not necessarily the sign of a performance issue.
%For example, enabling encryption would likely slow down some components of the system.
%For example, some behavior that takes a long time to execute was enabled.
%and do not provide any information about how the issue is related to configurations.
%although profilers accurately report where a system spends the most time executing, code that runs for a long time is not necessarily the sign of a performance bug.
Additionally, 
traditional profilers only indicate the 
locations of the \emph{effect} of performance issues (i.e., where a system spends the most time executing) for one configuration at a time.
Developers are left to inspect the code to analyze the \emph{root cause} of the performance issues and to determine how the issues relate to configurations.
%(i.e., locations in the code where the \emph{effect} of the unexpected performance behavior is observed).
%In the context of configurable software systems, this information is often not sufficient to find the \emph{root cause} of an issue, as options are typically not directly used where they influence the performance of a system~\cite{VJSSAK:ASEJ20, VJSAK:ICSE21}.
%In these situations, researchers often suggest to use techniques to identify inefficient code patterns that cause performance slowdowns~\cite{NSML:ICSE13, SL:ICSE17, NCRL:ICSE15, BT:OOPSLA18}.
%However, there is limited empirical evidence of the extent that these techniques help developers debug in the scenarios that we target: the only visible \emph{effect} of a performance issue is an unexpected long execution time and developers need to debug the system to determine whether there is a performance bug or the system was misconfigured.
\looseness=-1

With an example scenario in Figure~\ref{cause-effect-fig}, we illustrate the kind of performance challenge that developers may face in configurable software systems and that we seek to support:
%in configurable software systems.
A user executes a configuration of this system with $50$ configuration options, which results in an \emph{unexpected} $20\times$ slowdown.
The \emph{only visible effect} 
%in this scenario 
is the excessive execution time.
While, in some situations, developers might be able to change some options to work around the problem, users might not know which options cause the problem, and may want to select certain options to satisfy specific needs (e.g., enable encryption, use a specific transformation algorithm, or set a specific cache size). 
In these situations, developers need to determine whether the system has a potential \emph{bug}, is \emph{misconfigured}, or \emph{works as intended}
%, but the user has a \emph{different expectation} about what the performance of the system should be under the configuration that they execute.
To determine the cause of the potentially problematic performance behavior, developers would need to debug the system and, most likely, the \emph{implementation} to identify which options or interactions in this configuration of $50$ options are the \emph{root cause} of the \emph{unexpected performance behavior} (e.g., the system works as expected, setting a specific \textsc{Cache\_Size} results in a misconfiguration, or setting the options \textsc{Transactions} and \textsc{Duplicates} to \texttt{true} results in a bug).
\looseness=-1

When performance issues such as in our example occur, there are numerous techniques that developers could use to determine whether there is a performance bug
%in a configurable software system 
or the system was misconfigured.
In addition to off-the-shelf profilers~\cite{JPROFILER10, NS:PLDI07, VVM}, developers could use more targeted profiling techniques~\cite{CB:ATC16, ALZM:EUROSYS17, YP:ISSTA16, YP:EMSE18, CAPPJ:TACO15}, visualize performance behavior~\cite{G:CAMC16, BPG:SANER15, SBB:VISSOFT19, CLBKMG:ICSECP18, AH:SOFTVIS10, TDT:ICPC13}, search for inefficient coding patterns~\cite{NSML:ICSE13, SL:ICSE17, NCRL:ICSE15, BT:OOPSLA18, LXC:ICSE14}, use information-flow analyses~\cite{ZE:ICSE14,  TD:ECOOP16, LKB:TSE18, XJHZLJP:OSDI16, WMLK:OOPSLA18, MWKTS:ASE16, LWHL:EUROSYS20}, or model the performance of the systems in terms of its options and interactions~\cite{SGAK:ESECFSE15, HZ:ICSME19, VJSSAK:ASEJ20, KSKGA:SOSYM18, VJSAK:ICSE21, WAS:ICSE21}.
Likewise, developers could use established program debugging techniques, such as delta debugging~\citep{Z:SIGSOFTNotes99}, program slicing~\citep{W:ICSE81, AH:ASP90, KL:IPL88}, and statistical debugging~\cite{AMLZ:ECML07, SL:OOPSLA14} for some part of the debugging process.
One reason why we cannot reliably suggest developers to use any of the above techniques is that there is limited empirical evidence of how useful the techniques are to help developers debug the performance of configurable software systems; the techniques typically solve a specific technical challenge that is usually evaluated in terms of accuracy, not usability~\cite{PO:ISSTA11}.
Hence, we could only, at best, speculate which techniques might support developers' needs to debug unexpected performance behaviors in configurable software systems.
%determine whether there is a performance bug in a configurable software system or the system was misconfigured.
\looseness=-1

%performance optimization~\cite{OBMS:ESECFSE17, ZLGBMLSY:SOCC17, GCASW:ASE13, NMSA:ESECFSE17}

%~\cite{HJLXYYWL:ASE20}

%One of the reasons that existing techniques are not fully applicable to situations such as Figure~\ref{cause-effect-fig} is that we have limited evidence about the information that developers need and how they approach the task of debugging the performance of configurable software systems; particularly, in situations in which developers do not even know, to begin with, which options are causing an unexpected performance behavior. 
%\looseness=-1

%In this paper, we \emph{identify the existing performance debugging concepts that can support developers' needs when debugging the performance of configurable systems}.
%In this paper, we \emph{identify how to support the information needs that developers have and overcome the barriers that they face when debugging the performance of configurable software systems}.

\begin{figure}[t]
\begin{center}
    \includegraphics[width=\columnwidth]{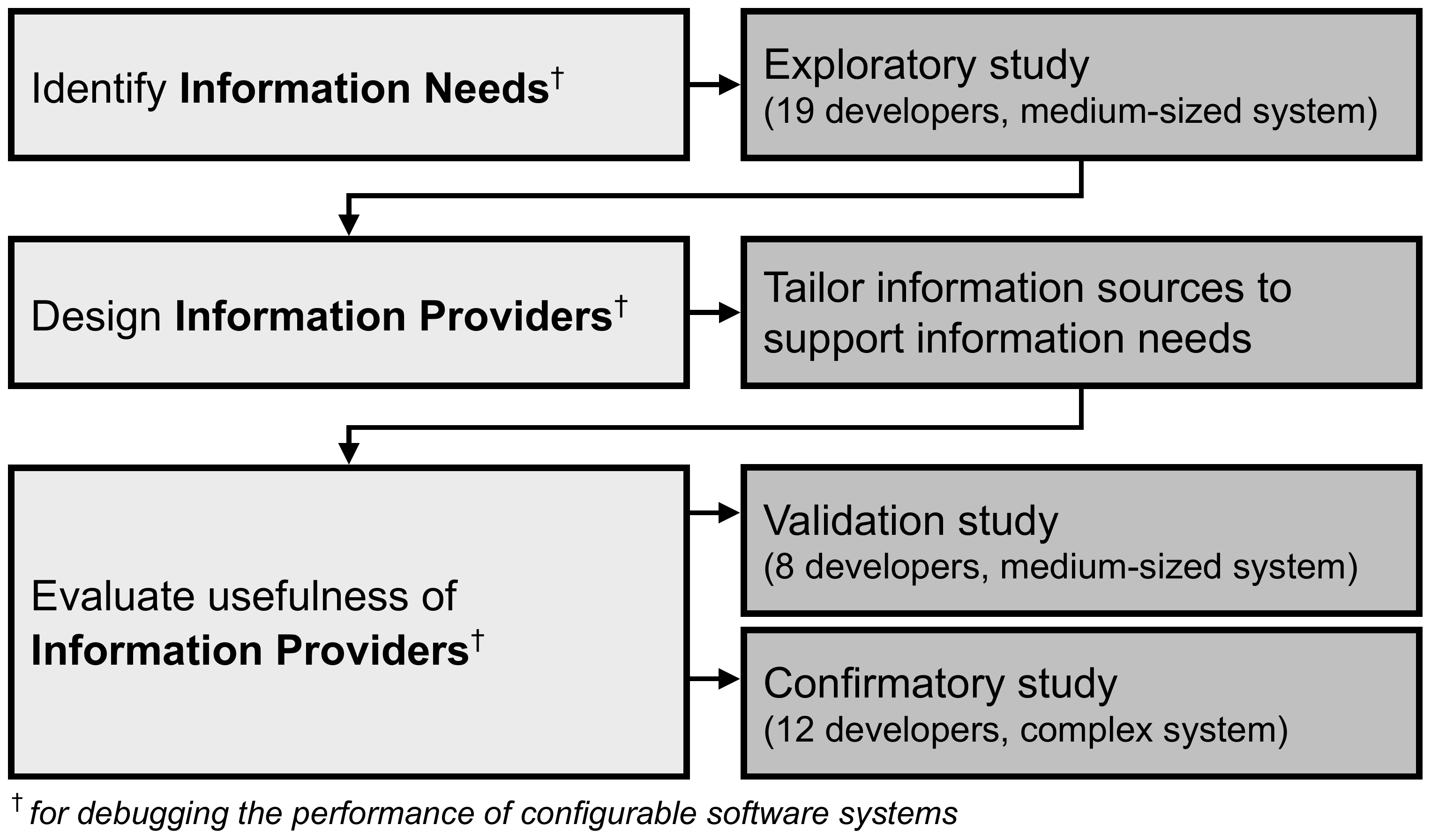}
\end{center}
\caption{
Overview of our human-centered approach to 
%explore and 
%provide
support 
%for 
the information needs that developers have when debugging the performance of configurable software systems.
\looseness=-1
}
\label{overview-fig}
\end{figure}

In this paper, we take a human-centered approach~\cite{MKLY:C16, FRJ:CCODE19} to \emph{identify}, \emph{design}, \emph{implement}, and \emph{evaluate} a solution to support developers in the process of debugging the performance of configurable software systems; particularly, in situations such as our example in Figure~\ref{cause-effect-fig}.
%, in which developers do not even know, to begin with, which options are causing an unexpected performance behavior. 
Our human-centered research design consists of three steps, summarized in Figure~\ref{overview-fig}:
We first conduct an \emph{exploratory user study} to \emph{identify the information needs} that developers have 
%and the barriers that they face 
when debugging the performance of configurable software systems.
Our 
%exploratory user 
study reveals that developers \emph{struggle} to find relevant information to 
(a)~identify \hlyellow{influencing options}; the options or interactions \emph{causing} an unexpected performance behavior, 
(b)~locate \hlblue{option hotspots}; the methods \emph{where} options affect the performance of the system, and 
(c)~trace the \hlred{cause-effect chain}; \emph{how} \IO\ are used in the implementation to directly and indirectly affect the performance of \OH.
Subsequently, we \emph{design} and \emph{implement information providers} to support developers' needs, adapting and tailoring global and local performance-influence models, CPU profiling, and program slicing,
%\GL\ and \LL\ \PIM, \CP, and \PS, 
in a tool called \TOOL.
Finally, we conduct 
two user studies to \emph{validate} and \emph{confirm} that the designed information providers are useful to developers when debugging the performance of complex configurable software systems, in terms of supporting their information needs and speeding up the process.
%conduct two user studies to evaluate the usefulness of the designed information studies; the first study to \emph{validate} that the designed information providers can support the information needs that we identified in our exploratory study, and the second study to \emph{confirm} that the designed information providers can potentially generalize to support the information needs that developers have when debugging the performance of complex configurable software systems.
\looseness=-1

In summary, we make the following contributions:
\begin{itemize}[noitemsep,topsep=0pt]
\item The information needs -- \IO, \OH, and \CC\ -- that developers have when debugging the performance of configurable software systems; 
%We identify these information needs using an exploratory 
%user 
%study with $19$ participants.
\looseness=-1
\item The design of information providers, adapted from global and local performance-influence models, CPU profiling, and program slicing,
%\GL\ and \LL\ \PIM, \CP, and \PS, 
to support
%developers' information
the above needs;
%the information needs that developers have when debugging the performance of configurable software systems.
\looseness=-1
\item Two empirical evaluations to demonstrate the usefulness of the designed information providers;
\looseness=-1
\item A prototype tool, \TOOL, that implements the designed information providers to help developers debug
the performance of configurable software systems.
\looseness=-1
\end{itemize}

\section{Performance Debugging in Configurable Software Systems}
%\section{Related Work}

There is substantial literature on debugging the performance of software systems~\cite[e.g.,][]{HDGZX:ICSE12, NSML:ICSE13, SL:OOPSLA14, LXC:ICSE14, HJLXYYWL:ASE20}.
Our goal is to support developers 
in the process of debugging 
%debug 
the performance of configurable software systems; 
%particularly,
in particular, 
when developers do not even know which options or interactions in their 
current 
configuration cause an unexpected performance behavior.
\looseness=-1

When performance issues occur in software systems, developers need to 
%analyze and 
identify relevant information to debug the unexpected performance behaviors~\cite{BPSZ:CSCW10, HYL:ASE18, NJT:MSR13, CLZMDMBN:ESECFSE17}.
%of a system~\cite{BPSZ:CSCW10, HYL:ASE18, NJT:MSR13, CLZMDMBN:ESECFSE17}.
%There are numerous techniques that developers could use for this task.
%For instance, 
For this task, in addition to using off-the-shelf profilers~\cite{JPROFILER10, NS:PLDI07, VVM}, some researchers suggest using more targeted profiling techniques~\cite{CAPPJ:TACO15, YP:EMSE18, CLBKMG:ICSECP18, G:CAMC16, CB:ATC16}
and 
%some of which provide 
%different 
visualizations~\cite{G:CAMC16, BPG:SANER15, SBB:VISSOFT19, CLBKMG:ICSECP18, AH:SOFTVIS10, TDT:ICPC13} to 
identify and 
analyze the locations of performance bottlenecks.
Alternatively, some researchers 
%argue that developers could use 
%argue
suggest
using techniques to 
search 
%searching 
for inefficient coding patterns~\cite{NSML:ICSE13, SL:ICSE17, NCRL:ICSE15, BT:OOPSLA18, LXC:ICSE14, CB:ATC16}.
%While the above techniques have been evaluated in terms of their technical accuracy, 
While these techniques are quite useful, there is limited evidence of their usefulness when debugging the performance of configurable software systems; particularly, to determine how performance issues are related to options and their interactions.
\looseness=-1

In addition to 
%specific 
performance debugging techniques, there are several
%While there are several  established program debugging techniques that might also be useful in the context of configurable systems, there is limited empirical evidence of how useful they actually are for helping developers debug the performance of configurable software systems.
%For instance, there are several 
established \emph{program debugging techniques}, such as delta debugging~\citep{Z:SIGSOFTNotes99}, program slicing~\citep{W:ICSE81, AH:ASP90, KL:IPL88}, and statistical debugging~\cite{AMLZ:ECML07, SL:OOPSLA14}, that can help developers 
%narrow down 
%and 
isolate 
relevant parts of a system to focus their debugging efforts.
For 
%these reasons, 
this reason, the techniques have been implemented in the backend of tools to help developers debug~\citep{KM:CHI04, LM:VLHCC11, BBKE:UIST13, PTP:OOPSLA07}.
For example, Whyline~\citep{KM:CHI04} 
combines static and dynamic 
%uses 
program slicing to allow developers to ask 
%``why did'' and why did not" 
questions 
about a system's output.
%Similarly, Reacher~\citep{LM:VLHCC11} uses static data-flow analysis to help developers answer reachability questions.
%%as they navigate call graphs.
While these 
%human-centered 
tools have been 
%designed
%, implemented, 
%and 
evaluated in terms of their technical accuracy and some also in terms of usability with user studies, there is limited evidence of which and how 
%these tools and, in particular, 
these \emph{program debugging techniques} can be used or adopted for debugging the performance of configurable software systems.
%have not been evaluated in terms of their usefulness in debugging performance; specially, in configurable software systems.
\looseness=-1

\paragraph*{Performance issues in configurable software systems:}

Performance issues are often caused by misconfigurations or software bugs, both of which impair user experience.
Misconfigurations are errors in which the system and the input are correct, but the system does not behave as desired, because the \emph{user}-selected configuration is inconsistent or does not match the intended  behavior~\cite{XZHZSYZP:SOSP13, ZTZGBXZ:ASPLOS14, ZE:ISSTA15}.
By contrast, a software bug is a programming error by a \emph{developer} that degrades a system's behavior or functionality~\cite{Z:WPF09, Z:SIGSOFTNotes99, AMLZ:ECML07}.
Regardless of the root cause of the unexpected behavior (misconfigurations or software bugs), systems often misbehave with similar symptoms, such as crashes,
%missing functionality, 
incorrect results~\cite{Z:SIGSOFTNotes99, Z:WPF09, AMLZ:ECML07, MWKS:ARXIV18}, and, in terms of performance, long execution times or increased energy consumption~\cite{HY:ESEM16, HJLXYYWL:ASE20, JSSSL:PLDI12, SL:ICSE17, LLGH:ICSE16, WRGPA:GCC13, IKJRK:EUROSYS22}.
\looseness=-1

Research has repeatedly found that configuration-related performance issues are  common and complex to fix
%~\cite{HY:ESEM16, HYL:ASE18, HJLXYYWL:ASE20, WLHLSK:ASPLOS18, KIJRJ:CADET20}.
in software systems~\cite{HY:ESEM16, HYL:ASE18, HJLXYYWL:ASE20, WLHLSK:ASPLOS18, KIJRJ:CADET20}.
For example, \citet{HY:ESEM16} found that $59$~percent of performance issues 
%are related to 
concern configurations, $72$~percent of 
%these issues involved 
which involved one option, and $88$~percent 
%of all performance issues 
require fixing the code.
Similarly, \citet{KIJRJ:CADET20} found that $61$~percent of performance issues take an average of $5$ weeks to fix, and \citet{WLHLSK:ASPLOS18} found that $50$~percent of patches in open-source cloud systems and $30$~percent of questions in forums are related to configurations and performance issues.  
%For example, \citet{HY:ESEM16} found that $59\%$ of performance bugs are related to configuration issues and $72\%$ of these bugs involved one option.
Performance issues were primarily caused by incorrect implementation of configurations,
synchronization issues, 
and misconfigurations.
%In $88\%$ of these bugs, developers needed to change the implementation to fix the performance issues.
\looseness=-1

\paragraph*{Debugging performance in configurable software systems:}

Similar to debugging performance in 
%software systems, 
general,
identifying relevant information is key when debugging unexpected performance behaviors in configurable software systems.
%Incomplete information to make debugging the performance of any software system a challenging task.
%When performance issues occur, developers need to 
%%analyze and 
%identify relevant information to debug the unexpected performance behavior of a system~\cite{BPSZ:CSCW10, HYL:ASE18, NJT:MSR13, CLZMDMBN:ESECFSE17}.
%Having relevant information helps developers debug performance issues in configurable software systems, or more generally, When performance issues occur in configurable software systems,  issue occur, including performance issues, in any system, including configurable software systems, 
%Relevant infomration is needed to help developers debug any type of iusse, not only performance issues, Lacking relevant information increases the challenge of debugging any software system~\cite{BPSZ:CSCW10, HYL:ASE18, NJT:MSR13, CLZMDMBN:ESECFSE17}.
Ideally, %In configurable software systems, 
developers would 
%ideally 
have relevant information to 
debug 
%determine 
how 
%unexpected performance behaviors 
performance issues
are related to specific options and their interactions. % of the system.
%the option or interaction that cause the problematic performance behavior to narrow down the search space that they need to focus on for debugging.
%Unfortunately, developers often only know the \emph{effect} of the performance issue (e.g., long execution times) and some contextual information to reproduce the issue (e.g., hardware, inputs, and configuration).
Unfortunately, there are 
%some 
situations in which developers only know the \emph{effect} of an 
unexpected performance behavior
%performance issue 
(e.g., a long execution time as in Figure~\ref{cause-effect-fig}).
In these situations, developers need to debug the system to determine whether the system has a potential \emph{bug}, is \emph{misconfigured}, or \emph{works correctly}, but the user has a \emph{different expectation} about the performance behavior of the system.
%This limited information complicates the developers' debugging process who need to find the root cause of how configurations are cause a problematic performance behavior.
%For these reasons, we seek to support developers in the process of debugging when they have limited information about the cause a problematic performance behavior.
For these reasons, our goal is to support developers in \emph{finding relevant information} to debug unexpected performance behaviors in configurable software systems.
\looseness=-1

%While there is limited empirical evidence that performance and established program debugging techniques can support the needs that developers have when debugging the performance of configurable systems, 
%In the context of configurable software systems, 
There are 
some 
%areas of research 
research areas
that 
aim to 
%can
help developers understand how options
%and their interactions 
affect a system's behavior.
%, including performance.
%\looseness=-1
%In our context,
%For these situations, 
For instance, 
%several 
some researchers argue that information-flow analyses can help developers understand how options
%and their interactions 
affect a system's behavior~\cite{ZE:ICSE14, TD:ECOOP16, LKB:TSE18, XJHZLJP:OSDI16, WMLK:OOPSLA18, MWKTS:ASE16, LWHL:EUROSYS20}.
For example, Lotrack~\cite{LKB:TSE18} used 
static 
taint analysis to identify under which configurations 
particular 
code fragments are executed.
%Likewise, Staccato~\cite{TD:ECOOP16} used dynamic taint analysis to identify the use of stale configuration data.
While these techniques 
%have shown that they 
can solve specific 
%technical 
challenges, there is limited 
%empirical 
evidence of
the usefulness of
%how useful 
these techniques,
%are for developers; 
%in particular, 
particularly, 
for debugging the performance of configurable software systems.
\looseness=-1

%Likewise, Toddler~\citep{NSML:ICSE13} detects performance bugs by identifying repetitive memory read sequences across loop iterations.
%Furthermore, Coz~\citep{CB:ATC16} introduced causal profiling to help developers identify which components in their concurrent system they should optimize to improve performance.
%These techniques, however, do not consider configurability or how options might cause problematic performance behaviors.
%\looseness=-1

%Recent research has focused on specific techniques for debugging the performance of configurable software systems~\cite{LWHL:EUROSYS20, HJLXYYWL:ASE20}.
%For instance, \citet{HJLXYYWL:ASE20} suggested using developers' expected performance behavior of individual and pairs of options as a testing oracle for identifying the incorrect implementation of configurations.
%This research, however, assumes that developers know the options that are causing a problematic performance behavior; in contrast to situations when developers do not even know this information to begin with.
%\looseness=-1

%\paragraph*{Performance-influence modeling:}

%More closely related to understand the performance behavior of configurable software systems
In terms of understanding performance, some researchers suggest that performance-influence models can help developers debug unexpected performance behaviors~\cite{SGAK:ESECFSE15, HZ:ICSME19, VJSSAK:ASEJ20, KSKGA:SOSYM18, VJSAK:ICSE21, WAS:ICSE21}, as the models describe a system's performance in terms of its options and their interactions. %influence the 
%%end-to-end 
%performance of a system.
For example, the model $8 \cdot \textsc{A} \cdot \textsc{B} + 5 \cdot \textsc{C}$ explains the influence of the options \textsc{A}, \textsc{B}, and \textsc{C}, and their interactions on the performance of a system; in this example, enabling \textsc{C} increases the execution time
%of the system 
by $5$ seconds, and enabling \textsc{A} and \textsc{B}, \emph{together}, further increase the execution time by $8$ seconds.
%The model also shows that other options do not influence the performance of the system.
%Additionally, 
%recent 
Some approaches 
%to build 
%performance-influence 
%these models first 
also model the performance of individual methods~\cite{VJSAK:ICSE21, WAS:ICSE21, VJSSAK:ASEJ20, HYP:ICPE21},
%in terms of options and interactions~\cite{VJSAK:ICSE21, WAS:ICSE21, VJSSAK:ASEJ20, HYP:ICPE21}.
%Researchers argue that such \emph{local} models 
%for individual methods 
which
can be useful to locate \emph{where} options affect a system's performance.
However, while performance-influence models have been evaluated in terms of  accuracy~\cite{SKKABRS:ICSE12, SRKKAS:SQJ12, SGAK:ESECFSE15, KGSGA:ICSE19, VJSAK:ICSE21, VJSSAK:ASEJ20} and optimizing performance~\cite{OBMS:ESECFSE17, ZLGBMLSY:SOCC17, GCASW:ASE13, NMSA:ESECFSE17}, they have not been evaluated in terms of usability; in particular, to support developers' needs when debugging the performance of configurable software systems.
\looseness=-1

%While there are several  established program debugging techniques that might also be useful in the context of configurable software systems, there is limited empirical evidence of how useful they actually are for helping developers debug the performance of configurable software systems.
%For instance, there are several established program debugging techniques, such as delta debugging~\citep{Z:SIGSOFTNotes99}, program slicing~\citep{W:ICSE81, AH:ASP90, KL:IPL88}, and statistical debugging~\cite{AMLZ:ECML07, SL:OOPSLA14}, that can help developers narrow down and isolate relevant parts of a system to focus their debugging efforts.
%For these reasons, the techniques have been implemented in the backend of tools to help developers debug~\citep{KM:CHI04, LM:VLHCC11, BBKE:UIST13, PTP:OOPSLA07}.
%For example, the Whyline~\citep{KM:CHI04} combines static and dynamic slicing to allow developers to ask "why did " and "why did not" questions directly about a system's output.
%Similarly, Reacher~\citep{LM:VLHCC11} uses static data-flow analysis to help developers answer reachability questions as they navigate call graphs.
%While these human-centered tools have been designed, implemented, and evaluated in terms of their technical accuracy and usability, with numerous user studies, these tools and, in particular, the techniques have not been evaluated in terms of their usefulness in debugging the performance of configurable software systems.
%\looseness=-1

% \paragraph*{\nopunct}
\paragraph*{Contributions:}

%In this paper, we identify the existing performance debugging concepts that can support developers' needs when debugging the performance of configurable software system. 
%In this paper, we 
We take a human-centered approach~\cite{MKLY:C16, FRJ:CCODE19} to identify, design, implement, and evaluate a solution to support developers %in the process of 
when debugging the performance of configurable software systems.
%Our human-centered research design consists of three steps, summarized in Figure~\ref{overview-fig}.
We first conduct an exploratory user study to \emph{identify} the \emph{information needs} that developers have
%and the barriers that they face 
when debugging the performance of configurable software systems. 
%(Sec.~\ref{study-1-sec}).
Afterwards, we \emph{design} and \emph{implement information providers} to 
%help address those needs (Sec.~\ref{concepts-sec}).
support developers' needs, adapting techniques from prior work.
Finally, we conduct a \emph{validation} user study
%(Sec.~\ref{study-2-sec}) 
and a \emph{confirmatory} user study 
%(Sec.~\ref{study-3-sec}) 
to evaluate that the designed information providers actually support developers' information needs and speed up the process of debugging the performance of complex configurable software systems.
\looseness=-1

%\section{Explore Information Needs (Study I)}
\section{Exploring Information Needs}
\label{study-1-sec}

%\todo[inline]{The mapping between research question and studies could be clearer in the beginning, and it gets even more lost while presenting the results.
%Some times a matrix helps: RQs x Studies}

While there are numerous performance and program debugging techniques, there is limited 
empirical 
evidence of how useful the techniques are to support developers'
%actual 
needs when debugging the performance of configurable software systems.
%, there is limited empirical evidence about the information that developers need when working on this task.
Hence, 
%To determine the information we can provide developers to support them in the process of debugging the performance of configurable software systems, we 
we first 
%set out to 
investigate the \emph{information needs} that developers have and the process that they follow to debug the performance of configurable software systems.
%to understand how options affect the performance of configurable software systems in the implementation. 
Specifically, we answer the following research questions:
\looseness=-1

\vspace{2ex}

\noindent
{\textbf{RQ1: What information do developers look for when debugging the performance of configurable software systems?}}
\looseness=-1

\vspace{2ex}

\noindent
{\textbf{RQ2: What is the process that developers follow and the activities that they perform to obtain this information?}}
\looseness=-1

\vspace{2ex}

\noindent
% \textit{\textbf{RQ3: What barriers do developers face to find this information?}}
{\textbf{RQ3: What barriers do developers face during this process?}}
\looseness=-1

\vspace{2ex}

%Answering these questions will help us 
%%to understand the \emph{information needs} that developers have when debugging the performance of configurable software systems and will help us to identify existing techniques and tools that can provide relevant information.
%determine the information we can provide developers when we design tailored tool support in Section~\ref{concepts-sec}.
%\looseness=-1

\subsection{Method}
\label{study-1-method-sec}

We conducted an exploratory user study to \emph{identify} the \emph{information needs} that developers have when debugging the performance of configurable software systems.
%understanding how options and their interactions affect the performance in the implementation. 
Using Zeller's terminology~\cite{Z:WPF09}, we want to understand how developers \emph{find possible infection origins}: where
%, in the code, 
options affect performance, and \emph{analyze the infection chain}: what are the causes of an unexpected performance behavior, when debugging the performance of configurable software systems. 
\looseness=-1

\paragraph*{Study design} 
We conducted the exploratory user study, combining a think-aloud protocol~\cite{J:TAP10} and a Wizard of Oz approach~\cite{DJA:KBS93}, to observe how participants debug a performance issue for 50 minutes:
We encourage participants to verbalize what they are doing or trying to do (i.e., think-aloud component), while the experimenter plays the role of some tool that can provide performance behavior information, such as performance profiles and execution time of
specific configurations, on demand (i.e., Wizard of Oz component),
thus avoiding overhead from finding or learning specific tools.
% After the task, we conducted a brief semi-structured interview to discuss the participant's experience during the debugging process.
\looseness=-1

We decided to provide additional information to participants halfway through the study, after we found, in a pilot study with $4$ graduate students from our personal network, that participants spend an 
extremely 
long time (${\sim}60$ minutes in a 
relatively 
small system) 
%in the early stages 
just identifying relevant options and methods.
To additionally explore how participants search for the cause of performance issues once they have identified options and methods, we told participants, after $25$ minutes, 
%(a)~
which options cause the 
%unexpected performance behavior 
performance issue
and 
%(b)~
the methods where the options influence performance. 
In this way, we can both observe how participants start addressing the problem and 
%how they 
analyze how options affect the performance in the implementation.

After the task, we conducted a brief semi-structured interview to discuss the participants’ experience in debugging the system, as well as the information that they found useful and would like to have when debugging the performance of configurable software systems.
%\looseness=-1

Due to the COVID-$19$ pandemic, we conducted the studies remotely over Zoom. 
We asked participants to download and import the source code of the subject system to their favorite IDE, to avoid struggles with using an unfamiliar environment. 
We also asked participants to share their screen. 
With the participants' permission, we recorded audio and video of the sessions for subsequent analysis.
%\looseness=-1

\begin{table*}[t]
\caption{Information needs, activities, and information sources for debugging the performance of configurable software systems.}
\includegraphics[width=\textwidth]{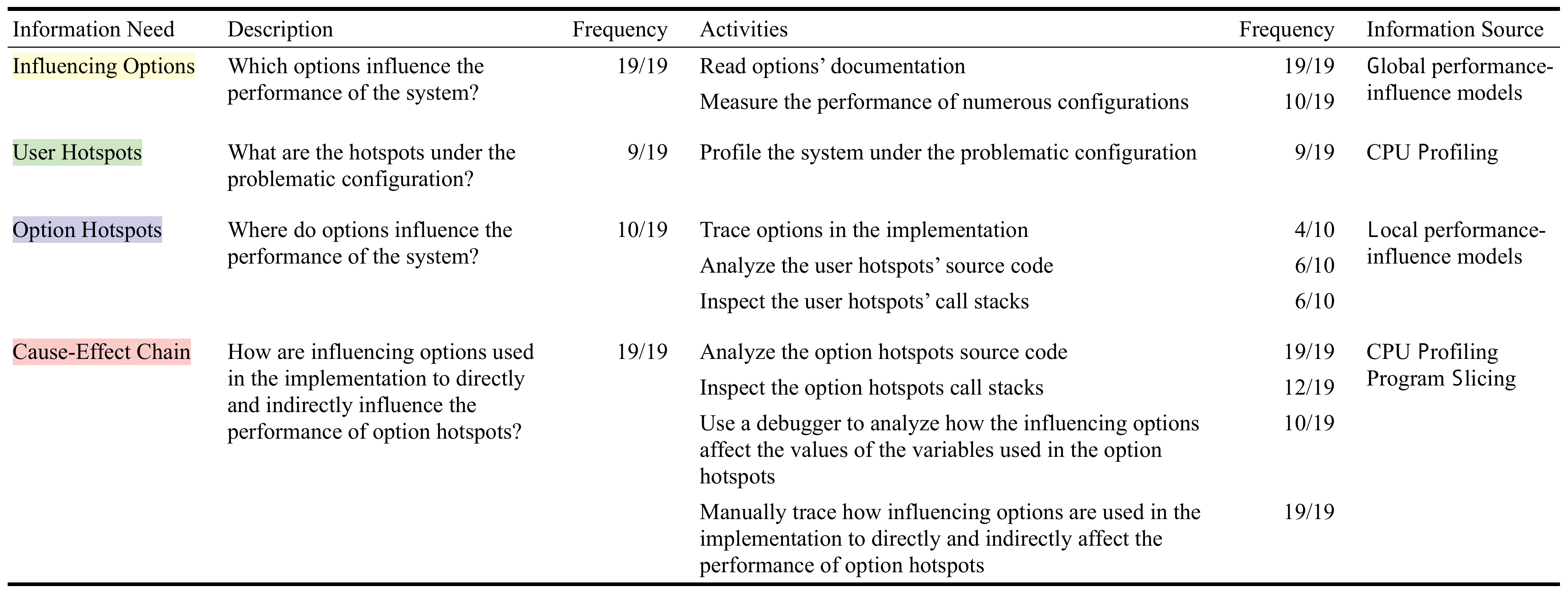}
\label{info-needs-tbl}
\end{table*}

\paragraph*{Task and subject system}

Based on past studies~\cite{MBW:ICSE16, MNHBW:ICPC17, MWKS:ARXIV18, PO:ISSTA11} that have shown how time-consuming debugging even 
%very 
small configurable software systems is, we prepared 
%a single 
one performance debugging task for one configurable software system of moderate size and complexity.
%\looseness=-1
We selected \textit{Density Converter} as the subject system, which transforms images to different dimensions and formats.
We selected this Java system because it is medium-sized, yet non-trivial, with over $49$K SLOC and $22$ binary and non-binary options, and has many options that influence its performance behavior; execution time on the same workload ranged from few seconds to a couple of minutes, depending on the configuration.
%, ($3$) the global and local performance behavior of the system has been studied in prior work~\cite{WAS:ICSE21, VJSSAK:ASEJ20, VJSAK:ICSE21}, and ($4$) the system has been analyzed alongside other configurable software systems that exhibit representative characteristics of numerous configurable software systems~\cite{WAS:ICSE21, VJSSAK:ASEJ20, VJSAK:ICSE21}.
%\looseness=-1
The task involved a user-defined configuration that spends an excessive amount of time executing.
We introduced a bug caused by the incorrect implementation of one option, representative of bugs reported in past research~\cite{AMSBRW:TOSEM18, HY:ESEM16, JSSSL:PLDI12} (the system was spending a long time to transform and output a JPEG image).
%drastically increased the execution time of the system. 
Participants were asked to identify and explain which and how options caused the 
unexpected performance behavior.
We, however, did not explicitly tell participants that the performance problem was related to configurations.
%excessive execution time
%in the implementation.
\looseness=-1

% While we wanted participants to debug a non-trivial configurable software system, our pilot study and similar studies~\cite{MBW:ICSE16, MNHBW:ICPC17, MWKS:ARXIV18, PO:ISSTA11} have shown that understanding how options influence the behavior of systems is a complex and time-consuming task. 
%Therefore, we introduced a bug caused by the incorrect implementation of one option\footnote{
%The system was spending a long time to transform and output a JPEG image.
%\citet{HY:ESEM16} showed that $72$\% of 
%%the studied 
%configuration-related performance bugs involved one option. 
%Several studies have shown that a non-trivial number of performance bugs are caused by algorithm issues~\cite{AMSBRW:TOSEM18, HY:ESEM16, JSSSL:PLDI12}.} to simplify the task and keep the study under $60$ minutes. 
%We expected participants to identify the problematic option and explain how the option is causing the unexpected performance behavior.
\looseness=-1

\paragraph*{Participants} 
We recruited $14$ graduate students and $5$ professional software engineers with extensive experience analyzing the performance of configurable Java systems. 
We stopped recruiting 
%participants 
when we observed similar information needs 
%that the participants had 
and 
%recurring 
patterns in the 
debugging
process.
%that they followed when approaching the task. 
%The participants were 
We 
%recruited using 
used our professional network and LinkedIn for recruiting. 
The graduate students had a median of $6.5$ years of programming experience, a median of $5$ years in Java, a median of $3$ years 
analyzing performance
%of performance analysis experience, 
and a median of $4.5$ years working with configurable software systems. 
The software engineers had a median of $13$ years of programming experience, a median of $13$ years in Java, a median of $5$ years 
%of performance analysis experience, 
analyzing performance
and a median of $5$ years working with configurable software systems. 
%The participants were not compensated for participating in the study.
\looseness=-1

\paragraph*{Analysis} 
We analyzed transcripts of the audio and video recordings of the debugging task and interviews using standard qualitative research methods~\cite{S:CMQR15}.
The first author conducted the study and coded the sessions using open and descriptive coding, 
summarizing
%Specifically, the author derived the descriptive codes that summarized
observations, discussions, and trends~\cite{VJSAK:ICSE22SM}.
%during the debugging task and the interviews~\cite{MV:ICSE22SM}.
All authors met weekly to discuss the codes and observations. 
When codes were updated, previously analyzed sessions were reanalyzed to update the 
%sessions' 
coding.
%of such sessions.
\looseness=-1

\paragraph*{Threats to Validity and Credibility}

%In the setting of our study, 
We observe how developers debug the performance of a system that they had not used before.
%All participants were unfamiliar with \textit{Density Converter}.
%, although a few participants mentioned having experience with photography and image transformation. 
Developers who are familiar with a system might have different needs or follow different 
%debugging 
processes.
While readers should be careful when generalizing our findings, the needs help us identify the information that, at the very least, developers want to find when debugging the performance of 
%support can help developers debug the performance of
unfamiliar
configurable software systems.
%with which they do not have extensive experience.
\looseness=-1

Conducting 
%our exploratory user 
a study with 
%a single 
one system in which one option causes a performance issue has the potential to overfit the findings to this scenario, even though the scenario mirrors common problems in practice~\cite{HY:ESEM16}.
While our later study intentionally varies some aspects of the design to observe whether our solutions generalize to other tasks,
%We mitigate this threat by triangulation~\cite{CP:QIRD16, ESSDSSS:SEMSER08} to validate that our findings can potentially generalize to scenarios beyond the study in this section.
%In particular, we conducted a confirmatory study with a more complex configurable software system
%, in which we observed similar information needs 
%(Sec.~\ref{study-3-sec}).
%Specifically, we will first conduct a follow-up study, with some of our participants, to corroborate that the performance analysis and program debugging concepts that we identified provide relevant information to help participants debug a comparable scenario (see Sec.~\ref{study-2-sec}).
%Subsequently, we will conduct a new user study to show that the concepts can help developers debug the performance of a more complex system in which an interaction of multiple options cause a performance bug (see Sec.~\ref{study-2-sec}).
%Nevertheless, 
generalizations about our results should be done with care.
\looseness=-1

\subsection{Results}

%We observed Finding relevant information was very challenging 

We observed that participants struggle for a 
%substantial amount of 
long time looking for relevant information to debug the performance of the configurable subject system.
In fact, 
%none of the 
no participant
%s were
was able to finish debugging the system within 50 minutes!
In what follows, we present the information needs that participants had, the process that they followed, and the barriers that they faced during the debugging process. %when debugging the performance of configurable software systems.
\looseness=-1

\paragraph*{RQ1: Information Needs}

Table~\ref{info-needs-tbl} lists the four information needs that we identified and the number of participants that recognizably demonstrated each need.
We refer to the information needs as \hlyellow{influencing options}, \hlblue{option hotspots}, \hlred{cause-effect chain}, and \hlgreen{user hotspots}.
The participants referred to these needs using varying terms.
\looseness=-1

When participants 
%first 
faced a non-trivial configuration space, they all tried to identify the \hlyellow{influencing options} -- the option or interaction causing the unexpected performance behavior. 
More specifically, the participants tried to identify \emph{which options} in the \emph{problematic configuration} caused the unexpected performance behavior.
%out of all the selections of options in the problematic configuration  tried to identify which selections in the problematic configuration where cause unexpected behavior.
\looseness=-1

Some participants 
%(who did not run out of time)
%in the first part of the study
tried locating \hlblue{option hotspots} -- the methods where options affect the performance of the system.
More specifically, the participants tried to \emph{locate} 
%the \emph{methods}
%in the system 
where 
%in the implementation 
the \emph{effect} of the problematic configuration could be observed; the methods whose execution time increased under the problematic configuration.
%These findings confirm our hypothesis that developers need to ($1$) identify performance-influencing options and ($2$) locate where options influence performance when debugging the performance of configurable software systems (see Sec.~\ref{study-1-method-sec}).
\looseness=-1

When we told participants (a)~which options cause the unexpected performance behavior (i.e., the \IO) and (b)~the methods where these options influence performance (i.e., the \OH), all participants tried tracing the \hlred{cause-effect chain} -- how \IO\ are used in the implementation to directly and indirectly affect the performance of \OH. 
More specifically, as the participants 
%as the participants observed the effect of the problematic behavior in specifc parts of the program, and knew wchih options \emph{caused} that effect, 
(a)~knew which options were \emph{causing} an unexpected performance behavior and 
%the which options in the problematic configuration caused the unexpected performance behavior and 
(b)~had observed the \emph{effect} of those options on the system's performance,
%slowdown in specific methods, 
the participants tried to find the \emph{root cause} of the unexpected performance behavior.
%, the participants tried to find and understand \emph{how} the \IO\ were \emph{used} in the implementation to directly and indirectly cause the unexpected performance behavior in the \OH.
%reason for the options to causes the unexpected performance  cause of the options  find the root cause, in the implemetation for the observed unexpected performance behavior. 
%This finding confirms our hypothesis that developers, at some point in the debugging process, need to analyze how options are used in the implementation to affect the performance of the system (see Sec.~\ref{study-1-method-sec}).
\looseness=-1

%In addition to the above information needs, 
Some participants also looked for \hlgreen{user hotspots} -- 
the 
methods that spend a long time executing under the user-defined problematic configuration. 
However, as we will discuss in RQ$2$, these participants looked for this information 
%with the goal of locating
trying to locate \OH\ (i.e., how options might affect 
the execution time of 
the expensive methods%).
under the user-defined configuration). 

\vspace{1em}

\begin{mdframed}[backgroundcolor=gray!20, innerleftmargin=5pt, innerrightmargin=5pt] 
	\noindent{\textit{RQ1: Developers look for information to (1)~identify \IO, (2)~locate \OH, and (3)~trace the \CC\ of how options 
	    %directly and indirectly 
	    influence performance in the implementation. 
	    %Some developers also look for hotspots under the user-defined configuration as a means to locate where options influence performance.
	    \looseness=-1
	    }
	}
\end{mdframed}

\paragraph*{RQ2: Process and Activities}

Table~\ref{info-needs-tbl} lists the activities that participants performed when looking for relevant information and the number of participants that performed each activity.
Overall, all participants \emph{compared} the problematic configuration to the default configuration,
%(i.e., a non-problematic baseline configuration), 
to understand the causes of the unexpected performance behavior.
%(i.e., excessive execution time).
In particular, the participants compared the values selected for each option and 
%tried various activities,
%, and how the differences affected various performance behaviors and code execution properties of the system.
%the participants compared the execution time of each configuration and the values selected for each option. 
%This comparison allowed participants 
%trying to 
analyzed how the 
%values in the problematic configuration 
changes were affecting the performance of the system in the implementation.
%; that is, finding the \CC. 
\looseness=-1

When looking for the \IO, the participants mainly read 
%the 
documentation and executed the system under multiple configurations, primarily \emph{comparing} 
%the 
execution times.
%of each configuration.
With these approaches, the participants tried 
%trying 
to identify \emph{which options} in the problematic configuration were causing the unexpected behavior.
\looseness=-1

When looking for \OH, the participants mainly \emph{profiled} the system under the problematic configuration, and analyzed the \emph{call stacks} and source code of hotspots,
%During these activities, the participants tried 
trying to \emph{locate the methods} where \emph{options} might be affecting the performance of hotspots.
\looseness=-1

When looking for the \CC, some participants analyzed the option hotspots' \emph{source code}, whereas others used a debugger;
%In both approaches, the participants tried 
trying
to understand \emph{how} the \IO\ are \emph{used} in the implementation to affect the performance of \OH.
Several participants also \emph{compared} the hotspots' \emph{call stacks} under the problematic and default configurations, trying to understand how the \IO\ affected how the \OH\ were called.
Ultimately, all participants tried to \emph{manually trace} how the \IO\ were being used
%and propagated
%through 
in the implementation to directly and indirectly affect the performance of the \OH.
\looseness=-1

While identifying the \IO\ and locating the \OH\ is needed to trace the \CC, the order in which the first two pieces of information was acquired did not affect the debugging process.
%In fact,
%%While all participants first tried to find either the \IO\ or the \OH, 
%knowing one of these pieces of information is not necessary to find the other.
For instance, nine participants started looking for \IO, but gave up trying after a while.
Then, the participants looked for and were able to identify \UH. 
Six of these participants 
%(who did run out of time) 
subsequently started looking for \OH\ (i.e., how options might affect the 
execution time of the
expensive methods
under the user-defined configuration).
\looseness=-1

\vspace{1em}

\begin{mdframed}[backgroundcolor=gray!20, innerleftmargin=5pt, innerrightmargin=5pt] 
	\noindent{\textit{RQ2: Overall, developers compare the problematic configuration to a non-problematic baseline configuration
	    %(e.g., the default configuration) 
	    to understand the causes of an unexpected performance behavior.
	    Initially, developers compare execution times to identify \IO, and analyze call stacks and source code to locate \OH.
	    These two pieces of information are necessary to trace the \CC\ of how \IO\ are used in the implementation to 
	    directly and indirectly 
	    influence the performance of \OH.
	    \looseness=-1
	    }
	}
\end{mdframed}

\paragraph*{RQ3: Barriers}

Our participants struggled for a 
%substantial amount of 
long time trying to find relevant information to debug how options influence the performance of the system in the implementation.
%The majority of 
Most participants discussed the \emph{``tedious and manual''} process of executing multiple configurations when looking for \IO. 
For instance, only $10$ out of the $19$ participants identified the \IO. 
%The strategy that these participants followed was to revert the changes made in the problematic configuration to the default values. 
While we told participants the system's execution time under any configuration that they wanted,
%to execute, 
several participants mentioned that finding the problematic option would have \emph{``taken me hours.''} 
\looseness=-1

Most participants also mentioned the struggle to locate \OH. 
In fact, no participant found any option hotspot! 
Several participants mentioned that \emph{locating} these methods is challenging since options are not 
typically
%[typically] 
directly used in expensive methods.
\looseness=-1

%Several participants mentioned that identifying the Influencing Options and Option Hotspots would have taken them a long time if we did not provide this information for the second part of the study.

The participants struggled the most when trying to trace the \CC.
In fact, 
%none of the participants were able to debug the system!
%N
no participant could establish the \CC, even when 
%(a)~
our task consisted of tracing \emph{a single} influencing option and
%(b)~
we explicitly told participants the influencing option and the \OH\ 
%that 
they needed to analyze.
Most participants mentioned that manually tracing even one
option
%\hlyellow{influencing option} 
through a relatively small system is \emph{``error-prone.''}
Additionally, some participants discussed that \emph{identifying differences} in the \OH’ \emph{call stacks} was difficult
%, which prevented them from 
for determining whether the \IO\ were affecting how the \OH\ were called.
As mentioned by several participants: Variables used in \OH\ are often a \emph{``result of several computations''} involving \IO.
Since the \IO\ are used in various parts of the system, \emph{``tracing which paths to follow is very challenging.''}
\looseness=-1

%While analyzing the \CC, some participants mentioned that \emph{identifying differences} in the \OH’ \emph{call stacks} was difficult. 
%%The participants mainly used the call stacks to trace which methods were executed.
%%However, 
%Several participants stated that they wanted to compare the call stacks to determine whether the \IO\ were affecting how the \OH\ were called.%, which might be a reason for the performance differences.
%%the performance differences in the option hotspots were caused by different calling contexts based on the \IO.
%%were affecting how the option hotspots were called to affect the performance of the system.
%%While participants mentioned that the call stacks were useful to navigate the code, it was difficult to spot different to know whtehter the \IO\ were changing how the hotspots were called to affect performance.
%\looseness=-1

\vspace{1em}

\begin{mdframed}[backgroundcolor=gray!20, innerleftmargin=5pt, innerrightmargin=5pt] 
	\noindent{\textit{RQ3: Developers struggle for a substantial amount of time looking for relevant information to identify \IO, locate \OH, and trace the \CC.
	    \looseness=-1
	    }
	}
\end{mdframed}

\section{Supporting Information Needs}
\label{concepts-sec}

%After identifying the \emph{information needs} that developers have and the barriers that they face when debugging the performance of configurable software systems, 
We aim to support developers in 
%(a)~
identifying \IO, 
%(b)~
locating \OH, and 
%(c)~
tracing the \CC.
To this end, we design \emph{information providers}, adapting \emph{information sources}, to support the above needs.
We implement the designed information providers in a tailored
and cohesive 
prototype called \TOOL~\cite{VJSAK:ICSE22SM}, which can assist developers to debug the performance of configurable software systems.
Table~\ref{info-needs-tbl} shows which information needs are supported by the information sources global and local performance-influence models, CPU profiling, and program slicing
%\GL\ and \LL\ \PIM, \CP, and \PS\ 
that we adapt for designing information providers.
\looseness=-1

%While in the following sections we discuss how the concepts and our prototype help developers address the information needs that

%Concepts to Support Info Needs

%How information that the prototype provides supports the info needs.

%Prototype user interface and information is collected for the backend with a CI server.
%\subsection{Identifying \protect\colorbox{myyellow}{Influencing Options}}
\subsection{Identifying Influencing Options}
\label{global-models-sec}

To help developers 
%in reasoning over the entire configuration space 
identify the \IO\ that cause an unexpected performance behavior,
%To this end, 
we 
%present developers 
select %\GL\ \PIM
global performance-influence models~\cite{SGAK:ESECFSE15, HZ:ICSME19, VJSSAK:ASEJ20, KSKGA:SOSYM18, VJSAK:ICSE21, WAS:ICSE21}, which describe a system's performance in terms of its options and their interactions.
For instance, the model $4.6 + 54.7 \cdot \textsc{Duplicates} \cdot \textsc{Transactions} + 8.9 \cdot \textsc{Evict} + 3.5 \cdot \textsc{Temporary}$ explains the influence of 
%the options 
\textsc{Duplicates}, \textsc{Transactions}, \textsc{Evict}, and \textsc{Temporary}, and their interactions on the performance of Berkeley DB.
%; for instance, enabling \textsc{Evict} increases the execution time
%of the system 
%by $8.9$ seconds, and enabling \textsc{Duplicates} and \textsc{Transactions}, \emph{together}, further increase the execution time by $54.7$ seconds.
%The model also shows that other options 
%and 
%their 
%interactions 
%(e.g., \textsc{Replicated}) do not influence the performance of the system.
\looseness=-1

%More specifically, we help developers \emph{compare any two configurations} (e.g., a problematic and a non-problematic configuration) by \emph{showing how specific changes made between the configurations
%(e.g., the problematic configuration) 
%influence the performance of the system}.
%In the context of debugging the performance of configurable software systems, 
We adapt this information source to design an \emph{information provider} that shows developers \IO; specifically, \emph{which and how differences} between configurations (e.g., a problematic and a non-problematic configuration) influence a system's performance.
%To help developers focus on the relevant information when comparing configurations, our prototype has the option to only show the influencing changes between two configurations.
%In this setting, if additional changes between the configurations are not shown, then the changes do not have an influence on the performance of the system.
In 
%\TOOL,
our implementation,
this information provider highlights the differences in the values of options selected between two configurations, and shows the \IO\ between the configurations.
If changes between the configurations are not shown, then the changes do not
%have an 
influence 
%on 
the performance of the system.\footnote{
Any performance-influence model is shown relative to one configuration (e.g., the default configuration), which explains the impact of changes to that configuration.
In 
%\TOOL, 
our tool, 
developers can select that one configuration.
%Our prototype allows developers to select a single configuration to analyze the global performance-influence model of the system.
\looseness=-1
}
%while global performance-influence models describe the influence options on the performance of the system, our prototype shows the changes actually made between the configurations. 
%That is, 
%While there might be other influential configuration changes in a system, \TOOL\ only shows the changes between the configurations that the developer is comparing.
\looseness=-1

%\paragraph*{\nopunct}

Global performance-influence models are typically built by measuring 
%the 
execution time 
%of a system 
under different configurations~\cite{SGAK:ESECFSE15}.
The models can be built using white-box techniques~\cite{VJSSAK:ASEJ20, WAS:ICSE21, VJSAK:ICSE21}, machine-learning approaches~\cite{GCASW:ASE13, GSA:PPSCNSB19, KGSA:IEEESOFT20, HZ:ICSME19, HZ:ICSE19}, or a brute-force approach.
Details on 
%each of 
these techniques are beyond the scope of this paper.
%In \TOOL,
In our implementation,
we generate the models using white-box techniques, as these approaches first generate local performance-influence models (which our tool also uses) to obtain the global model for the system~\cite{VJSSAK:ASEJ20, WAS:ICSE21, VJSAK:ICSE21}.
\looseness=-1

%\begin{figure}[t]
%\begin{center}
%    \includegraphics[width=0.78\columnwidth]{configs-fig-glps.pdf}
%\end{center}
%%\setlength{\textfloatsep}{5pt}
%\caption{
%\TOOL\ highlights the differences in the values of options selected between two configurations.
%\looseness=-1
%}
%\label{config-diff-fig}
%\end{figure}
%%\setlength{\textfloatsep}{5pt}

%\begin{figure}[t]
%\begin{center}
%    \includegraphics[width=0.78\columnwidth]{global-model-fig-glps.pdf}
%\end{center}
%%\setlength{\textfloatsep}{5pt}
%\caption{
%\TOOL\ shows the \IO\ when changing values 
%%of options 
%from one configuration to ($\blacktriangleright$) another configuration.
%%This information is derived from \GT lobal performance-influence models.
%\looseness=-1
%}
%\label{global-model-fig}
%\end{figure}
%\setlength{\textfloatsep}{5pt}

\begin{figure}[t]
    \begin{center}
    \begin{minipage}[t]{\columnwidth}
    \begin{center}
        \includegraphics[width=\columnwidth]{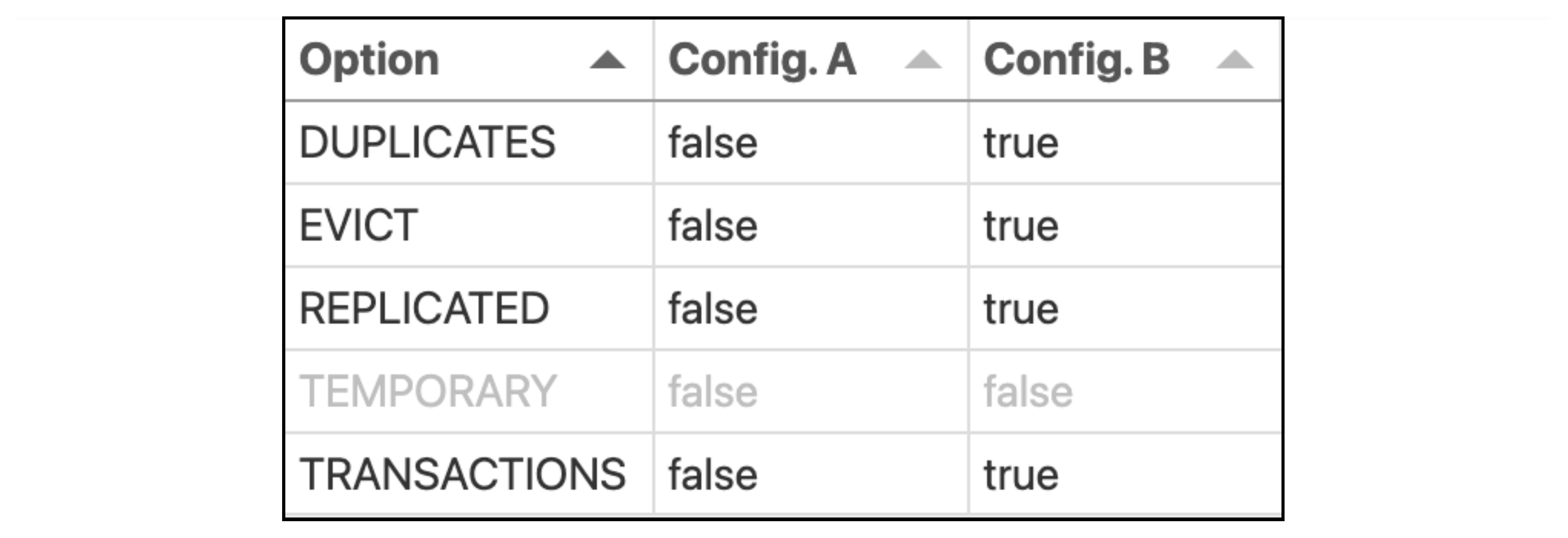}
    \end{center}
    \smallskip
    \end{minipage}
    \begin{minipage}[t]{\columnwidth}
    \begin{center}
        \includegraphics[width=\columnwidth]{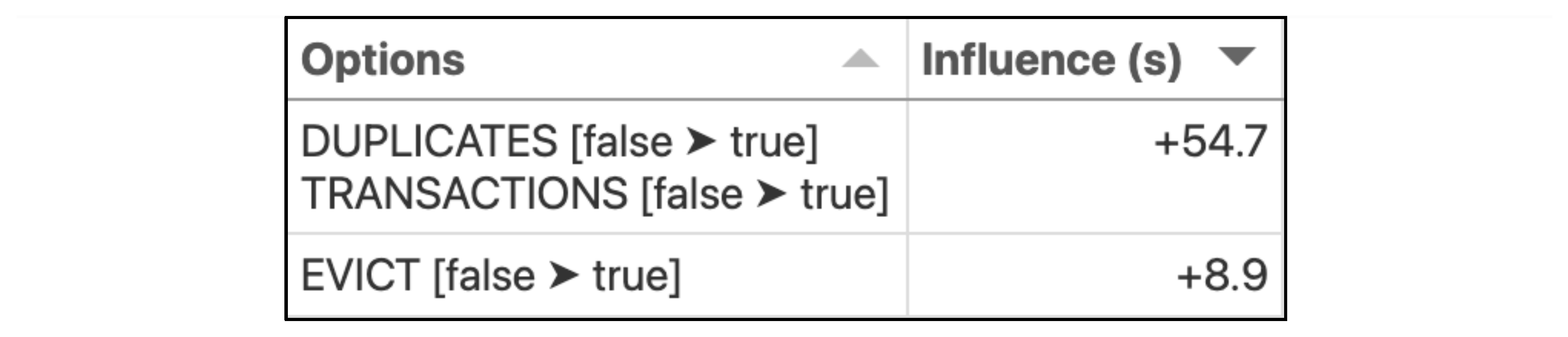}
    \end{center}
    \end{minipage}
    \end{center}
    \caption{
    %\TOOL\ 
    Our tool highlights 
    %the 
    differences in the 
    %values of 
    options selected between two configurations, and shows the \IO\ 
    %when changing values 
    of the changes
    %of options 
    from one configuration to ($\blacktriangleright$) another configuration.
    %This information is derived from \GT lobal performance-influence models.
    \looseness=-1}
    \label{config-diff-fig}
\end{figure}

\paragraph*{Example}
Figure~\ref{config-diff-fig} shows a screenshot of our tool highlighting the differences 
%in the values of 
of the
options selected between two configurations (e.g., a non-problematic and problematic configurations).
%Only the value for the option \textsc{Temporary} is the same between the two configurations.
%Figure~\ref{global-model-fig} 
%Figure~\ref{config-diff-fig} 
Our tool
%also 
shows the \IO\ between these two configurations.
For instance, changing 
both 
options 
\textsc{Duplicates} and \textsc{Transactions} from \texttt{false} to \texttt{true} results in an interaction that increased the execution time
%of the system 
by $54.7$ seconds.
%Likewise, the change in \textsc{Evict}
%%from \texttt{false} to \texttt{true} 
%further increased the execution time by $8.9$ seconds.
Based on this information, most developers would consider \textsc{Duplicates} and \textsc{Transactions} as 
%the 
\IO\ 
that are causing an unexpected performance behavior.
\looseness=-1

Note that individual changes to \textsc{Duplicates} and \textsc{Transactions} did not influence the performance of the system; only the \emph{interaction} increased the execution time.
Additionally, any other changes between the configurations -- \textsc{Replicated} -- do not influence the performance of the system.
%, as they are not present in Figure~\ref{config-diff-fig}.
%the 
%performance-influence model above and are not shown in the 
%table above.
Likewise, the influence of \textsc{Temporary} is not shown,
%in Figure~\ref{config-diff-fig}, 
as both configurations selected the same value.
%for this option.
%(see Figure~\ref{config-diff-fig}).
\looseness=-1

%\subsection{Locating \protect\colorbox{myblue}{Option Hotspots}}
\subsection{Locating Option Hotspots}
After helping developers identify the \IO,
%\ that cause an unexpected performance behavior, 
we help developers locate the \OH\ where these options 
%influence the performance of a system.
cause an unexpected performance behavior.
%and cause an unexpected performance behavior.
To this end, we 
%present developers 
select local performance-influence models~\cite{VJSSAK:ASEJ20, VJSAK:ICSE21, WAS:ICSE21}. 
Analogous to how global performance-influence models describe the influence of options and interactions on a \emph{system}'s performance, local models describe the influence of options on the performance of \emph{individual methods}.
Hence, local models indicate \emph{where} options affect the performance 
%of the system 
in the implementation~\cite{VJSSAK:ASEJ20, VJSAK:ICSE21, WAS:ICSE21}.
For instance, 
%derived from the global performance-influence model in Sec.~\ref{global-models-sec},
the local model of a method $0.9 + 42.9 \cdot \textsc{Duplicates} \cdot \textsc{Transactions}$ explains the influence of \textsc{Duplicates} and \textsc{Transactions} on the performance of the specific method, rather than the entire system.
%\texttt{Cursor.putInternal}.
%Likewise, the sparse-linear model $0.2 + 10.3 \cdot \textsc{Duplicates} \cdot \textsc{Transactions}$ explains \emph{how} the options influence the performance of the method \texttt{FileManager.read}.
%To this end, we use \emph{local performance-influence models}.
%Similar to global performance-influence models (see Sec.~\ref{global-models-sec}), local performance-influence models describe performance in terms of options and interactions. 
%However, local performance-influence models also indicate \emph{where} options influence the performance of the system (e.g., method or function).
\looseness=-1

We adapt this information source to design another \emph{information provider} that shows developers \OH; specifically, \emph{where} and \emph{by how much} options influence the performance of the system.
In 
%\TOOL, 
our implementation, this information provider shows (a)~the \emph{methods} whose performance is influenced by changes made between configurations (e.g., a problematic and a non-problematic configuration) and (b)~the 
%actual 
\emph{influence} of the changes on each method's performance.\footnote{
Our tool
%\TOOL\ 
also allows developers to select a single configuration to analyze individual local performance-influence models.
}
\looseness=-1

%\paragraph*{\nopunct}

Local performance-influence models are usually built with white-box approaches as by-products of 
%building 
global models~\cite{VJSAK:ICSE21, WAS:ICSE21, VJSSAK:ASEJ20}.
%In \TOOL, 
In our implementation,
we generate local models using these approaches.
\looseness=-1

%Something about methods influenced by other options, but if those changes were not made in the configurations that are compared, then it will not show does
%Our prototype shows the influencing changes between two configurations.
%If additional changes between the configurations are not shown, then the changes do not have an influence on the performance of the system.
%Likewise, 
%%while global performance-influence models describe the influence options on the performance of the system, our prototype shows the changes actually made between the configurations. 
%That is, 
%while there might be other influential changes, the prototype shows the changes actually made between %the configurations that the developer is comparing.\footnote{Our prototype allows developers to select a single configuration to analyze the global performance-influence model of the system.
%\looseness=-1}
%\looseness=-1

\begin{figure}[t]
\begin{center}
    \includegraphics[width=\columnwidth]{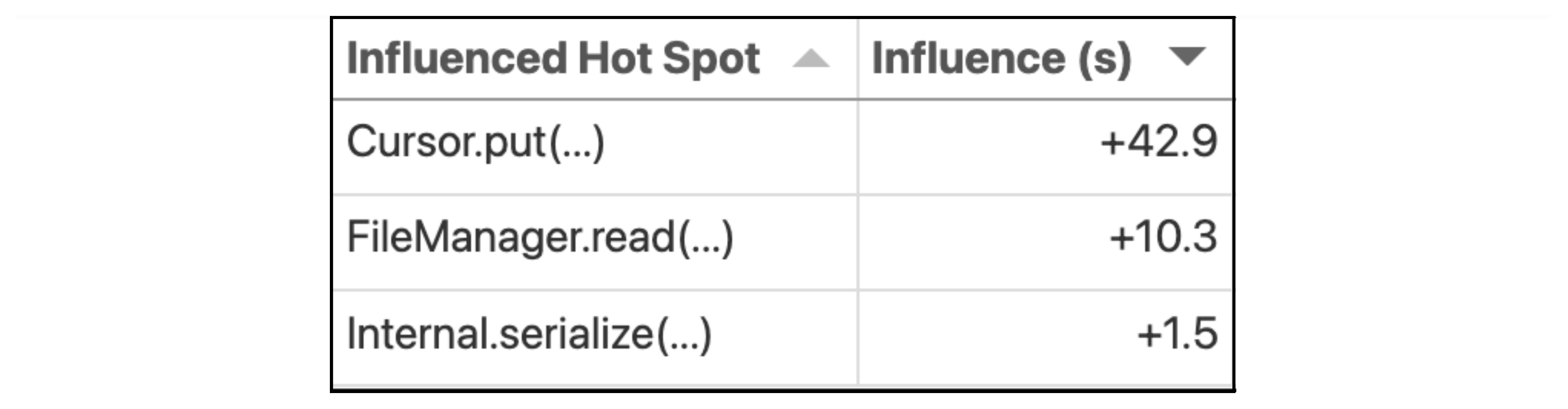}
\end{center}
\caption{
%\TOOL\ 
Our tool shows \OH\ 
affected by 
%the 
\IO,
%\textsc{Duplicates} and \textsc{Transactions},
as well as the influence on performance in each method.
%This information is derived from \LT ocal performance-influence models.
\looseness=-1
}
\label{local-model-fig}
\end{figure}

\paragraph*{Example}
Figure~\ref{local-model-fig} shows as screenshot of our tool indicating the \OH\ where the \IO\ \textsc{Duplicates} and \textsc{Transactions} affect the system's performance.
Note that the influence on all methods equals the influence in the entire system (see Figure~\ref{config-diff-fig}).
Based on this information, most developers would consider \texttt{Cursor.put}
%and \texttt{FileManager.read} 
as an 
%\OH
option hotspot; the location where the \emph{effect} of the \IO\ is observed.
\looseness=-1

%\subsection{Tracing the \protect\colorbox{myred}{Cause-Effect Chain}}
\subsection{Tracing the Cause-Effect Chain}

After helping developers 
%(a)~
identify the \IO\ 
%which options cause an unexpected performance behavior 
and 
%(b)~
locate the \OH,
%methods where these options influence the performance of a system, 
we help developers trace the \CC.
%how options influence the performance of a system in the implementation.
To this end, we 
%present developers 
select 
%(a)~
CPU profiling and 
%(b)~
program slicing.
%(a) \emph{a comparison of the hotspot view of performance profiles} and (b) \emph{the program slice, specifically the chop, from the point where \IO\ are loaded into the system to the option hotspots}.
\looseness=-1

%\subsubsection{\CP\nopunct}
\subsubsection{CPU Profiling}

% \paragraph{\nopunct}
We select CPU profiling to collect the hotspot view of the problematic configuration and a non-problematic configuration.
The hotspot view is the inverse of a call tree:  A list of all methods sorted by their total execution time, cumulated from all different call stacks, and with back traces that show how the methods were called.
\looseness=-1

We adapt this information source to design another \emph{information provider} that helps developers trace the \CC; specifically, \emph{compare} the hotspot view of two configurations 
(e.g., a non-problematic and a problematic configuration) 
to help developers determine whether the \IO\ \emph{affect} how \OH\ are called.
%In \TOOL, 
In our implementation,
this information provider highlights \emph{differences} in the \OH' execution time and call stacks.\footnote{
Our tool also allows
%\TOOL, 
developers to analyze the CPU profile of one configuration.
\looseness=-1
}
\looseness=-1

%\paragraph*{\nopunct}

CPU profiles can be collected with most off-the-shelf profilers.
White-box approaches that build global and local performance-influence models collect these profiles~\cite{VJSSAK:ASEJ20, VJSAK:ICSE21, WAS:ICSE21}.
%In \TOOL, 
In our implementation,
we use the CPU profiles collected by
%white-box 
these approaches.
\looseness=-1

\begin{figure}[t]
\begin{center}
    \includegraphics[width=\columnwidth]{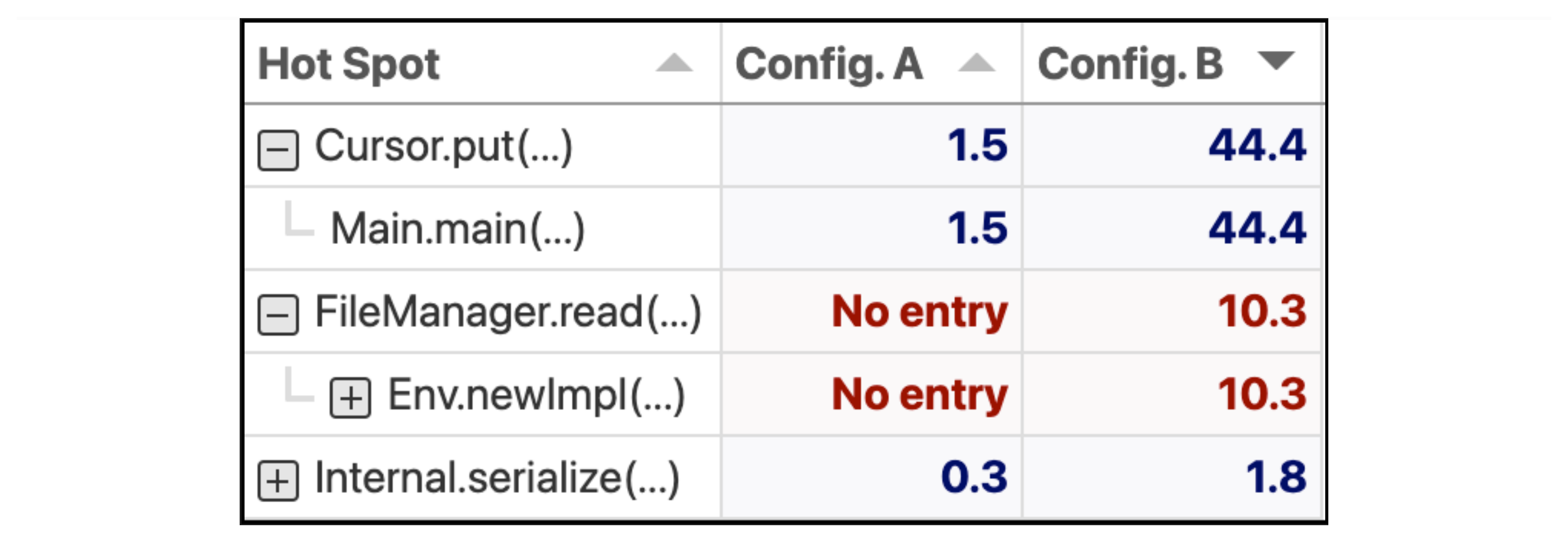}
\end{center}
\caption{
%\TOOL\ 
Our tool helps developers trace the \CC\ by highlighting the differences in the \OH' execution time and call stacks 
%based on 
affected by \IO.
%\ \textsc{Duplicates} and \textsc{Transactions} .
While the call stacks of \texttt{Cursor.put} are the same under both configurations, 
%the \emph{option hotspot} 
\texttt{FileManager.read} is only called under the second configuration.
%This information is derived from \PT rofiling the system under the problematic configuration and a non-problematic configuration.
\looseness=-1
}
\label{profile-diff-fig}
\end{figure}

\paragraph*{Example}
Figure~\ref{profile-diff-fig} shows a screenshot of our tool, which helps developers trace the \CC\ by highlighting 
%the 
differences in the \OH' execution time and call stacks based on the \IO\ \textsc{Duplicates} and \textsc{Transactions}.
For instance, the changes increased \texttt{Cursor.put}'s execution time, but did not affect how the method was called.
By contrast, \texttt{FileManager.read} is only executed under the problematic configuration.
This information can help developers understand how the \IO\ are used in the implementation to affect the \OH's performance.
\looseness=-1

%\subsubsection{Program \ST licing\nopunct}
\subsubsection{Program Slicing}

% \paragraph{\nopunct}

%TODO Here or in the implementation, what is the specific criteria used in the option hotspots?
 
 % Slice shows causes that happen in other places in the progran, Trace through the system what things are affected to influence the performance of options.
We %present developers 
select program slicing~\cite{W:ICSE81, XQZWC:SEN05, K:SCAM03, Z:WPF09}; an approach to compute 
%the 
relevant fragments of a system based on a criterion.
Several debugging tools have been implemented on top of program slicers~\cite{KM:CHI04, LM:VLHCC11, FCL:FSE20, XQZWC:SEN05} to help developers narrow down 
%and isolate 
relevant 
%inputs and 
parts of a system where developers should focus their debugging efforts.
\looseness=-1

We adapt this information source to design another \emph{information provider} that helps developers trace the \CC; specifically, \emph{tracking} \emph{how} \IO\ are \emph{used in the implementation} to directly and indirectly \emph{influence} the performance of \OH.
%In \TOOL,
In our implementation,
this information provider slices
(chops) 
a system from 
%the point 
where \IO\ are first loaded into the system to the \OH, and shows (a)~a \emph{method-level dependence graph} and (b)~\emph{highlighted statements} of the slice in the source code.
%to help developers trace how \IO\ affect the performance of option hotspots}.
%This information would help developers trace the causes of the unexpected performance behavior effects observed in \OH.
\looseness=-1

%\paragraph*{\nopunct}

\begin{figure}[t]
\begin{center}
    \includegraphics[width=\columnwidth]{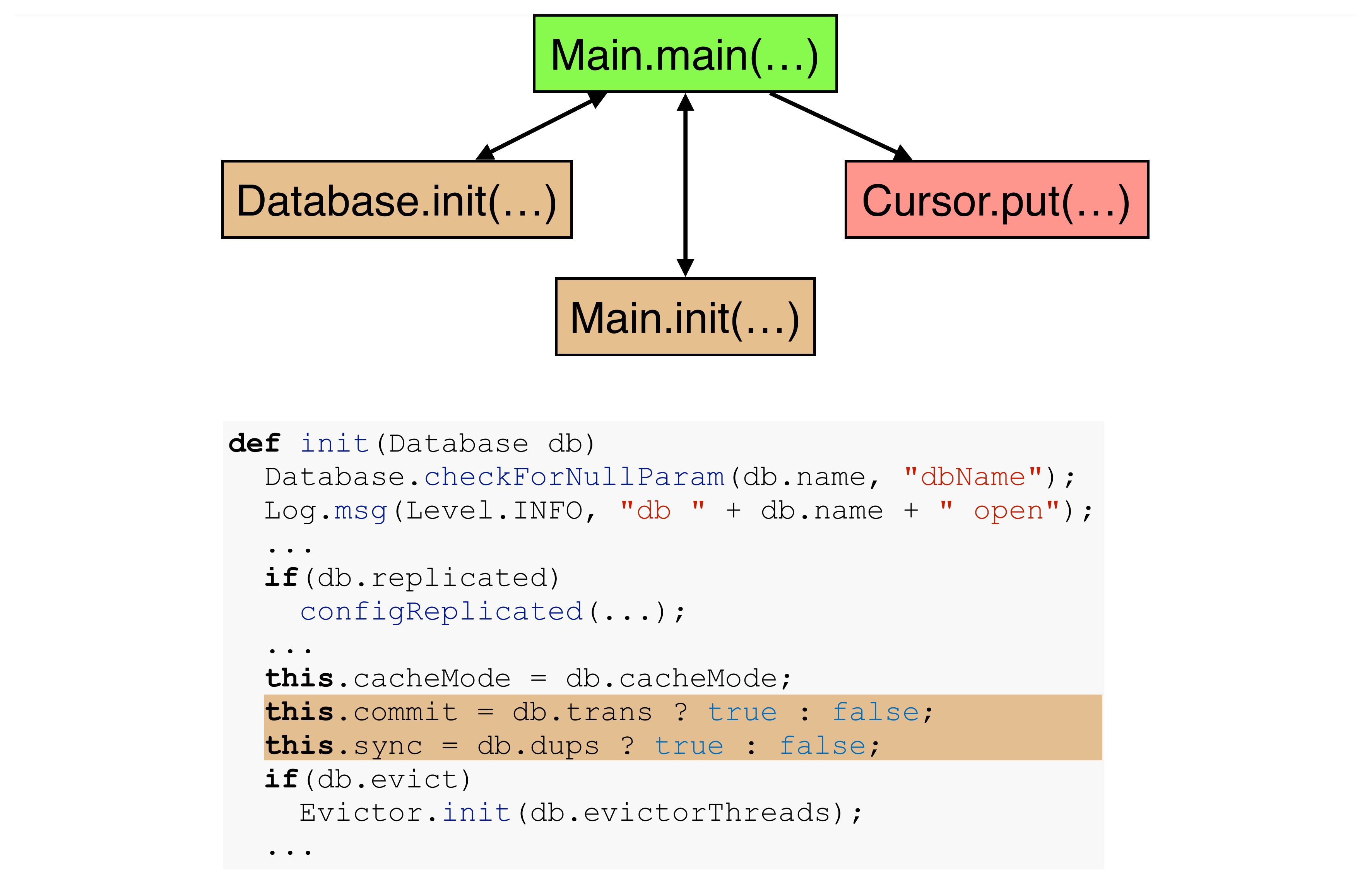}
\end{center}
\caption{
%\TOOL\ 
Our tool
helps developers trace the \CC\ by displaying a method-level dependence graph from the method where the \IO\ 
%\textsc{Duplicates} and \textsc{Transactions} 
are first loaded into the system (green box) to 
%the 
and option hotspot
%\texttt{Cursor.put} 
(red box). 
Other relevant methods 
%that are part of the dependence graph 
are shown in brown boxes.
When clicking on a box,
%method,
%on the graph, 
\TOOL\ opens the file with the method and highlights the statements of the slice.
The position of the nodes in the graph (left to right and top to bottom) does not represent the order of execution of methods.
%in the source code.
%This information is derived from \ST licing the system from the \IO\ to an \hlyellow{option hotspot}.
\looseness=-1
}
\label{highlight-statements-fig}
\end{figure}

\paragraph*{Example}
Figure~\ref{highlight-statements-fig} shows a screenshot of our tool, which helps developers trace the \CC\ by showing a method-level dependence graph from the \texttt{main} method, in which the \IO\ \textsc{Duplicates} and \textsc{Transactions} are first loaded into the system, to the option hotspot \texttt{Cursor.put}.
The graph can help developers 
%obtain an overview of the methods of the slice and 
track dependences across methods in the system.
%The position of the nodes in the graph (left to right and top to bottom) does not represent the order of execution of methods.
%\looseness=-1
When clicking on a method on the graph, 
%\TOOL\ 
our implementation
opens the file with the method and highlights the statements in the slice, such as in Figure~\ref{highlight-statements-fig}.
The highlighted statements can help developers trace the \CC\ by tracking 
%the
%highlighted
%statements 
%of the slice 
%that explain 
how \IO\ are used in the implementation to 
directly and indirectly 
cause an unexpected performance behavior in \OH.
\looseness=-1

%These two pieces of information can help developers trace the \CC.
%Developers can use this information to trace how \IO\ are used in the implementation to directly and indirectly cause an unexpected behavior in \OH.
%\looseness=-1

\subsection{Implementation}

We implemented the information providers in a Visual Studio Code extension 
%prototype 
called \TOOL~\cite{VJSAK:ICSE22SM}.
Our prototype adapts
%(a)~
global\ and 
%(b)~
local performance-influence models, 
%(c)~
CPU profiles, and 
%(d)~
a program slicer.
The first three items are collected prior to debugging, 
%the performance of 
%a system, 
using
%, for example, 
an infrastructure where developers configure and run the system.
%to collect data.
Subsequently, developers use \TOOL\ to 
%help them 
identify \IO,
%\ (from \GL\ \PIM), 
locate \OH,
%\ (from \LL\ \PIM), 
and trace the \CC.
%\ (using CPU \PT rofiles and \PS).
\looseness=-1

\TOOL\ is agnostic to the information sources used to adapt and implement information providers.
In fact, \TOOL\ is entirely built on existing infrastructure of information sources for global and local performance-influence modeling, CPU profiling, and program slicing.
The novelty to \TOOL\ is in the \emph{design} and \emph{integration of information providers}, from multiple \emph{information sources}, into a \emph{cohesive} infrastructure and user interface, which can help developers debug the performance of configurable software systems. 
%Rather, \TOOL\ collects and presents relevant information using off-the-shelf and state-of-the-art tools from past research.
%In the following paragraphs, we describe how we collected the relevant information, though we could have built our implementation also on different tools, including different modeling techniques~\cite{GCASW:ASE13, GSA:PPSCNSB19, HZ:ICSME19, HZ:ICSE19}, different profilers, and different slicing tools.
\looseness=-1

Our implementation uses Comprex~\cite{VJSAK:ICSE21} to build the global and local performance-influence models and uses JProfiler~\cite{JPROFILER10} to collect the CPU profiles.

Ideally, we would slice the program dynamically, as we are analyzing a system's dynamic behavior and to avoid approximations in the results.
However, after exploring various dynamic and static slicing research tools, we settled on the 
state-of-the-art 
static slicer provided by JOANA~\cite{GHM:KIT12}, as it is the most
%feasible and 
mature option. 
%the state-of-the-art static program slicing tool, to slice our programs.
%\looseness=-1
Our implementation uses JOANA to slice the system 
%with \TOOL\ 
from \IO\ to \OH,
%\TOOL\ 
%Our implementation
%uses 
using a fixed-point chopper algorithm, which first computes a backward slice 
%using 
from the \OH,
%\ as the slicing criteria, 
and then computes a forward slice, on the backward slice, 
%using 
from the \IO~\cite{G:SQJ11}.
For scalability and to reduce approximations,
%in the results, 
we modified JOANA to consider code coverage under the problematic and a non-problematic configurations.
%Any other approaches for collecting this information could have been used.
\looseness=-1

\section{Evaluating Usefulness of Information Providers}
\label{eval-sec}

We evaluate the usefulness of our designed information providers
%that we designed 
to help developers debug the performance of configurable software systems.
Specifically, we answer the following research question:
\looseness=-1

\vspace{2ex}

\noindent
{\textbf{RQ4: To what extent do the designed information providers help developers debug the performance of configurable software systems?}}
\looseness=-1

\vspace{2ex}

We answer this research question with two user studies using different designs.
We first evaluate the extent that our designed information providers
%(Sec.~\ref{concepts-sec}) 
support the information needs that we identified in our exploratory study.
%(Sec.~\ref{study-1-sec}).
To this end, we conduct a \emph{validation} study, in which we ask the participants of our exploratory study to debug a \emph{comparable} unexpected performance behavior using \TOOL\ on the same subject system (Sec~\ref{study-2-sec}).
Afterwards, we replicate the study, intentionally \emph{varying} some aspects of the designs (theoretical replication~\cite{JG:ESEV11, S:RGP09}), to evaluate to which extent our information providers generalize for a more complex task with an interaction in a larger system.
Specifically, we conduct a \emph{confirmatory} study, in which we ask a \emph{new set of participants} to debug a \emph{more complex task} on a \emph{more complex subject system}
(Sec~\ref{study-3-sec}).
The validation and confirmatory studies, \emph{together}, provide evidence that our information providers help developers debug the performance of complex configurable software systems \emph{because} the information providers \emph{support} the information needs that developers have in this process.
\looseness=-1

%Answering this question will help us to (a)~validate that the performance analysis and program debugging concepts that we identified
%(see Sec.~\ref{concepts-sec}) 
%provide relevant information to support the needs that we identified in our first study (see Sec.~\ref{study-1-sec}) and (b)~show that the concepts can help developers debug the performance of more complex configurable software systems.
%\looseness=-1

\subsection{Validating Usefulness of Information Providers}
\label{study-2-sec}

%We first conducted a validation study to demonstrate the extent that the information providers that we designed can support the information needs that developers have when debugging the performance of configurable software systems.

%\todo[inline]{Why did we do this study? Why use the same system? What was the same or different from the previous study?}

%\todo[inline]{That is in the first study we see whether people with the tool still look for the same information whereas in the second one we see whether our tool is effective if people look for that specific information?}

We first conducted a \emph{validation} study to evaluate the extent that the designed information providers support the information needs that we identified in our exploratory study.
\looseness=-1

\subsubsection{Method\nopunct}
\label{study-2-method-sec}

\paragraph*{Study design}

We invited the participants from our exploratory study, after $5$ months, to solve another problem in the same system, but now with the help of our information providers. % available through out tool.
This design can be considered as a within-subject study, where subjects perform tasks both in the control and in the treatment condition:
Specifically, we consider our exploratory study
%(Sec.~\ref{study-1-sec}) 
as the \emph{control} condition, in which participants debugged
a system 
\emph{without} \TOOL, and consider the new study as the \emph{treatment} condition, in which participants debug a \emph{comparable} 
%unexpected performance behavior 
performance issue for $50$ minutes in the same subject system with \TOOL.
Similar to the exploratory study, we use think-aloud protocol~\cite{J:TAP10} to identify whether our information providers actually support the information needs that developers have when debugging the performance of the subject system.
%in which we encourage participants to verbalize what they are doing (or trying to do), while they use our tool to obtain information.
\looseness=-1

Prior to the task, participants worked on a warm-up task for $20$ minutes using \TOOL.
%to learn how to use the information providers. 
We tested the time for the warm-up task, as well as \TOOL's design and implementation in a pilot study with $4$ graduate students from our personal network.
\looseness=-1

After the task, we conducted a brief semi-structured interview to discuss the
participants’ experience in debugging the system, as well as the usefulness of the information providers,
%by \TOOL, 
and to contrast their experience to debugging without the information providers.
%\TOOL.
%participants’ experience in debugging the system.
%In particular, we asked participants about the usefulness of the information provided by \TOOL\ and to contrast their experience to debugging without \TOOL.
\looseness=-1

Due to the COVID-$19$ pandemic, we conducted the studies remotely over Zoom. 
Participants used Visual Studio Code through their preferred Web browser.
The IDE was running on a remote server and was configured with \TOOL.
We asked participants to share their screen. 
With the participants' permission, we recorded audio and video of the sessions for subsequent analysis.
\looseness=-1

\paragraph*{Task and subject system}
We prepared a comparable, \emph{but different}, debugging task to the
%task that we used 
task in our exploratory study.
Similar to the exploratory study, the task involved a user-defined configuration in \textit{Density Converter} that spends an excessive amount of time executing. 
We introduced a bug caused by the incorrect implementation of one option
%The unexpected performance behavior was caused by a bug that we introduced as a result of the incorrect implementation of one option.
(%During transformation, 
the system was using a 
%significantly 
larger scale of the input image instead of using a fraction).
Participants were asked to identify and explain which and how options caused the unexpected performance behavior.
%in
%the implementation.
%Similar to the exploratory study, we did not explicitly tell participants that the performance problem was related to configurations.
In contrast to the exploratory study, the user-defined configuration, bug, and problematic option, were \emph{different}.
\looseness=-1

\paragraph*{Participants}
We invited the participants from our exploratory study to work on our task.
After conducting the study with $8$ participants, we observed a \emph{massive effect size} between debugging with and without our information providers (correctly debugging within $19$ minutes compared to failing after $50$ minutes, see Sec.~\ref{study-2-res-sec}).
Hence, we did not invite the remaining participants.
%performed the study, 
%We attempted to recruit all $19$ participants from our exploratory study.
%to debug with \TOOL. 
%Eight out of the $14$ graduate students agreed to participate again.
%Since we observed a \emph{massive effect size} with these $8$ participants between debugging with and without our tool (see Sec.~\ref{study-2-res-sec}), 
%we did not attempt to incentivize the remaining participants to participate again.
%Despite recruiting a subset of the $14$ graduate students, the debugging experience of the $8$ participants is similar to the debugging experience of all $14$ graduate students.
%The $8$ graduate students had a median of $7$ (vs. $6.5$) years of programming experience, the same median of $5$ years in Java, the same median of $3$ years of performance analysis experience, and a median of $3.5$ (vs. $4.5$) years working with configurable software systems. 
\looseness=-1

\paragraph*{Analysis}
We analyzed and compared transcripts of 
the 
audio and video recordings of the exploratory and validation studies to measure the time participants spend working on the task and their success rates.
Based on our exploratory study, participants were required to identify 
%the 
\IO, locate 
%the 
\OH, and trace the \CC\ to 
%successfully 
correctly debug the system.
%Additionally, w
We 
also
analyzed the interviews using standard qualitative research methods~\cite{S:CMQR15}.
The first author  conducted the study and analyzed the sessions independently, summarizing observations, discussions, and trends during the 
%debugging 
task and
%the 
interviews. 
All authors met weekly to discuss the observations. 
\looseness=-1

\paragraph*{Threats to Validity and Credibility}

We 
%purposely 
invited the same participants and use the same subject system as in our exploratory study.
%, to validate that the designed information providers support the information needs that we identified in our exploratory study.
Such a design might only validate the information needs 
%that participants have 
when 
%of 
debugging performance
%the specific task 
in the selected subject system.
Additionally, the exploratory and validation studies were conducted $5$ months apart, which might result in learning effects that help 
%the 
%$8$ 
participants in the 
%validation 
latter
study.
Furthermore, there is the threat that the task in the validation study is simpler.
While our later study 
%intentionally 
varies these aspects to observe whether our solutions generalize to other tasks, generalizations about our results
%in this study 
should be done with care.
%We mitigate this threat by conducting a confirmatory study on a more complex configurable software system with a new set of participants (Sec.~\ref{study-3-sec}).
%The confirmatory study will help us demonstrate that the concepts that we identified can potentially generalize to debugging the performance of other systems.
%As we will show in the results, these participants described how the information that \TOOL\ provides helps address the information needs that they had during the debugging process.
\looseness=-1

%We were able to recruit $8$ out of the $19$ participants from our exploratory study to debug with \TOOL.
%We can only speculate that we might observe similar results in the remaining participants.
%\looseness=-1

%While the exploratory and validation studies were conducted $5$ months apart, there could be some learning effects
%on how to approach our debugging task in our subject system, 
%which might help the 
%%$8$ 
%participants in the validation study.
%\looseness=-1

%The 
%debugging tasks of 
%exploratory and validation studies'
%tasks
%are comparable, but different.
%The massive effect size in our results makes a threat of the task in the validation study being simpler unlikely.
%While there is the threat that the 
%latter
%task 
%in the validation study 
%is simpler, the massive effect size in our results makes this threat %unlikely.
%\looseness=-1

% The participants were nowhere close to finishing when we stopped them at 50 minutes, so we strongly believe that they would not finish debugging if they were given more time.

%To further demonstrate that the concepts that we identified can potentially generalize to debugging the performance of other systems, we conducted a confirmatory study on a more complex configurable software system (Sec.~\ref{study-3-sec}).
%\looseness=-1

\subsubsection{Results\nopunct}
\label{study-2-res-sec}

\paragraph*{RQ4: Validating Usefulness of Information Providers.}

\begin{figure}[t]
\begin{center}
    \includegraphics[width=\columnwidth]{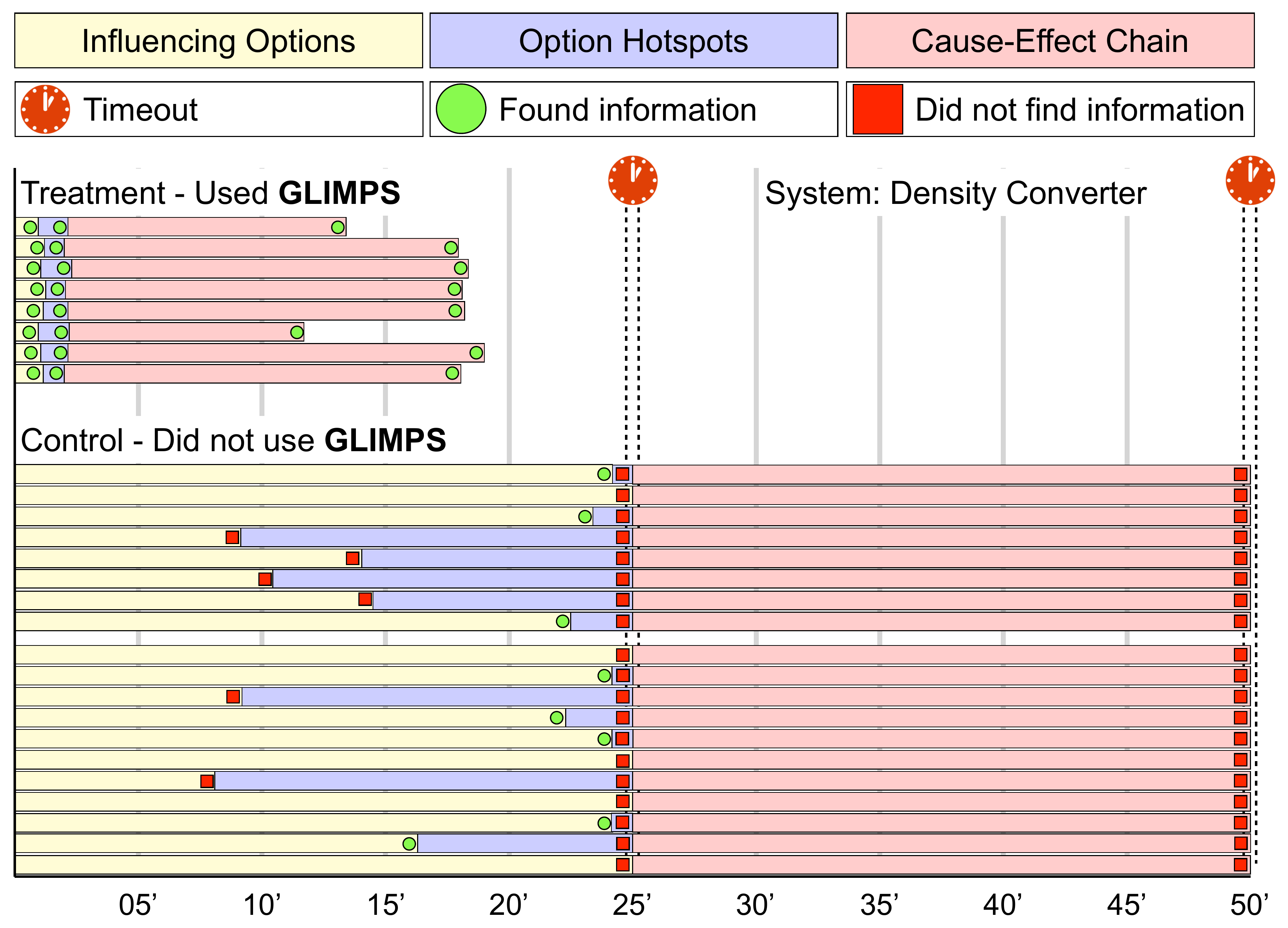}
\end{center}
\caption{
Time 
%spent 
looking for each piece of information when debugging 
%the 
performance 
%of \textit{Density Converter} 
with 
%(treatment) 
and without 
%(control) 
\TOOL.
The first $8$ participants who did not use \TOOL\ are the same participants who used our tool.
The data for the other participants who did not use \TOOL\ is included for reference.
\looseness=-1
}
\label{density-study-fig}
\end{figure}

Figure~\ref{density-study-fig} shows the time that each participant spent looking for each piece of information while debugging the performance of \textit{Density Converter} with (treatment) and without (control) \TOOL.
Overall, all participants who used our information providers identified the \IO, located \OH, and traced the \CC, and correctly explained the root cause of the 
%unexpected performance behavior 
performance issue 
in less than $19$ minutes.
By contrast, 
%as discussed in our exploratory study (Sec.~\ref{study-1-sec}), 
the $8$ participants could debug the unexpected behavior without our information providers
%\TOOL\ 
%within 
in 
$50$ minutes, when they struggled to find relevant information.\footnote{
We did not conduct a statistical significance test, since comparing completion rates is obvious: All participants correctly debugged with our tool, but no participants correctly debugged without our tool; comparing completion times cannot be done since nobody completed the task without our tool.
\looseness=-1
}
\looseness=-1

All $8$ participants who used our information providers
%from \TOOL\ 
identified the \IO\ and located the \OH\ in a few minutes.
%After finding these two pieces of information, 
Afterwards,
all participants traced the \CC\ and explained how the \IO\ caused the unexpected performance behavior in the \OH.
\looseness=-1

%By contrast, in the time that all participants who used the information from \TOOL\ correctly debugged the performance of the system, $18$ out of $19$ participants could not find a single piece of information when they did not use \TOOL.
When these $8$ participants did not use our information providers,
%\TOOL, 
%$18$ out of $19$ participants could not 
no participant found a single piece of information in the same 
%amount of 
timeframe as they did when using our information providers.
%they correctly debugged the system
%'s performance 
%when using 
%our information providers.
%\TOOL.
In fact, during a $25$-minute window, only $3$ 
%out of the $8$ 
participants found the \IO, and 
%none of the participants 
no participant 
found any \OH.
Furthermore, as described in our exploratory study, even when we \emph{explicitly told} participants (a)~the one influencing option that was causing the unexpected performance behavior and (b)~the \OH\ whose execution times drastically increased as a result of the problematic option, 
no participant
%none of the participants 
could trace the \CC\ and find the root cause of the unexpected behavior within $25$ minutes.
\looseness=-1

%\subsubsection{Discussion}

%\paragraph{\nopunct}

%\setlength{\textfloatsep}{5pt}

%The concepts that we identified and implemented in \TOOL\ provided participants with relevant information to satisfy the needs that they had while debugging the performance of configurable software systems.
After completing the task, the participants discussed how the information providers
%in \TOOL\ 
helped them debug the performance of the system, and contrasted their experience to debugging without our tool.
%debugging the subject system with and without our tool
%\TOOL.
%In particular, participants discussed how the different components of \TOOL\ helped them identify the \IO, locate the \OH, and trace the \CC.
%and contrasted their experiences to the time that they did not use \TOOL.
%\looseness=-1
All participants mentioned that the information providers
%in \TOOL\ 
helped them obtain relevant information for debugging.
%identify the \IO, locate the \OH, and trace the \CC.
The consensus 
%among participants
was that the information providers
%in \TOOL
\emph{``helped me focus on the relevant parts of the system''} to debug the unexpected performance behavior.
The participants contrasted this experience to the struggles that they faced when debugging without
our tool.
%\TOOL.
In particular, some participants remembered \emph{``being lost''} on what methods to follow or knowing \emph{``which parts of the program are relevant.''}
\looseness=-1

\vspace{1em}

\begin{mdframed}[backgroundcolor=gray!20, innerleftmargin=5pt, innerrightmargin=5pt] 
	\noindent{
	    \textit{RQ4: The validation study provides evidence that the designed information providers support the information needs that
	    we identified in our exploratory study.
	    %developers have when debugging the performance of a configurable software system.
	    Specifically, the information providers support developers' needs to (a)~identify \IO, (b)~locate \OH, and (c)~trace the \CC.
	    %To what extent do the concepts that we identified satisfy the information needs that developers have when debugging %the performance of configurable software systems
	    \looseness=-1
	    }
	}
\end{mdframed}

\subsection{Confirming Usefulness of Information Providers}
\label{study-3-sec}

%\todo[inline]{There are a lot of similarities between the confirmatory and validation studies. Mention that they are the same in some respects and different in others; explaining why we vary things. What is the goal of changing some aspects?}

After validating that our information providers support the information needs that we identified, we conducted a \emph{confirmatory} study to evaluate the extent that the information providers can potentially generalize to support 
the 
information needs 
%that developers have when
of
debugging the performance of complex configurable software systems.
\looseness=-1

\subsubsection{Method\nopunct}

\paragraph*{Study design}

We replicated the  validation study, intentionally \emph{varying} some aspects of the designs: We used a \emph{between-subject design} where we ask a \emph{new set of participants} to debug a \emph{more complex task} on a \emph{more complex subject system}, all working on the same task but using \emph{different tool support}.
With these variations, we evaluate that the results and \emph{massive effect size} in our previous studies are not due to, for example, a simpler task or learning effects,
%that benefited the participants in the validation study, 
but rather, that the information providers help developers debug the performance of complex configurable software systems, \emph{because} the information providers support the needs that developers have in this process.
%\looseness=-1

%With these variations, we can evaluate that the results are not because the task in our evaluation study was simpler or there were some learning effects that helped participants in the validation study, but rather that the information providers help debugging bevause they support the information needs that developers ahve in this process.

%To this end, we conducted a \emph{confirmatory} study, in which we replicate the exploratory and validation studies, \emph{varying} some aspects of the designs; particularly, we ask a \emph{new set of participants} to debug a \emph{more complex task} on a \emph{more complex subject system}.

As in our prior study, we conducted the confirmatory study using a think-aloud protocol~\cite{J:TAP10}, to compare how two new sets of participants
%, who had not been exposed to our performance debugging task, 
debug the performance of a complex configurable software system using different tool support in $60$ minutes.
The treatment group used \TOOL, while the control group used a simple plugin, which profiles and provides the execution time of the system under any configuration.
This information is the same that we gave participants in our exploratory study using a Wizard of Oz approach.
%(Sec.~\ref{study-1-sec}.
For this confirmatory study, however, we did not use a Wizard of Oz approach, as we wanted both groups to access information for debugging using a tool and the same IDE.
%For this confirmatory study, however, we did not use a Wizard of Oz approach, as we wanted both groups to debug the subject system using tool support.
%In this setting, both groups accessed information for debugging using a tool and the same IDE.
\looseness=-1

%Prior to the task, all participants from both groups were introduced to the information needs that we identified in our exploratory study.
%(Sec.~\ref{study-1-sec}). 
%We provided this information,
%to all participants,
%as the warm-up task using \TOOL\ presented the information providers, which describe relevant information needed for the task.
%the information needs.
%Hence, we wanted participants in both groups to know the information that might help them in the debugging process.
Similar to our prior study, participants worked on a warm-up task for $20$ minutes using either \TOOL\ or the simple plugin to learn how to use the information providers or the components that provided performance 
behavior 
information, respectively. 
We tested the simple plugin's design and implementation in a pilot study with $4$ graduate students from our personal network.
\looseness=-1

As in our prior study, we conducted a brief semi-structured interview, after the task, to discuss the participants’ experience in debugging the system.
%toi know whether \TOOL\ provides enoguh support to debug complex systems or there are more things that we should add to the tool, more info providers.
%Likewise, we wanted to know if the control group had any new needs in more complex systems.
In particular, we asked participants in the treatment group about the usefulness of the information providers 
%by \TOOL\ 
and whether there was additional information that they would like to have in the debugging process.
Similarly, we asked participants in the control group for the information that they would like to have when debugging the performance of configurable software systems.
%\looseness=-1

Due to the COVID-$19$ pandemic, we conducted the studies remotely over Zoom. 
Participants used Visual Studio Code through their preferred Web browser, which
%The IDE 
was running on a remote server and was configured with \TOOL\ and the simple plugin.
We asked participants to share their screen. 
With the participants' permission, we recorded audio and video of the sessions for subsequent analysis.
%\looseness=-1

\paragraph*{Task and subject system}

We prepared a more complex performance debugging task for a more complex configurable software system than the task and subject system in our exploratory and validation studies.
Similar to the prior studies, the task involved a user-defined configuration that spends an excessive amount of time executing.
%drastically increased the execution time of the system. 
Participants were asked to identify and explain which and how options caused the 
unexpected performance behavior.
%Similar to the previous studies, we did not explicitly tell participants that the performance problem was related to configurations.
In contrast to the previous studies, we introduced a bug caused by the incorrect implementation of \emph{an interaction of two options} 
%\footnote{
(The system spent a long time inserting duplicate data using transactions).
% 
%excessive execution time
%in the implementation.
%\looseness=-1
We selected \textit{Berkeley DB} as the subject system for the following reasons: ($1$) the system is implemented in Java, is open source, and is more complex than \textit{Density Converter} (over $150$K SLOC and over $30$ binary and non-binary options) and ($2$) the system has a complex performance behavior (execution time ranges from a couple of seconds to a few minutes, depending on the configuration).
%, ($3$) the global and local performance behavior of the system has been studied in prior work, and ($4$) the system has been analyzed alongside other configurable software systems that exhibit representative characteristics of numerous configurable software systems~\cite{VJSSAK:ASEJ20, VJSAK:ICSE21}.
\looseness=-1

\paragraph*{Participants}

We recruited $12$ graduate students, \emph{independent} of our exploratory and validation studies, with extensive experience analyzing the performance of configurable Java systems.
\looseness=-1

When determining the number of participants for the control group, we made some ethical considerations, while also ensuring that we obtain reliable results.
In our exploratory study, we observed $19$ \emph{experienced} researchers and professional software engineers who \emph{could not debug} the performance of a \emph{medium-sized} system with a performance bug caused by a \emph{single option} within $50$ minutes (see Figure~\ref{density-study-fig}).
With \textit{Berkeley DB}, we want to observe how participants debug
%aim to confirm that our designed information providers can potentially generalize to support the information needs that developers have when 
%debugging 
the performance of \emph{a more complex system}; a significantly larger system, in terms of SLOC and configuration space size, in which the unexpected behavior is caused by an \emph{interaction of two options}.
Based on (a)~the fact that we have \emph{strong empirical evidence} that debugging the performance of configurable software systems without relevant information is frustrating and is highly likely to not be completed under $60$ minutes and (b)~the massive effect size in our validation study between debugging with and without our information providers,
%in \TOOL, 
we decided to minimize the number of participants that we expect to struggle and fail to complete the task, while still having a reasonable number participants in the control group.
%\looseness=-1

Ultimately, we randomly assigned $4$ out of the $12$ participants to the control group,
%and the remaining participants to the treatment group, 
making sure to balance the groups in terms of the participants' debugging experience:
The median programming experience for both groups is $6$ years, a median of $3.5$ years in Java, a median of $2.2$ years of performance analysis experience, and a median of $2.7$ years working with configurable software systems.
%Participants in the control group had a median of $6$ years of programming experience, a median of $3.5$ years in Java, a median of $2.3$ years of performance analysis experience, and a median of $3$ years working with configurable software systems. 
%Participants in the treatment group had a median of $6$ years of programming experience, a median of $4$ years in Java, a median of $2$ years of performance analysis experience, and a median of $2.5$ years working with configurable software systems. 
%\looseness=-1

\paragraph*{Analysis}
We analyzed transcripts of the audio and video recordings to measure the time participants spend working on the task and their success rates.
Based on our exploratory and validation studies, participants needed to identify \IO, locate \OH, and trace the \CC\ to successfully debug the system.
Additionally, we analyzed the interviews using standard qualitative research methods~\cite{S:CMQR15}.
The first author conducted the study and analyzed the sessions independently, summarizing observations, discussions, and trends during the debugging task and the interviews. 
All authors met weekly to discuss the observations. 
%\looseness=-1

\paragraph*{Threats to Validity and Credibility}

%As we will show in the results, the participants in the control group mentioned the same struggles and barriers as the participants in our exploratory study.
%Similarly, the participants in the treatment group made the same observations about the usefulness of the information provided by \TOOL.
%\looseness=-1
While we aimed to increase the complexity of the performance debugging task,
%debugging task with \textit{Berkeley DB} and a performance bug caused by the interaction of two options, 
readers should be careful when generalizing our results to other complex systems.
\looseness=-1

Our control group consisted of $4$ participants.
As argued previously, we did not recruit more participants due to the struggles that we observed in our exploratory study on a simpler system and the massive effect size in our validation study between debugging with and without our information providers.
%in \TOOL.
Nevertheless, readers should be careful when generalizing our results.
\looseness=-1

While the control group had access to the IDE's debugger and used a simple plugin,
%that provided performance behavior information, 
we might obtain different results if the participants had used other debugging tools and techniques. 
\looseness=-1

%Specifically, we introduced a performance bug such that the bug could be identified with the information that we provide within $30$ minutes.
%Nevertheless, our results demonstrate that the concepts that we identified provide relevant information to guide developers in the process of debugging the performance of configurable software systems
%and (b)~has the potential compared to not having access to such information.
%that the information that we show helps developers to (a)~concentrate on the relevant parts and information to debug the performance of configurable software systems and (b)~can help them debug the performance of configurable software systems faster compare to developers who do not have access to this relevant information.
\looseness=-1

\begin{figure}[t]
\begin{center}
    \includegraphics[width=\columnwidth]{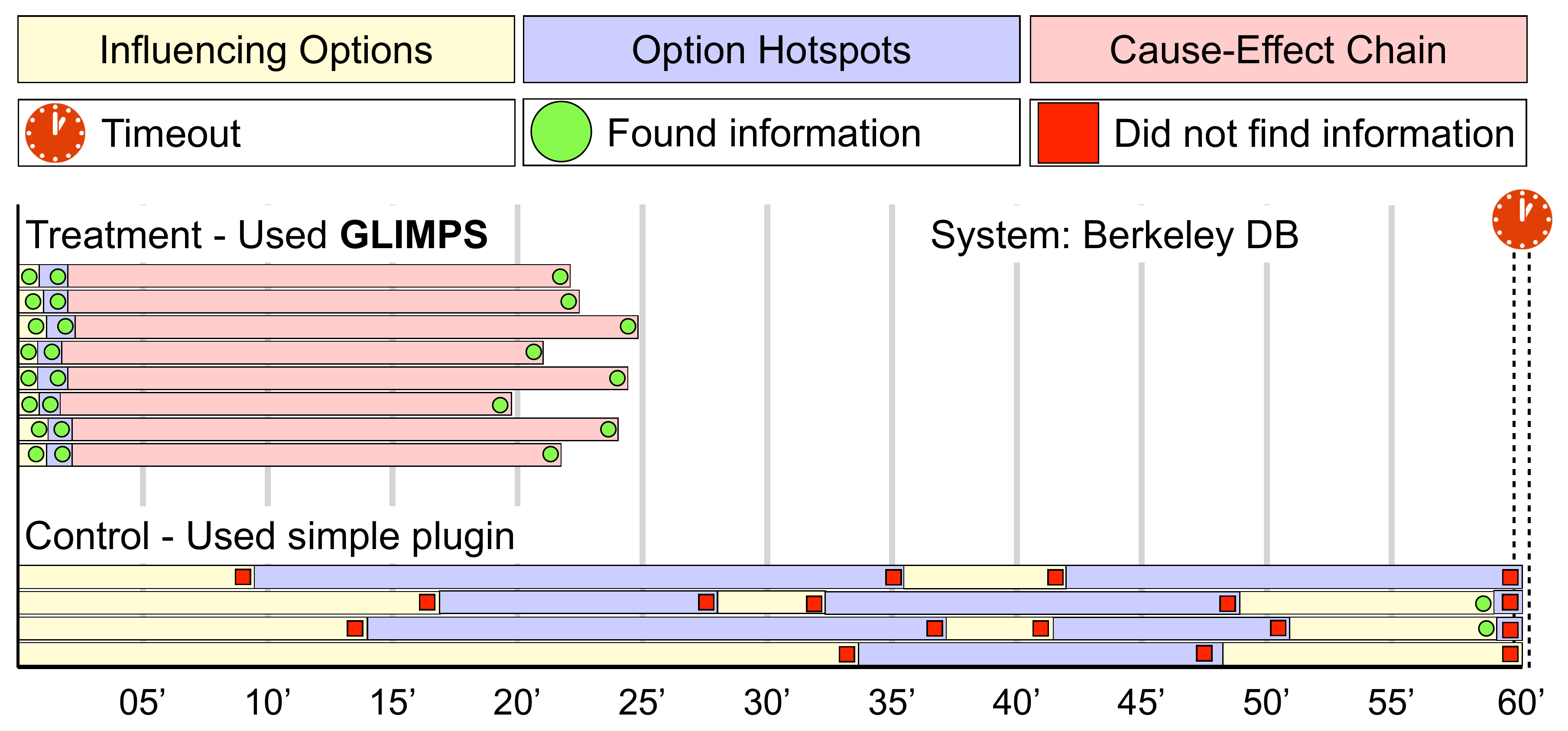}
\end{center}
\caption{
Time 
%spent 
looking for each piece of information when debugging
%the 
performance 
%of \textit{Berkeley DB} 
with different tool support.
%(treatment) and without (control) \TOOL.
\looseness=-1
}
\label{berkeley-study-fig}
\end{figure}

\subsubsection{Results\nopunct}

\paragraph*{RQ4: Confirming Usefulness of Information Providers.}

Figure~\ref{berkeley-study-fig} shows the time 
%that 
each participant spent looking for each piece of information while debugging the performance of \textit{Berkeley DB} with \TOOL\ (treatment) and the simple plugin (control).
Similar to 
%the results of 
our validation study,
%(Sec.~\ref{study-2-sec}), 
all participants who used our information providers identified the \IO, located the \OH, and traced the \CC\
%, and correctly explained the root cause of the unexpected behavior 
in less than $25$ minutes.
By contrast, the participants who did not use 
our information providers
%\TOOL\ 
struggled for $60$ minutes and could not debug the 
%unexpected behavior in the 
%performance of the
system.\footnote{
Again, we did not conduct a statistical significance test, since comparing completion rates is obvious: All participants in the treatment group correctly debugged the system, but no participant in the control group did; comparing completion times cannot be done since nobody in the control group completed the task.
\looseness=-1
}
\looseness=-1

%All participants who used the information from \TOOL\ identified the \IO, located the \OH, and traced the \CC.
%Furthermore, all participants explained how the interaction of \IO\ affected the performance of the option hotspots to cause the unexpected performance behavior.
%\looseness=-1

%%By contrast, in the time that all participants who used the information from \TOOL\ correctly debugged the performance of the system, $18$ out of $19$ participants could not find a single piece of information when they did not use \TOOL.
%By contrast, none of the participants who did not use \TOOL\ could debug the performance of the system in $60$.
%Only $2$ participants identified the \IO\ after a couple of iterations of trying to find the \IO\ and the \OH.
%\looseness=-1

%\subsubsection{Discussion}

%\paragraph{\nopunct}
While working on the task, we observed the participants in the treatment group looking for the same information as in Table~\ref{info-needs-tbl} and using our information providers
%\TOOL\ 
similarly to the participants in the validation study to find 
%relevant 
information to debug 
%the performance in 
the subject system.
%to debug the performance of the subject system.
%information they performed similar activities as those performed by the %participants in our exploratory
%%(Sec.~\ref{study-1-sec}) 
%and validation 
%%(Sec.~\ref{study-2-sec}) 
%studies.
%For instance, participants in the treatment group used the various components of \TOOL\ to identify the \IO, locate the \OH, and trace the \CC.
Likewise, the participants in the control group struggled while performing the same activities as those listed in Table~\ref{info-needs-tbl} when trying to identify the \IO\ and locate the \OH.
\looseness=-1

%\paragraph{\nopunct}

After working on the task, all participants discussed their experience in debugging the performance of the system using tool support.
%\looseness=-1
Similar to the discussion in our validation study,
%(Sec.~\ref{study-2-sec}), 
all participants in the treatment group commented how the information providers
%in \TOOL\ 
helped them identify \IO, locate \OH, and trace the \CC.
%\looseness=-1
Likewise, the participants who used the simple plugin described similar struggles and barriers as those mentioned 
%by the participants
in our exploratory study.
%(Sec.~\ref{study-1-sec}). 
All participants in this group mentioned that identifying the \IO\ that cause the expected behavior is \emph{``difficult''} and locating the \OH\ is \emph{``challenging.''}
However, none of the participants in this group commented on tracing the \CC, as they never got to that point in the debugging process.
\looseness=-1

%Since none of the participants in this group traced the \CC, as they did not identify the \IO\ and locate the \OH, none of them discussed their experiences tracing the \CC.
%However, we conjecture that 

%Talk aobut that particpants spent about the same amoutn of time finding the \IO\ and option hotspots when given the information from \TOOL. 
%Where the two groups differ is on the time that they spend finding the \CC.
%While both groups had the same information, it shows that the second group was analyzing a more complicated system, as it took them a longer time to understand how \emph{the interaction of two options} was causing the unexpected performance beavhior.

%\section{Other Scenarios}?

\vspace{1em}

\begin{mdframed}[backgroundcolor=gray!20, innerleftmargin=5pt, innerrightmargin=5pt] 
	\noindent{\textit{RQ4: The confirmatory study provides evidence that the designed information providers help developers debug the performance of complex configurable software systems because the information providers support the needs that developers have in this process.
        %To what extent do the concepts that we identified satisfy the information needs that developers have when debugging %the performance of configurable software systems
	    \looseness=-1
	    }
	}
\end{mdframed}

%\section{Related Work}

%\todo[inline]{Is this section needed?}

%\todo[inline]{Is this section needed?}

\section{Conclusion}

We identified the information needs -- \hlyellow{influencing options}, \hlblue{option hotspots}, and \hlred{cause-effect chain} -- that developers have when debugging the performance of configurable software systems.
Subsequently, we designed and implemented information providers, adapted from global and local performance-influence models, CPU profiling, and program slicing, that support the above needs.
Two user studies, with a total of $20$ developers, validate and confirm that our designed information providers help developers debug the performance of complex configurable software systems.
%\GT lobal performance-influence models help answer "which options influence performance?", \LT ocal performance-influence models help answer "where options influence performance?", and CPU \PT rofiling and program \ST licing help answer "how do options influence performance in the implementation?".
\looseness=-1

\section{Acknowledgments}
We want to thank James Herbsleb, for his guidance on conducting a Wizard of Oz experiment, Thomas LaToza, for his advice on how to analyze and code sessions, and Rohan Padhye, Janet Siegmund, and Bogdan Vasilescu, for their feedback on the design of our confirmatory user study.
This work was supported in part by the NSF (Awards 2007202, 2038080, and 2107463), NASA (Awards 80NSSC20K1720 and 521418-SC), the Software Engineering Institute, the German Research Foundation (SI 2171/2, SI 2171/3-1, and AP 206/11-1, and Grant 389792660 as part of TRR 248 -- CPEC), and the German Federal Ministry of Education and Research (AgileAI: 01IS19059A and 01IS18026B) by funding the competence center for Big Data and AI "ScaDS.AI Dresden/Leipzig".

\balance
\bibliographystyle{ACM-Reference-Format}
\bibliography{bibliography.bib}


\begin{thebibliography}{90}

%%% the Web address.

\ifx \showCODEN    \undefined \def \showCODEN     #1{\unskip}     \fi
\ifx \showDOI      \undefined \def \showDOI       #1{#1}\fi
\ifx \showISBNx    \undefined \def \showISBNx     #1{\unskip}     \fi
\ifx \showISBNxiii \undefined \def \showISBNxiii  #1{\unskip}     \fi
\ifx \showISSN     \undefined \def \showISSN      #1{\unskip}     \fi
\ifx \showLCCN     \undefined \def \showLCCN      #1{\unskip}     \fi
\ifx \shownote     \undefined \def \shownote      #1{#1}          \fi
\ifx \showarticletitle \undefined \def \showarticletitle #1{#1}   \fi
\ifx \showURL      \undefined \def \showURL       {\relax}        \fi
% invisible to TeX
\providecommand\bibfield[2]{#2}
\providecommand\bibinfo[2]{#2}
\providecommand\natexlab[1]{#1}
\providecommand\showeprint[2][]{arXiv:#2}

\bibitem[\protect\citeauthoryear{Abal, Melo, St\u{a}nciulescu, Brabrand,
  Ribeiro, and W\k{a}sowski}{Abal et~al\mbox{.}}{2018}]%
        {AMSBRW:TOSEM18}
\bibfield{author}{\bibinfo{person}{Iago Abal}, \bibinfo{person}{Jean Melo},
  \bibinfo{person}{\c{S}tefan St\u{a}nciulescu}, \bibinfo{person}{Claus
  Brabrand}, \bibinfo{person}{M\'{a}rcio Ribeiro}, {and}
  \bibinfo{person}{Andrzej W\k{a}sowski}.} \bibinfo{year}{2018}\natexlab{}.
\newblock \showarticletitle{Variability Bugs in Highly Configurable Systems: A
  Qualitative Analysis}.
\newblock \bibinfo{journal}{\emph{ACM Trans. Softw. Eng. Methodol. (TOSEM)}}
  \bibinfo{volume}{26}, \bibinfo{number}{3}, Article \bibinfo{articleno}{10}
  (\bibinfo{date}{Jan.} \bibinfo{year}{2018}), \bibinfo{numpages}{34}~pages.
\newblock


\bibitem[\protect\citeauthoryear{Adamoli and Hauswirth}{Adamoli and
  Hauswirth}{2010}]%
        {AH:SOFTVIS10}
\bibfield{author}{\bibinfo{person}{Andrea Adamoli} {and}
  \bibinfo{person}{Matthias Hauswirth}.} \bibinfo{year}{2010}\natexlab{}.
\newblock \showarticletitle{Trevis: A Context Tree Visualization and Analysis
  Framework and Its Use for Classifying Performance Failure Reports}. In
  \bibinfo{booktitle}{\emph{Proc. Int'l Symposium Software Visualization
  (SOFTVIS)}} (Salt Lake City, UT, USA). \bibinfo{publisher}{ACM},
  \bibinfo{address}{New York, NY, USA}, \bibinfo{pages}{73--82}.
\newblock


\bibitem[\protect\citeauthoryear{Agrawal and Horgan}{Agrawal and
  Horgan}{1990}]%
        {AH:ASP90}
\bibfield{author}{\bibinfo{person}{Hiralal Agrawal} {and}
  \bibinfo{person}{Joseph~R Horgan}.} \bibinfo{year}{1990}\natexlab{}.
\newblock \showarticletitle{Dynamic {P}rogram {S}licing}.
\newblock \bibinfo{journal}{\emph{ACM SIGPlan Notices}} \bibinfo{volume}{25},
  \bibinfo{number}{6} (\bibinfo{year}{1990}), \bibinfo{pages}{246--256}.
\newblock


\bibitem[\protect\citeauthoryear{Alam, Liu, Zeng, and Muzahid}{Alam
  et~al\mbox{.}}{2017}]%
        {ALZM:EUROSYS17}
\bibfield{author}{\bibinfo{person}{Mohammad Mejbah~ul Alam},
  \bibinfo{person}{Tongping Liu}, \bibinfo{person}{Guangming Zeng}, {and}
  \bibinfo{person}{Abdullah Muzahid}.} \bibinfo{year}{2017}\natexlab{}.
\newblock \showarticletitle{SyncPerf: Categorizing, Detecting, and Diagnosing
  Synchronization Performance Bugs}. In \bibinfo{booktitle}{\emph{Proc.
  European Conference on Computer Systems (EuroSys)}} (Belgrade, Serbia).
  \bibinfo{publisher}{ACM}, \bibinfo{address}{New York, NY, USA},
  \bibinfo{pages}{298--313}.
\newblock


\bibitem[\protect\citeauthoryear{Andrzejewski, Mulhern, Liblit, and
  Zhu}{Andrzejewski et~al\mbox{.}}{2007}]%
        {AMLZ:ECML07}
\bibfield{author}{\bibinfo{person}{David Andrzejewski}, \bibinfo{person}{Anne
  Mulhern}, \bibinfo{person}{Ben Liblit}, {and} \bibinfo{person}{Xiaojin Zhu}.}
  \bibinfo{year}{2007}\natexlab{}.
\newblock \showarticletitle{Statistical Debugging Using Latent Topic Models}.
  In \bibinfo{booktitle}{\emph{Proc. European Conf. Machine Learning}} (Warsaw,
  Poland). \bibinfo{publisher}{Springer-Verlag}, \bibinfo{address}{Berlin,
  Heidelberg}, \bibinfo{pages}{6--17}.
\newblock


\bibitem[\protect\citeauthoryear{Bezemer, Pouwelse, and Gregg}{Bezemer
  et~al\mbox{.}}{2015}]%
        {BPG:SANER15}
\bibfield{author}{\bibinfo{person}{Cor-Paul Bezemer}, \bibinfo{person}{J.A.
  Pouwelse}, {and} \bibinfo{person}{Brendan Gregg}.}
  \bibinfo{year}{2015}\natexlab{}.
\newblock \showarticletitle{Understanding Software {P}erformance {R}egressions
  {U}sing {D}ifferential {F}lame {G}raphs}. In \bibinfo{booktitle}{\emph{Int'l
  Conf. Software Analysis, Evolution, and Reengineering (SANER)}} (Montreal,
  Canada). \bibinfo{publisher}{IEEE}, \bibinfo{address}{Los Alamitos, CA, USA},
  \bibinfo{pages}{535--539}.
\newblock


\bibitem[\protect\citeauthoryear{Bornholt and Torlak}{Bornholt and
  Torlak}{2018}]%
        {BT:OOPSLA18}
\bibfield{author}{\bibinfo{person}{James Bornholt} {and} \bibinfo{person}{Emina
  Torlak}.} \bibinfo{year}{2018}\natexlab{}.
\newblock \showarticletitle{Finding Code That Explodes Under Symbolic
  Evaluation}.
\newblock \bibinfo{journal}{\emph{Proc. Int'l Conf. Object-Oriented
  Programming, Systems, Languages and Applications (OOPSLA)}}
  \bibinfo{volume}{2}, Article \bibinfo{articleno}{149} (\bibinfo{date}{Oct.}
  \bibinfo{year}{2018}), \bibinfo{numpages}{26}~pages.
\newblock


\bibitem[\protect\citeauthoryear{Breu, Premraj, Sillito, and Zimmermann}{Breu
  et~al\mbox{.}}{2010}]%
        {BPSZ:CSCW10}
\bibfield{author}{\bibinfo{person}{Silvia Breu}, \bibinfo{person}{Rahul
  Premraj}, \bibinfo{person}{Jonathan Sillito}, {and} \bibinfo{person}{Thomas
  Zimmermann}.} \bibinfo{year}{2010}\natexlab{}.
\newblock \showarticletitle{Information Needs in Bug Reports: Improving
  Cooperation between Developers and Users}. In \bibinfo{booktitle}{\emph{Proc.
  Conf. Computer Supported Cooperative Work (CSCW)}} (Savannah, GA, USA).
  \bibinfo{publisher}{ACM}, \bibinfo{address}{New York, NY, USA},
  \bibinfo{pages}{301--310}.
\newblock


\bibitem[\protect\citeauthoryear{Burg, Bailey, Ko, and Ernst}{Burg
  et~al\mbox{.}}{2013}]%
        {BBKE:UIST13}
\bibfield{author}{\bibinfo{person}{Brian Burg}, \bibinfo{person}{Richard
  Bailey}, \bibinfo{person}{Andrew~J. Ko}, {and} \bibinfo{person}{Michael~D.
  Ernst}.} \bibinfo{year}{2013}\natexlab{}.
\newblock \showarticletitle{Interactive Record/Replay for Web Application
  Debugging}. In \bibinfo{booktitle}{\emph{Proc. Symposium User Interface
  Software and Technology (UIST)}} (St. Andrews, Scotland, United Kingdom).
  \bibinfo{publisher}{ACM}, \bibinfo{address}{New York, NY, USA},
  \bibinfo{pages}{473--484}.
\newblock


\bibitem[\protect\citeauthoryear{Castro, Akel, Petit, Popov, and Jalby}{Castro
  et~al\mbox{.}}{2015}]%
        {CAPPJ:TACO15}
\bibfield{author}{\bibinfo{person}{Pablo De~Oliveira Castro},
  \bibinfo{person}{Chadi Akel}, \bibinfo{person}{Eric Petit},
  \bibinfo{person}{Mihail Popov}, {and} \bibinfo{person}{William Jalby}.}
  \bibinfo{year}{2015}\natexlab{}.
\newblock \showarticletitle{CERE: LLVM-Based Codelet Extractor and REplayer for
  Piecewise Benchmarking and Optimization}.
\newblock \bibinfo{journal}{\emph{ACM Trans. Archit. Code Optim. (TACO)}}
  \bibinfo{volume}{12}, \bibinfo{number}{1}, Article \bibinfo{articleno}{6}
  (\bibinfo{date}{April} \bibinfo{year}{2015}), \bibinfo{numpages}{24}~pages.
\newblock


\bibitem[\protect\citeauthoryear{Chaparro, Lu, Zampetti, Moreno, Di~Penta,
  Marcus, Bavota, and Ng}{Chaparro et~al\mbox{.}}{2017}]%
        {CLZMDMBN:ESECFSE17}
\bibfield{author}{\bibinfo{person}{Oscar Chaparro}, \bibinfo{person}{Jing Lu},
  \bibinfo{person}{Fiorella Zampetti}, \bibinfo{person}{Laura Moreno},
  \bibinfo{person}{Massimiliano Di~Penta}, \bibinfo{person}{Andrian Marcus},
  \bibinfo{person}{Gabriele Bavota}, {and} \bibinfo{person}{Vincent Ng}.}
  \bibinfo{year}{2017}\natexlab{}.
\newblock \showarticletitle{Detecting Missing Information in Bug Descriptions}.
  In \bibinfo{booktitle}{\emph{Proc. Europ. Software Engineering Conf.
  Foundations of Software Engineering (ESEC/FSE)}} (Paderborn, Germany).
  \bibinfo{publisher}{ACM}, \bibinfo{address}{New York, NY, USA},
  \bibinfo{pages}{396--407}.
\newblock


\bibitem[\protect\citeauthoryear{Cito, Leitner, Bosshard, Knecht, Mazlami, and
  Gall}{Cito et~al\mbox{.}}{2018}]%
        {CLBKMG:ICSECP18}
\bibfield{author}{\bibinfo{person}{J\"{u}rgen Cito}, \bibinfo{person}{Philipp
  Leitner}, \bibinfo{person}{Christian Bosshard}, \bibinfo{person}{Markus
  Knecht}, \bibinfo{person}{Genc Mazlami}, {and} \bibinfo{person}{Harald~C.
  Gall}.} \bibinfo{year}{2018}\natexlab{}.
\newblock \showarticletitle{Performance{H}at: Augmenting Source Code with
  Runtime Performance Traces in the {IDE}}. In \bibinfo{booktitle}{\emph{Proc.
  Int'l Conf. Software Engineering: Companion Proceeedings}} (Gothenburg,
  Sweden). \bibinfo{publisher}{ACM}, \bibinfo{address}{New York, NY, USA},
  \bibinfo{pages}{41--44}.
\newblock


\bibitem[\protect\citeauthoryear{Curtsinger and Berger}{Curtsinger and
  Berger}{2016}]%
        {CB:ATC16}
\bibfield{author}{\bibinfo{person}{Charlie Curtsinger} {and}
  \bibinfo{person}{Emery~D. Berger}.} \bibinfo{year}{2016}\natexlab{}.
\newblock \showarticletitle{{COZ}: Finding Code that Counts with Causal
  Profiling}. In \bibinfo{booktitle}{\emph{USENIX Annual Technical Conference
  (ATC)}}. \bibinfo{publisher}{{USENIX} Association}, \bibinfo{address}{Denver,
  CO, USA}, \bibinfo{pages}{184--197}.
\newblock


\bibitem[\protect\citeauthoryear{Dahlb{\"a}ck, J{\"o}nsson, and
  Ahrenberg}{Dahlb{\"a}ck et~al\mbox{.}}{1993}]%
        {DJA:KBS93}
\bibfield{author}{\bibinfo{person}{Nils Dahlb{\"a}ck}, \bibinfo{person}{Arne
  J{\"o}nsson}, {and} \bibinfo{person}{Lars Ahrenberg}.}
  \bibinfo{year}{1993}\natexlab{}.
\newblock \showarticletitle{Wizard of Oz {S}tudies—{W}hy and {H}ow}.
\newblock \bibinfo{journal}{\emph{Knowledge-Based Systems}}
  \bibinfo{volume}{6}, \bibinfo{number}{4} (\bibinfo{year}{1993}),
  \bibinfo{pages}{258--266}.
\newblock


\bibitem[\protect\citeauthoryear{EJ-technologies}{EJ-technologies}{2019}]%
        {JPROFILER10}
\bibfield{author}{\bibinfo{person}{EJ-technologies}.}
  \bibinfo{year}{2019}\natexlab{}.
\newblock \bibinfo{booktitle}{\emph{JProfiler 10}}.
\newblock EJ-technologies.
\newblock
\urldef\tempurl%
\url{https://www.ej-technologies.com/products/jprofiler/overview.html}
\showURL{%
Retrieved December 10, 2019 from \tempurl}


\bibitem[\protect\citeauthoryear{Farooqui, Rana, and Jafari}{Farooqui
  et~al\mbox{.}}{2019}]%
        {FRJ:CCODE19}
\bibfield{author}{\bibinfo{person}{Tayba Farooqui}, \bibinfo{person}{Tauseef
  Rana}, {and} \bibinfo{person}{Fakeeha Jafari}.}
  \bibinfo{year}{2019}\natexlab{}.
\newblock \showarticletitle{Impact of Human-Centered Design Process (HCDP) on
  Software Development Process}. In \bibinfo{booktitle}{\emph{Int'l Conf.
  Communication, Computing and Digital systems (C-CODE)}} (Islamabad,
  Pakistan). \bibinfo{publisher}{IEEE}, \bibinfo{address}{Los Alamitos, CA,
  USA}, \bibinfo{pages}{110--114}.
\newblock


\bibitem[\protect\citeauthoryear{Fu, Cai, and Li}{Fu et~al\mbox{.}}{2020}]%
        {FCL:FSE20}
\bibfield{author}{\bibinfo{person}{Xiaoqin Fu}, \bibinfo{person}{Haipeng Cai},
  {and} \bibinfo{person}{Li Li}.} \bibinfo{year}{2020}\natexlab{}.
\newblock \showarticletitle{Dads: Dynamic Slicing Continuously-Running
  Distributed Programs with Budget Constraints}. In
  \bibinfo{booktitle}{\emph{Proc. Int'l Symp. Foundations of Software
  Engineering (FSE)}} (Virtual Event, USA). \bibinfo{publisher}{ACM},
  \bibinfo{address}{New York, NY, USA}, \bibinfo{pages}{1566--1570}.
\newblock


\bibitem[\protect\citeauthoryear{Giffhorn}{Giffhorn}{2011}]%
        {G:SQJ11}
\bibfield{author}{\bibinfo{person}{Dennis Giffhorn}.}
  \bibinfo{year}{2011}\natexlab{}.
\newblock \showarticletitle{Advanced {C}hopping of {S}equential and
  {C}oncurrent {P}rograms}.
\newblock \bibinfo{journal}{\emph{Software Quality Journal}}
  \bibinfo{volume}{19}, \bibinfo{number}{2} (\bibinfo{year}{2011}),
  \bibinfo{pages}{239--294}.
\newblock


\bibitem[\protect\citeauthoryear{Graf, Hecker, and Mohr}{Graf
  et~al\mbox{.}}{2012}]%
        {GHM:KIT12}
\bibfield{author}{\bibinfo{person}{J{\"u}rgen Graf}, \bibinfo{person}{Martin
  Hecker}, {and} \bibinfo{person}{Martin Mohr}.}
  \bibinfo{year}{2012}\natexlab{}.
\newblock \bibinfo{booktitle}{\emph{Using JOANA for Information Flow Control in
  Java Programs - A Practical Guide}}.
\newblock \bibinfo{type}{{T}echnical {R}eport}~24.
  \bibinfo{institution}{Karlsruhe Institute of Technology}.
\newblock


\bibitem[\protect\citeauthoryear{Grebhahn, Siegmund, and Apel}{Grebhahn
  et~al\mbox{.}}{2019}]%
        {GSA:PPSCNSB19}
\bibfield{author}{\bibinfo{person}{Alexander Grebhahn},
  \bibinfo{person}{Norbert Siegmund}, {and} \bibinfo{person}{Sven Apel}.}
  \bibinfo{year}{2019}\natexlab{}.
\newblock \bibinfo{title}{Predicting Performance of Software Configurations:
  There is no Silver Bullet}.
\newblock
\newblock
\showeprint[arxiv]{1911.12643}~[cs.SE]


\bibitem[\protect\citeauthoryear{Gregg}{Gregg}{2016}]%
        {G:CAMC16}
\bibfield{author}{\bibinfo{person}{Brendan Gregg}.}
  \bibinfo{year}{2016}\natexlab{}.
\newblock \showarticletitle{The Flame Graph}.
\newblock \bibinfo{journal}{\emph{Commun. ACM}} \bibinfo{volume}{59},
  \bibinfo{number}{6} (\bibinfo{date}{May} \bibinfo{year}{2016}),
  \bibinfo{pages}{48--57}.
\newblock


\bibitem[\protect\citeauthoryear{Guo, Czarnecki, Apel, Siegmund, and
  W{\k{a}}sowski}{Guo et~al\mbox{.}}{2013}]%
        {GCASW:ASE13}
\bibfield{author}{\bibinfo{person}{Jianmei Guo}, \bibinfo{person}{Krzysztof
  Czarnecki}, \bibinfo{person}{Sven Apel}, \bibinfo{person}{Norbert Siegmund},
  {and} \bibinfo{person}{Andrzej W{\k{a}}sowski}.}
  \bibinfo{year}{2013}\natexlab{}.
\newblock \showarticletitle{Variability-Aware Performance Prediction: A
  Statistical Learning Approach}. In \bibinfo{booktitle}{\emph{Proc. Int'l
  Conf. Automated Software Engineering (ASE)}} (Silicon Valley, CA, USA).
  \bibinfo{publisher}{ACM}, \bibinfo{address}{New York, NY, USA},
  \bibinfo{pages}{301--311}.
\newblock


\bibitem[\protect\citeauthoryear{Ha and Zhang}{Ha and Zhang}{2019}]%
        {HZ:ICSE19}
\bibfield{author}{\bibinfo{person}{Huong Ha} {and} \bibinfo{person}{Hongyu
  Zhang}.} \bibinfo{year}{2019}\natexlab{}.
\newblock \showarticletitle{Deep{P}erf: Performance Prediction for Configurable
  Software with Deep Sparse Neural Network}. In \bibinfo{booktitle}{\emph{Proc.
  Int'l Conf. Software Engineering (ICSE)}} (Montreal, Quebec, Canada).
  \bibinfo{publisher}{IEEE}, \bibinfo{address}{Los Alamitos, CA, USA},
  \bibinfo{pages}{1095--1106}.
\newblock


\bibitem[\protect\citeauthoryear{{Ha} and {Zhang}}{{Ha} and {Zhang}}{2019}]%
        {HZ:ICSME19}
\bibfield{author}{\bibinfo{person}{H. {Ha}} {and} \bibinfo{person}{H.
  {Zhang}}.} \bibinfo{year}{2019}\natexlab{}.
\newblock \showarticletitle{Performance-Influence Model for Highly Configurable
  Software with Fourier Learning and Lasso Regression}. In
  \bibinfo{booktitle}{\emph{Proc. Int'l Conf. Software Maintance and Evolution
  (ICSME)}}. \bibinfo{publisher}{IEEE}, \bibinfo{address}{Los Alamitos, CA,
  USA}, \bibinfo{pages}{470--480}.
\newblock


\bibitem[\protect\citeauthoryear{Han, Dang, Ge, Zhang, and Xie}{Han
  et~al\mbox{.}}{2012}]%
        {HDGZX:ICSE12}
\bibfield{author}{\bibinfo{person}{Shi Han}, \bibinfo{person}{Yingnong Dang},
  \bibinfo{person}{Song Ge}, \bibinfo{person}{Dongmei Zhang}, {and}
  \bibinfo{person}{Tao Xie}.} \bibinfo{year}{2012}\natexlab{}.
\newblock \showarticletitle{Performance Debugging in the Large via Mining
  Millions of Stack Traces}. In \bibinfo{booktitle}{\emph{Proc. Int'l Conf.
  Software Engineering (ICSE)}} (Zurich, Switzerland).
  \bibinfo{publisher}{IEEE}, \bibinfo{address}{Piscataway, NJ, USA},
  \bibinfo{pages}{145--155}.
\newblock


\bibitem[\protect\citeauthoryear{Han and Yu}{Han and Yu}{2016}]%
        {HY:ESEM16}
\bibfield{author}{\bibinfo{person}{Xue Han} {and} \bibinfo{person}{Tingting
  Yu}.} \bibinfo{year}{2016}\natexlab{}.
\newblock \showarticletitle{An Empirical Study on Performance Bugs for Highly
  Configurable Software Systems}. In \bibinfo{booktitle}{\emph{Proc. Int'l
  Symposium Empirical Software Engineering and Measurement (ESEM)}} (Ciudad
  Real, Spain). \bibinfo{publisher}{ACM}, \bibinfo{address}{New York, NY, USA},
  Article \bibinfo{articleno}{23}, \bibinfo{numpages}{10}~pages.
\newblock


\bibitem[\protect\citeauthoryear{Han, Yu, and Lo}{Han et~al\mbox{.}}{2018}]%
        {HYL:ASE18}
\bibfield{author}{\bibinfo{person}{Xue Han}, \bibinfo{person}{Tingting Yu},
  {and} \bibinfo{person}{David Lo}.} \bibinfo{year}{2018}\natexlab{}.
\newblock \showarticletitle{PerfLearner: Learning from Bug Reports to
  Understand and Generate Performance Test Frames}. In
  \bibinfo{booktitle}{\emph{Proc. Int'l Conf. Automated Software Engineering
  (ASE)}} (Montpellier, France). \bibinfo{publisher}{ACM},
  \bibinfo{address}{New York, NY, USA}, \bibinfo{pages}{17--28}.
\newblock


\bibitem[\protect\citeauthoryear{Han, Yu, and Pradel}{Han
  et~al\mbox{.}}{2021}]%
        {HYP:ICPE21}
\bibfield{author}{\bibinfo{person}{Xue Han}, \bibinfo{person}{Tingting Yu},
  {and} \bibinfo{person}{Michael Pradel}.} \bibinfo{year}{2021}\natexlab{}.
\newblock \showarticletitle{ConfProf: White-Box Performance Profiling of
  Configuration Options}. In \bibinfo{booktitle}{\emph{Proc. Int'l Conf.
  Performance Engineering (ICPE)}}. \bibinfo{publisher}{ACM},
  \bibinfo{address}{New York, NY, USA}, \bibinfo{pages}{1--8}.
\newblock


\bibitem[\protect\citeauthoryear{He, Jia, Li, Xu, Yu, Yu, Wang, and Liao}{He
  et~al\mbox{.}}{2020}]%
        {HJLXYYWL:ASE20}
\bibfield{author}{\bibinfo{person}{Haochen He}, \bibinfo{person}{Zhouyang Jia},
  \bibinfo{person}{Shanshan Li}, \bibinfo{person}{Erci Xu},
  \bibinfo{person}{Tingting Yu}, \bibinfo{person}{Yue Yu}, \bibinfo{person}{Ji
  Wang}, {and} \bibinfo{person}{Xiangke Liao}.}
  \bibinfo{year}{2020}\natexlab{}.
\newblock \showarticletitle{CP-Detector: Using Configuration-related
  Performance Properties to Expose Performance Bugs}. In
  \bibinfo{booktitle}{\emph{Proc. Int'l Conf. Automated Software Engineering
  (ASE)}}. \bibinfo{publisher}{ACM}, \bibinfo{address}{New York, NY, USA},
  \bibinfo{pages}{623--634}.
\newblock


\bibitem[\protect\citeauthoryear{Iqbal, Krishna, Javidian, Ray, and
  Jamshidi}{Iqbal et~al\mbox{.}}{2022}]%
        {IKJRK:EUROSYS22}
\bibfield{author}{\bibinfo{person}{Md~Shahriar Iqbal}, \bibinfo{person}{Rahul
  Krishna}, \bibinfo{person}{Mohammad~Ali Javidian}, \bibinfo{person}{Baishakhi
  Ray}, {and} \bibinfo{person}{Pooyan Jamshidi}.}
  \bibinfo{year}{2022}\natexlab{}.
\newblock \showarticletitle{Unicorn: Reasoning about Configurable System
  Performance through the lens of Causality}. In
  \bibinfo{booktitle}{\emph{Proc. Conf. Computer Systems (EuroSys)}} (Rennes,
  France). \bibinfo{publisher}{ACM}, \bibinfo{address}{New York, NY, USA}.
\newblock


\bibitem[\protect\citeauthoryear{J{\"a}{\"a}skel{\"a}inen}{J{\"a}{\"a}skel{\"a}inen}{2010}]%
        {J:TAP10}
\bibfield{author}{\bibinfo{person}{Riitta J{\"a}{\"a}skel{\"a}inen}.}
  \bibinfo{year}{2010}\natexlab{}.
\newblock \bibinfo{booktitle}{\emph{Think-aloud protocol}}.
\newblock \bibinfo{publisher}{John Benjamins Publishing
  Amsterdam/Philadelphia}, \bibinfo{address}{Amsterdam}. 371--374 pages.
\newblock


\bibitem[\protect\citeauthoryear{Jin, Song, Shi, Scherpelz, and Lu}{Jin
  et~al\mbox{.}}{2012}]%
        {JSSSL:PLDI12}
\bibfield{author}{\bibinfo{person}{Guoliang Jin}, \bibinfo{person}{Linhai
  Song}, \bibinfo{person}{Xiaoming Shi}, \bibinfo{person}{Joel Scherpelz},
  {and} \bibinfo{person}{Shan Lu}.} \bibinfo{year}{2012}\natexlab{}.
\newblock \showarticletitle{Understanding and Detecting Real-world Performance
  Bugs}. In \bibinfo{booktitle}{\emph{Proc. Conf. Programming Language Design
  and Implementation (PLDI)}} (Beijing, China). \bibinfo{publisher}{ACM},
  \bibinfo{address}{New York, NY, USA}, \bibinfo{pages}{77--88}.
\newblock


\bibitem[\protect\citeauthoryear{Jovic, Adamoli, and Hauswirth}{Jovic
  et~al\mbox{.}}{2011}]%
        {JAH:OOPSLA11}
\bibfield{author}{\bibinfo{person}{Milan Jovic}, \bibinfo{person}{Andrea
  Adamoli}, {and} \bibinfo{person}{Matthias Hauswirth}.}
  \bibinfo{year}{2011}\natexlab{}.
\newblock \showarticletitle{Catch Me If You Can: Performance Bug Detection in
  the Wild}. In \bibinfo{booktitle}{\emph{Proc. Int'l Conf. Object-Oriented
  Programming, Systems, Languages and Applications (OOPSLA)}} (Portland,
  Oregon, USA). \bibinfo{publisher}{ACM}, \bibinfo{address}{New York, NY, USA},
  \bibinfo{pages}{155--170}.
\newblock


\bibitem[\protect\citeauthoryear{Juristo and Gómez}{Juristo and
  Gómez}{2011}]%
        {JG:ESEV11}
\bibfield{author}{\bibinfo{person}{Natalia Juristo} {and}
  \bibinfo{person}{Omar~S. Gómez}.} \bibinfo{year}{2011}\natexlab{}.
\newblock \bibinfo{booktitle}{\emph{Replication of Software Engineering
  Experiments}}.
\newblock \bibinfo{publisher}{Springer Berlin Heidelberg},
  \bibinfo{address}{Berlin, Heidelberg}, \bibinfo{pages}{60--88}.
\newblock


\bibitem[\protect\citeauthoryear{{Kaltenecker}, {Grebhahn}, {Siegmund}, and
  {Apel}}{{Kaltenecker} et~al\mbox{.}}{2020}]%
        {KGSA:IEEESOFT20}
\bibfield{author}{\bibinfo{person}{C. {Kaltenecker}}, \bibinfo{person}{A.
  {Grebhahn}}, \bibinfo{person}{N. {Siegmund}}, {and} \bibinfo{person}{S.
  {Apel}}.} \bibinfo{year}{2020}\natexlab{}.
\newblock \showarticletitle{The Interplay of Sampling and Machine Learning for
  Software Performance Prediction}.
\newblock \bibinfo{journal}{\emph{IEEE Software}} \bibinfo{volume}{37},
  \bibinfo{number}{4} (\bibinfo{year}{2020}), \bibinfo{pages}{58--66}.
\newblock


\bibitem[\protect\citeauthoryear{Kaltenecker, Grebhahn, Siegmund, Guo, and
  Apel}{Kaltenecker et~al\mbox{.}}{2019}]%
        {KGSGA:ICSE19}
\bibfield{author}{\bibinfo{person}{Christian Kaltenecker},
  \bibinfo{person}{Alexander Grebhahn}, \bibinfo{person}{Norbert Siegmund},
  \bibinfo{person}{Jianmei Guo}, {and} \bibinfo{person}{Sven Apel}.}
  \bibinfo{year}{2019}\natexlab{}.
\newblock \showarticletitle{Distance-Based Sampling of Software Configuration
  Spaces}. In \bibinfo{booktitle}{\emph{Proc. Int'l Conf. Software Engineering
  (ICSE)}} (Montreal, Quebec, Canada). \bibinfo{publisher}{IEEE},
  \bibinfo{address}{Los Alamitos, CA, USA}, \bibinfo{pages}{21--31}.
\newblock


\bibitem[\protect\citeauthoryear{Ko and Myers}{Ko and Myers}{2004}]%
        {KM:CHI04}
\bibfield{author}{\bibinfo{person}{Andrew~J. Ko} {and} \bibinfo{person}{Brad~A.
  Myers}.} \bibinfo{year}{2004}\natexlab{}.
\newblock \showarticletitle{Designing the Whyline: A Debugging Interface for
  Asking Questions about Program Behavior}. In \bibinfo{booktitle}{\emph{Proc.
  Conf Human Factors in Computing Systems (CHI)}} (Vienna, Austria).
  \bibinfo{publisher}{ACM}, \bibinfo{address}{New York, NY, USA},
  \bibinfo{pages}{151--158}.
\newblock


\bibitem[\protect\citeauthoryear{Kolesnikov, Siegmund, K{\"a}stner, Grebhahn,
  and Apel}{Kolesnikov et~al\mbox{.}}{2019}]%
        {KSKGA:SOSYM18}
\bibfield{author}{\bibinfo{person}{Sergiy Kolesnikov}, \bibinfo{person}{Norbert
  Siegmund}, \bibinfo{person}{Christian K{\"a}stner},
  \bibinfo{person}{Alexander Grebhahn}, {and} \bibinfo{person}{Sven Apel}.}
  \bibinfo{year}{2019}\natexlab{}.
\newblock \showarticletitle{Tradeoffs in Modeling Performance of Highly
  Configurable Software Systems}.
\newblock \bibinfo{journal}{\emph{Software and System Modeling (SoSyM)}}
  \bibinfo{volume}{18}, \bibinfo{number}{3} (\bibinfo{year}{2019}),
  \bibinfo{pages}{2265--2283}.
\newblock


\bibitem[\protect\citeauthoryear{Korel and Laski}{Korel and Laski}{1988}]%
        {KL:IPL88}
\bibfield{author}{\bibinfo{person}{Bogdan Korel} {and} \bibinfo{person}{Janusz
  Laski}.} \bibinfo{year}{1988}\natexlab{}.
\newblock \showarticletitle{Dynamic Program Slicing}.
\newblock \bibinfo{journal}{\emph{Information processing letters}}
  \bibinfo{volume}{29}, \bibinfo{number}{3} (\bibinfo{year}{1988}),
  \bibinfo{pages}{155--163}.
\newblock


\bibitem[\protect\citeauthoryear{Krinke}{Krinke}{2003}]%
        {K:SCAM03}
\bibfield{author}{\bibinfo{person}{Jens Krinke}.}
  \bibinfo{year}{2003}\natexlab{}.
\newblock \showarticletitle{Barrier Slicing and Chopping}. In
  \bibinfo{booktitle}{\emph{Int'l Workshop Source Code Analysis and
  Manipulation (SCAM)}}. \bibinfo{publisher}{IEEE},
  \bibinfo{address}{Amsterdam, Netherlands}, \bibinfo{pages}{81--87}.
\newblock


\bibitem[\protect\citeauthoryear{Krishna, Iqbal, Javidian, Ray, and
  Jamshidi}{Krishna et~al\mbox{.}}{2020}]%
        {KIJRJ:CADET20}
\bibfield{author}{\bibinfo{person}{Rahul Krishna}, \bibinfo{person}{Md~Shahriar
  Iqbal}, \bibinfo{person}{Mohammad~Ali Javidian}, \bibinfo{person}{Baishakhi
  Ray}, {and} \bibinfo{person}{Pooyan Jamshidi}.}
  \bibinfo{year}{2020}\natexlab{}.
\newblock \bibinfo{title}{{CADET}: A Systematic Method For Debugging
  Misconfigurations using Counterfactual Reasoning}.
\newblock
\newblock
\showeprint[arxiv]{2010.06061}~[cs.SE]


\bibitem[\protect\citeauthoryear{{LaToza} and {Myers}}{{LaToza} and
  {Myers}}{2011}]%
        {LM:VLHCC11}
\bibfield{author}{\bibinfo{person}{T.~D. {LaToza}} {and} \bibinfo{person}{B.~A.
  {Myers}}.} \bibinfo{year}{2011}\natexlab{}.
\newblock \showarticletitle{Visualizing Call Graphs}. In
  \bibinfo{booktitle}{\emph{Symposium Visual Languages and Human-Centric
  Computing (VL/HCC)}}. \bibinfo{publisher}{IEEE}, \bibinfo{address}{Los
  Alamitos, CA, USA}, \bibinfo{pages}{117--124}.
\newblock


\bibitem[\protect\citeauthoryear{Li, Wang, Hoffmann, and Lu}{Li
  et~al\mbox{.}}{2020}]%
        {LWHL:EUROSYS20}
\bibfield{author}{\bibinfo{person}{Chi Li}, \bibinfo{person}{Shu Wang},
  \bibinfo{person}{Henry Hoffmann}, {and} \bibinfo{person}{Shan Lu}.}
  \bibinfo{year}{2020}\natexlab{}.
\newblock \showarticletitle{Statically Inferring Performance Properties of
  Software Configurations}. In \bibinfo{booktitle}{\emph{Proc. European Conf.
  Computer Systems (EuroSys)}} (Heraklion, Greece). \bibinfo{publisher}{ACM},
  \bibinfo{address}{New York, NY, USA}, Article \bibinfo{articleno}{10},
  \bibinfo{numpages}{10}~pages.
\newblock


\bibitem[\protect\citeauthoryear{Li, Lyu, Gui, and Halfond}{Li
  et~al\mbox{.}}{2016}]%
        {LLGH:ICSE16}
\bibfield{author}{\bibinfo{person}{Ding Li}, \bibinfo{person}{Yingjun Lyu},
  \bibinfo{person}{Jiaping Gui}, {and} \bibinfo{person}{William~G.J. Halfond}.}
  \bibinfo{year}{2016}\natexlab{}.
\newblock \showarticletitle{Automated Energy Optimization of HTTP Requests for
  Mobile Applications}. In \bibinfo{booktitle}{\emph{Proc. Int'l Conf. Software
  Engineering (ICSE)}} (Austin, TX, USA). \bibinfo{publisher}{ACM},
  \bibinfo{address}{New York, NY, USA}, \bibinfo{pages}{249--260}.
\newblock


\bibitem[\protect\citeauthoryear{Lillack, K\"{a}stner, and Bodden}{Lillack
  et~al\mbox{.}}{2018}]%
        {LKB:TSE18}
\bibfield{author}{\bibinfo{person}{Max Lillack}, \bibinfo{person}{Christian
  K\"{a}stner}, {and} \bibinfo{person}{Eric Bodden}.}
  \bibinfo{year}{2018}\natexlab{}.
\newblock \showarticletitle{Tracking Load-time Configuration Options}.
\newblock \bibinfo{journal}{\emph{IEEE Transactions on Software Engineering}}
  \bibinfo{volume}{44}, \bibinfo{number}{12} (\bibinfo{date}{12}
  \bibinfo{year}{2018}), \bibinfo{pages}{1269--1291}.
\newblock


\bibitem[\protect\citeauthoryear{Liu, Xu, and Cheung}{Liu
  et~al\mbox{.}}{2014}]%
        {LXC:ICSE14}
\bibfield{author}{\bibinfo{person}{Yepang Liu}, \bibinfo{person}{Chang Xu},
  {and} \bibinfo{person}{Shing-Chi Cheung}.} \bibinfo{year}{2014}\natexlab{}.
\newblock \showarticletitle{Characterizing and Detecting Performance Bugs for
  Smartphone Applications}. In \bibinfo{booktitle}{\emph{Proc. Int'l Conf.
  Software Engineering (ICSE)}} (Hyderabad, India) \emph{(\bibinfo{series}{ICSE
  2014})}. \bibinfo{publisher}{ACM}, \bibinfo{address}{New York, NY, USA},
  \bibinfo{pages}{1013--1024}.
\newblock


\bibitem[\protect\citeauthoryear{Meinicke, Wong, K{\"a}stner, and
  Saake}{Meinicke et~al\mbox{.}}{2018}]%
        {MWKS:ARXIV18}
\bibfield{author}{\bibinfo{person}{Jens Meinicke}, \bibinfo{person}{Chu-Pan
  Wong}, \bibinfo{person}{Christian K{\"a}stner}, {and} \bibinfo{person}{Gunter
  Saake}.} \bibinfo{year}{2018}\natexlab{}.
\newblock \bibinfo{booktitle}{\emph{Understanding Differences Among Executions
  with Variational Traces}}.
\newblock \bibinfo{type}{{T}echnical {R}eport} 1807.03837.
  \bibinfo{institution}{arXiv}.
\newblock


\bibitem[\protect\citeauthoryear{Meinicke, Wong, K\"{a}stner, Th\"{u}m, and
  Saake}{Meinicke et~al\mbox{.}}{2016}]%
        {MWKTS:ASE16}
\bibfield{author}{\bibinfo{person}{Jens Meinicke}, \bibinfo{person}{Chu-Pan
  Wong}, \bibinfo{person}{Christian K\"{a}stner}, \bibinfo{person}{Thomas
  Th\"{u}m}, {and} \bibinfo{person}{Gunter Saake}.}
  \bibinfo{year}{2016}\natexlab{}.
\newblock \showarticletitle{On Essential Configuration Complexity: Measuring
  Interactions in Highly-configurable Systems}. In
  \bibinfo{booktitle}{\emph{Proc. Int'l Conf. Automated Software Engineering
  (ASE)}} (Singapore, Singapore). \bibinfo{publisher}{ACM},
  \bibinfo{address}{New York, NY, USA}, \bibinfo{pages}{483--494}.
\newblock


\bibitem[\protect\citeauthoryear{Melo, Brabrand, and W\k{a}sowski}{Melo
  et~al\mbox{.}}{2016}]%
        {MBW:ICSE16}
\bibfield{author}{\bibinfo{person}{Jean Melo}, \bibinfo{person}{Claus
  Brabrand}, {and} \bibinfo{person}{Andrzej W\k{a}sowski}.}
  \bibinfo{year}{2016}\natexlab{}.
\newblock \showarticletitle{How Does the Degree of Variability Affect Bug
  Finding?}. In \bibinfo{booktitle}{\emph{Proc. Int'l Conf. Software
  Engineering (ICSE)}} (Austin, TX, USA). \bibinfo{publisher}{ACM},
  \bibinfo{address}{New York, NY, USA}, \bibinfo{pages}{679--690}.
\newblock


\bibitem[\protect\citeauthoryear{Melo, Narcizo, Hansen, Brabrand, and
  Wasowski}{Melo et~al\mbox{.}}{2017}]%
        {MNHBW:ICPC17}
\bibfield{author}{\bibinfo{person}{Jean Melo},
  \bibinfo{person}{Fabricio~Batista Narcizo}, \bibinfo{person}{Dan~Witzner
  Hansen}, \bibinfo{person}{Claus Brabrand}, {and} \bibinfo{person}{Andrzej
  Wasowski}.} \bibinfo{year}{2017}\natexlab{}.
\newblock \showarticletitle{Variability through the Eyes of the Programmer}. In
  \bibinfo{booktitle}{\emph{Proc. Int'l Conference Program Comprehension
  (ICPC)}} (Buenos Aires, Argentina). \bibinfo{publisher}{IEEE},
  \bibinfo{address}{Los Alamitos, CA, USA}, \bibinfo{pages}{34--44}.
\newblock


\bibitem[\protect\citeauthoryear{Myers, Ko, LaToza, and Yoon}{Myers
  et~al\mbox{.}}{2016}]%
        {MKLY:C16}
\bibfield{author}{\bibinfo{person}{Brad~A. Myers}, \bibinfo{person}{Andrew~J.
  Ko}, \bibinfo{person}{Thomas~D. LaToza}, {and} \bibinfo{person}{YoungSeok
  Yoon}.} \bibinfo{year}{2016}\natexlab{}.
\newblock \showarticletitle{Programmers Are Users Too: Human-Centered Methods
  for Improving Programming Tools}.
\newblock \bibinfo{journal}{\emph{Computer}} \bibinfo{volume}{49},
  \bibinfo{number}{7} (\bibinfo{date}{July} \bibinfo{year}{2016}),
  \bibinfo{pages}{44--52}.
\newblock


\bibitem[\protect\citeauthoryear{Nair, Menzies, Siegmund, and Apel}{Nair
  et~al\mbox{.}}{2017}]%
        {NMSA:ESECFSE17}
\bibfield{author}{\bibinfo{person}{Vivek Nair}, \bibinfo{person}{Tim Menzies},
  \bibinfo{person}{Norbert Siegmund}, {and} \bibinfo{person}{Sven Apel}.}
  \bibinfo{year}{2017}\natexlab{}.
\newblock \showarticletitle{Using Bad Learners to Find Good Configurations}. In
  \bibinfo{booktitle}{\emph{Proc. Europ. Software Engineering Conf. Foundations
  of Software Engineering (ESEC/FSE)}} (Paderborn, Germany)
  \emph{(\bibinfo{series}{ESEC/FSE 2017})}. \bibinfo{publisher}{ACM},
  \bibinfo{address}{New York, NY, USA}, \bibinfo{pages}{257--267}.
\newblock


\bibitem[\protect\citeauthoryear{Nethercote and Seward}{Nethercote and
  Seward}{2007}]%
        {NS:PLDI07}
\bibfield{author}{\bibinfo{person}{Nicholas Nethercote} {and}
  \bibinfo{person}{Julian Seward}.} \bibinfo{year}{2007}\natexlab{}.
\newblock \showarticletitle{Valgrind: A Framework for Heavyweight Dynamic
  Binary Instrumentation}. In \bibinfo{booktitle}{\emph{Proc. Conf. Programming
  Language Design and Implementation (PLDI)}} (San Diego, CA, USA).
  \bibinfo{publisher}{ACM}, \bibinfo{address}{New York, NY, USA},
  \bibinfo{pages}{89--100}.
\newblock


\bibitem[\protect\citeauthoryear{Nistor, Chang, Radoi, and Lu}{Nistor
  et~al\mbox{.}}{2015}]%
        {NCRL:ICSE15}
\bibfield{author}{\bibinfo{person}{Adrian Nistor}, \bibinfo{person}{Po-Chun
  Chang}, \bibinfo{person}{Cosmin Radoi}, {and} \bibinfo{person}{Shan Lu}.}
  \bibinfo{year}{2015}\natexlab{}.
\newblock \showarticletitle{Caramel: Detecting and Fixing Performance Problems
  That Have Non-intrusive Fixes}. In \bibinfo{booktitle}{\emph{Proc. Int'l
  Conf. Software Engineering (ICSE)}} (Florence, Italy).
  \bibinfo{publisher}{IEEE}, \bibinfo{address}{Piscataway, NJ, USA},
  \bibinfo{pages}{902--912}.
\newblock


\bibitem[\protect\citeauthoryear{Nistor, Jiang, and Tan}{Nistor
  et~al\mbox{.}}{2013a}]%
        {NJT:MSR13}
\bibfield{author}{\bibinfo{person}{Adrian Nistor}, \bibinfo{person}{Tian
  Jiang}, {and} \bibinfo{person}{Lin Tan}.} \bibinfo{year}{2013}\natexlab{a}.
\newblock \showarticletitle{Discovering, Reporting, and Fixing Performance
  Bugs}. In \bibinfo{booktitle}{\emph{Proc. Int'l Conf. Mining Software
  Repositories}} (San Francisco, CA, USA). \bibinfo{publisher}{IEEE},
  \bibinfo{address}{Piscataway, NJ, USA}, \bibinfo{pages}{237--246}.
\newblock


\bibitem[\protect\citeauthoryear{Nistor, Song, Marinov, and Lu}{Nistor
  et~al\mbox{.}}{2013b}]%
        {NSML:ICSE13}
\bibfield{author}{\bibinfo{person}{Adrian Nistor}, \bibinfo{person}{Linhai
  Song}, \bibinfo{person}{Darko Marinov}, {and} \bibinfo{person}{Shan Lu}.}
  \bibinfo{year}{2013}\natexlab{b}.
\newblock \showarticletitle{Toddler: Detecting Performance Problems via Similar
  Memory-access Patterns}. In \bibinfo{booktitle}{\emph{Proc. Int'l Conf.
  Software Engineering (ICSE)}} (San Francisco, CA, USA).
  \bibinfo{publisher}{IEEE}, \bibinfo{address}{Piscataway, NJ, USA},
  \bibinfo{pages}{562--571}.
\newblock


\bibitem[\protect\citeauthoryear{Oh, Batory, Myers, and Siegmund}{Oh
  et~al\mbox{.}}{2017}]%
        {OBMS:ESECFSE17}
\bibfield{author}{\bibinfo{person}{Jeho Oh}, \bibinfo{person}{Don Batory},
  \bibinfo{person}{Margaret Myers}, {and} \bibinfo{person}{Norbert Siegmund}.}
  \bibinfo{year}{2017}\natexlab{}.
\newblock \showarticletitle{Finding Near-optimal Configurations in Product
  Lines by Random Sampling}. In \bibinfo{booktitle}{\emph{Proc. Europ. Software
  Engineering Conf. Foundations of Software Engineering (ESEC/FSE)}}
  (Paderborn, Germany). \bibinfo{publisher}{ACM}, \bibinfo{address}{New York,
  NY, USA}, \bibinfo{pages}{61--71}.
\newblock


\bibitem[\protect\citeauthoryear{{Park}, {Kim}, {Ray}, and {Bae}}{{Park}
  et~al\mbox{.}}{2012}]%
        {PKRB:MSR12}
\bibfield{author}{\bibinfo{person}{J. {Park}}, \bibinfo{person}{M. {Kim}},
  \bibinfo{person}{B. {Ray}}, {and} \bibinfo{person}{D. {Bae}}.}
  \bibinfo{year}{2012}\natexlab{}.
\newblock \showarticletitle{An Empirical Study of Supplementary Bug Fixes}. In
  \bibinfo{booktitle}{\emph{Proc. Int'l Conf. Mining Software Repositories}}
  (Zurich, Switzerland). \bibinfo{publisher}{IEEE}, \bibinfo{address}{Los
  Alamitos, CA, USA}, \bibinfo{pages}{40--49}.
\newblock


\bibitem[\protect\citeauthoryear{Parnin and Orso}{Parnin and Orso}{2011}]%
        {PO:ISSTA11}
\bibfield{author}{\bibinfo{person}{Chris Parnin} {and}
  \bibinfo{person}{Alessandro Orso}.} \bibinfo{year}{2011}\natexlab{}.
\newblock \showarticletitle{Are Automated Debugging Techniques Actually Helping
  Programmers?}. In \bibinfo{booktitle}{\emph{Proc. Int'l Symp. Software
  Testing and Analysis (ISSTA)}} (Toronto, Canada). \bibinfo{publisher}{ACM},
  \bibinfo{address}{New York, NY, USA}, \bibinfo{pages}{199--209}.
\newblock


\bibitem[\protect\citeauthoryear{Pothier, Tanter, and Piquer}{Pothier
  et~al\mbox{.}}{2007}]%
        {PTP:OOPSLA07}
\bibfield{author}{\bibinfo{person}{Guillaume Pothier},
  \bibinfo{person}{\'{E}ric Tanter}, {and} \bibinfo{person}{Jos\'{e} Piquer}.}
  \bibinfo{year}{2007}\natexlab{}.
\newblock \showarticletitle{Scalable Omniscient Debugging}. In
  \bibinfo{booktitle}{\emph{Proc. Int'l Conf. Object-Oriented Programming,
  Systems, Languages and Applications (OOPSLA)}} (Montreal, Quebec, Canada).
  \bibinfo{publisher}{ACM}, \bibinfo{address}{New York, NY, USA},
  \bibinfo{pages}{535--552}.
\newblock


\bibitem[\protect\citeauthoryear{Salda{\~n}a}{Salda{\~n}a}{2015}]%
        {S:CMQR15}
\bibfield{author}{\bibinfo{person}{Johnny Salda{\~n}a}.}
  \bibinfo{year}{2015}\natexlab{}.
\newblock \bibinfo{booktitle}{\emph{The Coding Manual for Qualitative
  Researchers}}.
\newblock \bibinfo{publisher}{Sage}, \bibinfo{address}{London, England}.
\newblock


\bibitem[\protect\citeauthoryear{{Sandoval Alcocer}, {Beck}, and
  {Bergel}}{{Sandoval Alcocer} et~al\mbox{.}}{2019}]%
        {SBB:VISSOFT19}
\bibfield{author}{\bibinfo{person}{J.~P. {Sandoval Alcocer}},
  \bibinfo{person}{F. {Beck}}, {and} \bibinfo{person}{A. {Bergel}}.}
  \bibinfo{year}{2019}\natexlab{}.
\newblock \showarticletitle{Performance Evolution Matrix: Visualizing
  Performance Variations Along Software Versions}. In
  \bibinfo{booktitle}{\emph{Conf. Software Visualization (VISSOFT)}}.
  \bibinfo{publisher}{IEEE}, \bibinfo{address}{Los Alamitos, CA, USA},
  \bibinfo{pages}{1--11}.
\newblock


\bibitem[\protect\citeauthoryear{Schmidt}{Schmidt}{2009}]%
        {S:RGP09}
\bibfield{author}{\bibinfo{person}{Stefan Schmidt}.}
  \bibinfo{year}{2009}\natexlab{}.
\newblock \showarticletitle{Shall we Really do it Again? The Powerful Concept
  of Replication is Neglected in the Social Sciences}.
\newblock \bibinfo{journal}{\emph{Review of General Psychology}}
  \bibinfo{volume}{13}, \bibinfo{number}{2} (\bibinfo{year}{2009}),
  \bibinfo{pages}{90--100}.
\newblock


\bibitem[\protect\citeauthoryear{Siegmund, Grebhahn, Apel, and
  K\"{a}stner}{Siegmund et~al\mbox{.}}{2015}]%
        {SGAK:ESECFSE15}
\bibfield{author}{\bibinfo{person}{Norbert Siegmund},
  \bibinfo{person}{Alexander Grebhahn}, \bibinfo{person}{Sven Apel}, {and}
  \bibinfo{person}{Christian K\"{a}stner}.} \bibinfo{year}{2015}\natexlab{}.
\newblock \showarticletitle{Performance-influence Models for Highly
  Configurable Systems}. In \bibinfo{booktitle}{\emph{Proc. Europ. Software
  Engineering Conf. Foundations of Software Engineering (ESEC/FSE)}} (Bergamo,
  Italy). \bibinfo{publisher}{ACM}, \bibinfo{address}{New York, NY, USA},
  \bibinfo{pages}{284--294}.
\newblock


\bibitem[\protect\citeauthoryear{Siegmund, Kolesnikov, K\"{a}stner, Apel,
  Batory, Rosenm\"{u}ller, and Saake}{Siegmund et~al\mbox{.}}{2012a}]%
        {SKKABRS:ICSE12}
\bibfield{author}{\bibinfo{person}{Norbert Siegmund},
  \bibinfo{person}{Sergiy~S. Kolesnikov}, \bibinfo{person}{Christian
  K\"{a}stner}, \bibinfo{person}{Sven Apel}, \bibinfo{person}{Don Batory},
  \bibinfo{person}{Marko Rosenm\"{u}ller}, {and} \bibinfo{person}{Gunter
  Saake}.} \bibinfo{year}{2012}\natexlab{a}.
\newblock \showarticletitle{Predicting Performance via Automated
  Feature-interaction Detection}. In \bibinfo{booktitle}{\emph{Proc. Int'l
  Conf. Software Engineering (ICSE)}} (Zurich, Switzerland).
  \bibinfo{publisher}{IEEE}, \bibinfo{address}{Piscataway, NJ, USA},
  \bibinfo{pages}{167--177}.
\newblock


\bibitem[\protect\citeauthoryear{Siegmund, Rosenm\"{u}ller, Kuhlemann,
  K\"{a}stner, Apel, and Saake}{Siegmund et~al\mbox{.}}{2012b}]%
        {SRKKAS:SQJ12}
\bibfield{author}{\bibinfo{person}{Norbert Siegmund}, \bibinfo{person}{Marko
  Rosenm\"{u}ller}, \bibinfo{person}{Martin Kuhlemann},
  \bibinfo{person}{Christian K\"{a}stner}, \bibinfo{person}{Sven Apel}, {and}
  \bibinfo{person}{Gunter Saake}.} \bibinfo{year}{2012}\natexlab{b}.
\newblock \showarticletitle{{SPLC}onqueror: Toward Optimization of
  Non-functional Properties in Software Product Lines}.
\newblock \bibinfo{journal}{\emph{Software Quality Journal}}
  \bibinfo{volume}{20}, \bibinfo{number}{3-4} (\bibinfo{date}{Sept.}
  \bibinfo{year}{2012}), \bibinfo{pages}{487--517}.
\newblock


\bibitem[\protect\citeauthoryear{Song and Lu}{Song and Lu}{2014}]%
        {SL:OOPSLA14}
\bibfield{author}{\bibinfo{person}{Linhai Song} {and} \bibinfo{person}{Shan
  Lu}.} \bibinfo{year}{2014}\natexlab{}.
\newblock \showarticletitle{Statistical Debugging for Real-World Performance
  Problems}. In \bibinfo{booktitle}{\emph{Proc. Int'l Conf. Object-Oriented
  Programming, Systems, Languages and Applications (OOPSLA)}} (Portland, OR,
  USA). \bibinfo{publisher}{ACM}, \bibinfo{address}{New York, NY, USA},
  \bibinfo{pages}{561--578}.
\newblock


\bibitem[\protect\citeauthoryear{Song and Lu}{Song and Lu}{2017}]%
        {SL:ICSE17}
\bibfield{author}{\bibinfo{person}{Linhai Song} {and} \bibinfo{person}{Shan
  Lu}.} \bibinfo{year}{2017}\natexlab{}.
\newblock \showarticletitle{Performance Diagnosis for Inefficient Loops}. In
  \bibinfo{booktitle}{\emph{Proc. Int'l Conf. Software Engineering (ICSE)}}
  (Buenos Aires, Argentina). \bibinfo{publisher}{IEEE},
  \bibinfo{address}{Piscataway, NJ, USA}, \bibinfo{pages}{370--380}.
\newblock


\bibitem[\protect\citeauthoryear{Toman and Grossman}{Toman and
  Grossman}{2016}]%
        {TD:ECOOP16}
\bibfield{author}{\bibinfo{person}{John Toman} {and} \bibinfo{person}{Dan
  Grossman}.} \bibinfo{year}{2016}\natexlab{}.
\newblock \showarticletitle{Staccato: A Bug Finder for Dynamic Configuration
  Updates}. In \bibinfo{booktitle}{\emph{Proc. European Conf. Object-Oriented
  Programming (ECOOP)}}. \bibinfo{publisher}{Schloss Dagstuhl--Leibniz-Zentrum
  fuer Informatik}, \bibinfo{address}{Dagstuhl, Germany},
  \bibinfo{pages}{1--23}.
\newblock


\bibitem[\protect\citeauthoryear{Trümper, Döllner, and Telea}{Trümper
  et~al\mbox{.}}{2013}]%
        {TDT:ICPC13}
\bibfield{author}{\bibinfo{person}{Jonas Trümper}, \bibinfo{person}{Jürgen
  Döllner}, {and} \bibinfo{person}{Alexandru Telea}.}
  \bibinfo{year}{2013}\natexlab{}.
\newblock \showarticletitle{Multiscale Visual Comparison of Execution Traces}.
  In \bibinfo{booktitle}{\emph{Proc. Intl Conf. Program Comprehension (ICPC)}}
  (San Francisco, CA, USA). \bibinfo{publisher}{IEEE}, \bibinfo{address}{Los
  Alamitos, CA, USA}, \bibinfo{pages}{53--62}.
\newblock


\bibitem[\protect\citeauthoryear{Velez, Jamshidi, Sattler, Siegmund, Apel, and
  Kästner}{Velez et~al\mbox{.}}{2020}]%
        {VJSSAK:ASEJ20}
\bibfield{author}{\bibinfo{person}{Miguel Velez}, \bibinfo{person}{Pooyan
  Jamshidi}, \bibinfo{person}{Florian Sattler}, \bibinfo{person}{Norbert
  Siegmund}, \bibinfo{person}{Sven Apel}, {and} \bibinfo{person}{Christian
  Kästner}.} \bibinfo{year}{2020}\natexlab{}.
\newblock \showarticletitle{ConfigCrusher: Towards White-Box Performance
  Analysis for Configurable Systems}.
\newblock \bibinfo{journal}{\emph{Autom Softw Eng}} \bibinfo{volume}{27},
  \bibinfo{number}{3} (\bibinfo{year}{2020}), \bibinfo{pages}{265--300}.
\newblock


\bibitem[\protect\citeauthoryear{Velez, Jamshidi, Siegmund, Apel, and
  Kästner}{Velez et~al\mbox{.}}{2021}]%
        {VJSAK:ICSE21}
\bibfield{author}{\bibinfo{person}{Miguel Velez}, \bibinfo{person}{Pooyan
  Jamshidi}, \bibinfo{person}{Norbert Siegmund}, \bibinfo{person}{Sven Apel},
  {and} \bibinfo{person}{Christian Kästner}.} \bibinfo{year}{2021}\natexlab{}.
\newblock \showarticletitle{White-Box Analysis over Machine Learning: Modeling
  Performance of Configurable Systems}. In \bibinfo{booktitle}{\emph{Proc.
  Int'l Conf. Software Engineering (ICSE)}} (Madrid, Spain).
  \bibinfo{publisher}{IEEE}, \bibinfo{address}{Los Alamitos, CA, USA},
  \bibinfo{pages}{1072--1084}.
\newblock


\bibitem[\protect\citeauthoryear{Velez, Jamshidi, Siegmund, Apel, and
  Kästner}{Velez et~al\mbox{.}}{2022}]%
        {VJSAK:ICSE22SM}
\bibfield{author}{\bibinfo{person}{Miguel Velez}, \bibinfo{person}{Pooyan
  Jamshidi}, \bibinfo{person}{Norbert Siegmund}, \bibinfo{person}{Sven Apel},
  {and} \bibinfo{person}{Christian Kästner}.} \bibinfo{year}{2022}\natexlab{}.
\newblock \bibinfo{title}{On Debugging the Performance of Configurable Software
  Systems: Developer Needs and Tailored Tool Support - {S}upplementary
  {M}aterial - {https://bit.ly/35HUvl9}}.
\newblock
\newblock


\bibitem[\protect\citeauthoryear{VisualVM}{VisualVM}{2020}]%
        {VVM}
\bibfield{author}{\bibinfo{person}{VisualVM}.} \bibinfo{year}{2020}\natexlab{}.
\newblock \bibinfo{booktitle}{\emph{VisualVM}}.
\newblock VisualVM.
\newblock
\urldef\tempurl%
\url{https://visualvm.github.io/}
\showURL{%
Retrieved November 24, 2020 from \tempurl}


\bibitem[\protect\citeauthoryear{Wang, Li, Hoffmann, Lu, Sentosa, and
  Kistijantoro}{Wang et~al\mbox{.}}{2018}]%
        {WLHLSK:ASPLOS18}
\bibfield{author}{\bibinfo{person}{Shu Wang}, \bibinfo{person}{Chi Li},
  \bibinfo{person}{Henry Hoffmann}, \bibinfo{person}{Shan Lu},
  \bibinfo{person}{William Sentosa}, {and} \bibinfo{person}{Achmad~Imam
  Kistijantoro}.} \bibinfo{year}{2018}\natexlab{}.
\newblock \showarticletitle{Understanding and Auto-Adjusting
  Performance-Sensitive Configurations}. In \bibinfo{booktitle}{\emph{Proc.
  Int'l Conf. Architectural Support for Programming Languages and Operating
  Systems (ASPLOS)}} (Williamsburg, VA, USA). \bibinfo{publisher}{ACM},
  \bibinfo{address}{New York, NY, USA}, \bibinfo{pages}{154--168}.
\newblock


\bibitem[\protect\citeauthoryear{Weber, Apel, and Siegmund}{Weber
  et~al\mbox{.}}{2021}]%
        {WAS:ICSE21}
\bibfield{author}{\bibinfo{person}{Max Weber}, \bibinfo{person}{Sven Apel},
  {and} \bibinfo{person}{Norbert Siegmund}.} \bibinfo{year}{2021}\natexlab{}.
\newblock \showarticletitle{{W}hite-{B}ox {P}erformance-{I}nfluence Models: A
  Profiling and Learning Approach}. In \bibinfo{booktitle}{\emph{Proc. Int'l
  Conf. Software Engineering (ICSE)}} (Madrid, Spain).
  \bibinfo{publisher}{IEEE}, \bibinfo{address}{Los Alamitos, CA, USA},
  \bibinfo{pages}{232--233}.
\newblock


\bibitem[\protect\citeauthoryear{Weiser}{Weiser}{1981}]%
        {W:ICSE81}
\bibfield{author}{\bibinfo{person}{Mark Weiser}.}
  \bibinfo{year}{1981}\natexlab{}.
\newblock \showarticletitle{Program Slicing}. In
  \bibinfo{booktitle}{\emph{Proc. Int'l Conf. Software Engineering (ICSE)}}
  (San Diego, CA, USA). \bibinfo{publisher}{IEEE},
  \bibinfo{address}{Piscataway, NJ, USA}, \bibinfo{pages}{439--449}.
\newblock


\bibitem[\protect\citeauthoryear{Wilke, Richly, G\"{o}tz, Piechnick, and
  Amann}{Wilke et~al\mbox{.}}{2013}]%
        {WRGPA:GCC13}
\bibfield{author}{\bibinfo{person}{Claas Wilke}, \bibinfo{person}{Sebastian
  Richly}, \bibinfo{person}{Sebastian G\"{o}tz}, \bibinfo{person}{Christian
  Piechnick}, {and} \bibinfo{person}{Uwe Amann}.}
  \bibinfo{year}{2013}\natexlab{}.
\newblock \showarticletitle{Energy Consumption and Efficiency in Mobile
  Applications: A User Feedback Study}. In \bibinfo{booktitle}{\emph{Proc.
  Int'l Conf. Green Computing and Communications}}. \bibinfo{publisher}{IEEE},
  \bibinfo{address}{Los Alamitos, CA, USA}, \bibinfo{pages}{134--141}.
\newblock


\bibitem[\protect\citeauthoryear{Wong, Meinicke, Lazarek, and K\"{a}stner}{Wong
  et~al\mbox{.}}{2018}]%
        {WMLK:OOPSLA18}
\bibfield{author}{\bibinfo{person}{Chu-Pan Wong}, \bibinfo{person}{Jens
  Meinicke}, \bibinfo{person}{Lukas Lazarek}, {and} \bibinfo{person}{Christian
  K\"{a}stner}.} \bibinfo{year}{2018}\natexlab{}.
\newblock \showarticletitle{Faster Variational Execution with Transparent
  Bytecode Transformation}.
\newblock \bibinfo{journal}{\emph{Proc. Int'l Conf. Object-Oriented
  Programming, Systems, Languages and Applications (OOPSLA)}}
  \bibinfo{volume}{2}, Article \bibinfo{articleno}{117} (\bibinfo{date}{Oct.}
  \bibinfo{year}{2018}), \bibinfo{numpages}{30}~pages.
\newblock


\bibitem[\protect\citeauthoryear{Xu, Qian, Zhang, Wu, and Chen}{Xu
  et~al\mbox{.}}{2005}]%
        {XQZWC:SEN05}
\bibfield{author}{\bibinfo{person}{Baowen Xu}, \bibinfo{person}{Ju Qian},
  \bibinfo{person}{Xiaofang Zhang}, \bibinfo{person}{Zhongqiang Wu}, {and}
  \bibinfo{person}{Lin Chen}.} \bibinfo{year}{2005}\natexlab{}.
\newblock \showarticletitle{A Brief Survey of Program Slicing}.
\newblock \bibinfo{journal}{\emph{ACM SIGSOFT Software Engineering Notes}}
  \bibinfo{volume}{30}, \bibinfo{number}{2} (\bibinfo{year}{2005}),
  \bibinfo{pages}{1--36}.
\newblock


\bibitem[\protect\citeauthoryear{Xu, Jin, Huang, Zhou, Lu, Jin, and
  Pasupathy}{Xu et~al\mbox{.}}{2016}]%
        {XJHZLJP:OSDI16}
\bibfield{author}{\bibinfo{person}{Tianyin Xu}, \bibinfo{person}{Xinxin Jin},
  \bibinfo{person}{Peng Huang}, \bibinfo{person}{Yuanyuan Zhou},
  \bibinfo{person}{Shan Lu}, \bibinfo{person}{Long Jin}, {and}
  \bibinfo{person}{Shankar Pasupathy}.} \bibinfo{year}{2016}\natexlab{}.
\newblock \showarticletitle{Early Detection of Configuration Errors to Reduce
  Failure Damage}. In \bibinfo{booktitle}{\emph{Proc. Conf.Operating Systems
  Design and Implementation (OSDI)}} (Savannah, GA, USA).
  \bibinfo{publisher}{USENIX Association}, \bibinfo{address}{Berkeley, CA,
  USA}, \bibinfo{pages}{619--634}.
\newblock


\bibitem[\protect\citeauthoryear{Xu, Zhang, Huang, Zheng, Sheng, Yuan, Zhou,
  and Pasupathy}{Xu et~al\mbox{.}}{2013}]%
        {XZHZSYZP:SOSP13}
\bibfield{author}{\bibinfo{person}{Tianyin Xu}, \bibinfo{person}{Jiaqi Zhang},
  \bibinfo{person}{Peng Huang}, \bibinfo{person}{Jing Zheng},
  \bibinfo{person}{Tianwei Sheng}, \bibinfo{person}{Ding Yuan},
  \bibinfo{person}{Yuanyuan Zhou}, {and} \bibinfo{person}{Shankar Pasupathy}.}
  \bibinfo{year}{2013}\natexlab{}.
\newblock \showarticletitle{Do Not Blame Users for Misconfigurations}. In
  \bibinfo{booktitle}{\emph{Proc. Symp. Operating Systems Principles}}
  (Farminton, PA, USA). \bibinfo{publisher}{ACM}, \bibinfo{address}{New York,
  NY, USA}, \bibinfo{pages}{244--259}.
\newblock


\bibitem[\protect\citeauthoryear{Yu and Pradel}{Yu and Pradel}{2016}]%
        {YP:ISSTA16}
\bibfield{author}{\bibinfo{person}{Tingting Yu} {and} \bibinfo{person}{Michael
  Pradel}.} \bibinfo{year}{2016}\natexlab{}.
\newblock \showarticletitle{SyncProf: Detecting, Localizing, and Optimizing
  Synchronization Bottlenecks}. In \bibinfo{booktitle}{\emph{Proc. Int'l Symp.
  Software Testing and Analysis (ISSTA)}} (Saarbr\"{u}cken, Germany).
  \bibinfo{publisher}{ACM}, \bibinfo{address}{New York, NY, USA},
  \bibinfo{pages}{389--400}.
\newblock


\bibitem[\protect\citeauthoryear{Yu and Pradel}{Yu and Pradel}{2018}]%
        {YP:EMSE18}
\bibfield{author}{\bibinfo{person}{Tingting Yu} {and} \bibinfo{person}{Michael
  Pradel}.} \bibinfo{year}{2018}\natexlab{}.
\newblock \showarticletitle{Pinpointing and Repairing Performance Bottlenecks
  in Concurrent Programs}.
\newblock \bibinfo{journal}{\emph{Empirical Softw. Eng.}} \bibinfo{volume}{23},
  \bibinfo{number}{5} (\bibinfo{date}{Oct.} \bibinfo{year}{2018}),
  \bibinfo{pages}{3034--3071}.
\newblock


\bibitem[\protect\citeauthoryear{Zeller}{Zeller}{1999}]%
        {Z:SIGSOFTNotes99}
\bibfield{author}{\bibinfo{person}{Andreas Zeller}.}
  \bibinfo{year}{1999}\natexlab{}.
\newblock \showarticletitle{Yesterday, My Program Worked. Today, It Does Not.
  Why?}
\newblock \bibinfo{journal}{\emph{SIGSOFT Softw. Eng. Notes}}
  \bibinfo{volume}{24}, \bibinfo{number}{6} (\bibinfo{date}{Oct.}
  \bibinfo{year}{1999}), \bibinfo{pages}{253--267}.
\newblock


\bibitem[\protect\citeauthoryear{Zeller}{Zeller}{2009}]%
        {Z:WPF09}
\bibfield{author}{\bibinfo{person}{Andreas Zeller}.}
  \bibinfo{year}{2009}\natexlab{}.
\newblock \bibinfo{booktitle}{\emph{Why Programs Fail: A Guide to Systematic
  Debugging}}.
\newblock \bibinfo{publisher}{Elsevier}, \bibinfo{address}{Amsterdam, The
  Netherlands}.
\newblock


\bibitem[\protect\citeauthoryear{Zhang, Renganarayana, Zhang, Ge, Bala, Xu, and
  Zhou}{Zhang et~al\mbox{.}}{2014}]%
        {ZTZGBXZ:ASPLOS14}
\bibfield{author}{\bibinfo{person}{Jiaqi Zhang},
  \bibinfo{person}{Lakshminarayanan Renganarayana}, \bibinfo{person}{Xiaolan
  Zhang}, \bibinfo{person}{Niyu Ge}, \bibinfo{person}{Vasanth Bala},
  \bibinfo{person}{Tianyin Xu}, {and} \bibinfo{person}{Yuanyuan Zhou}.}
  \bibinfo{year}{2014}\natexlab{}.
\newblock \showarticletitle{EnCore: Exploiting System Environment and
  Correlation Information for Misconfiguration Detection}. In
  \bibinfo{booktitle}{\emph{Proc. Int'l Conf. Architectural Support for
  Programming Languages and Operating Systems (ASPLOS)}} (Salt Lake City, UT,
  USA). \bibinfo{publisher}{ACM}, \bibinfo{address}{New York, NY, USA},
  \bibinfo{pages}{687--700}.
\newblock


\bibitem[\protect\citeauthoryear{Zhang and Ernst}{Zhang and Ernst}{2014}]%
        {ZE:ICSE14}
\bibfield{author}{\bibinfo{person}{Sai Zhang} {and} \bibinfo{person}{Michael~D.
  Ernst}.} \bibinfo{year}{2014}\natexlab{}.
\newblock \showarticletitle{Which Configuration Option Should I Change?}. In
  \bibinfo{booktitle}{\emph{Proc. Int'l Conf. Software Engineering (ICSE)}}
  (Hyderabad, India). \bibinfo{publisher}{ACM}, \bibinfo{address}{New York, NY,
  USA}, \bibinfo{pages}{152--163}.
\newblock


\bibitem[\protect\citeauthoryear{Zhang and Ernst}{Zhang and Ernst}{2015}]%
        {ZE:ISSTA15}
\bibfield{author}{\bibinfo{person}{Sai Zhang} {and} \bibinfo{person}{Michael~D.
  Ernst}.} \bibinfo{year}{2015}\natexlab{}.
\newblock \showarticletitle{Proactive Detection of Inadequate Diagnostic
  Messages for Software Configuration Errors}. In
  \bibinfo{booktitle}{\emph{Proc. Int'l Symp. Software Testing and Analysis
  (ISSTA)}} (Baltimore, MD, USA). \bibinfo{publisher}{ACM},
  \bibinfo{address}{New York, NY, USA}, \bibinfo{pages}{12--23}.
\newblock


\bibitem[\protect\citeauthoryear{Zhu, Liu, Guo, Bao, Ma, Liu, Song, and
  Yang}{Zhu et~al\mbox{.}}{2017}]%
        {ZLGBMLSY:SOCC17}
\bibfield{author}{\bibinfo{person}{Yuqing Zhu}, \bibinfo{person}{Jianxun Liu},
  \bibinfo{person}{Mengying Guo}, \bibinfo{person}{Yungang Bao},
  \bibinfo{person}{Wenlong Ma}, \bibinfo{person}{Zhuoyue Liu},
  \bibinfo{person}{Kunpeng Song}, {and} \bibinfo{person}{Yingchun Yang}.}
  \bibinfo{year}{2017}\natexlab{}.
\newblock \showarticletitle{BestConfig: Tapping the Performance Potential of
  Systems via Automatic Configuration Tuning}. In
  \bibinfo{booktitle}{\emph{Proc. Symposium Cloud Computing (SoCC)}} (Santa
  Clara, CA, USA). \bibinfo{publisher}{ACM}, \bibinfo{address}{New York, NY,
  USA}, \bibinfo{pages}{338--350}.
\newblock


\end{thebibliography}

\end{document}